\documentclass[a4paper,11pt]{article}

\usepackage{jheppub}

\makeatletter

\newcommand{\Rmnum}[1]{\expandafter\@slowromancap\romannumeral #1@}

\makeatother

\usepackage{graphicx}
\usepackage{amsmath}
\usepackage{amssymb}
\usepackage{bm}
\usepackage{bbm}
\usepackage{color}
\usepackage{slashed}
\usepackage{soul}
\usepackage{ulem}
\usepackage{booktabs}
\usepackage{tabularx}
\usepackage{subcaption}
\usepackage{tabularx}
\usepackage{makecell}

\title{New Determinations of the Charm and Bottom Quark Masses Using QCD Quarkonium Sum Rules}

\author[a]{Qing Yu,}
\emailAdd{yuq@swust.edu.cn}

\author[a]{Hua Zhou,}
\emailAdd{zhouhua@swust.edu.cn}

\author[b,c]{and Xing-Gang Wu}
\emailAdd{wuxg@cqu.edu.cn}

\affiliation[a]{School of Mathematics and Physics, Southwest University of Science and Technology, \\Mianyang 621010, China}
\affiliation[b]{Department of Physics, Chongqing University, \\Chongqing 401331, People's Republic of China}
\affiliation[c]{Chongqing Key Laboratory for Strongly Coupled Physics, \\Chongqing 401331, P.R. China}

\abstract{We reanalyze the perturbative QCD (pQCD) corrections to quarkonium QCD sum rules and extract the heavy quark masses $\overline{m}_{q}(\overline{m}_{q})$ ($q=c,b$). At present, the pQCD corrections to the correlation functions of two heavy-quark pseudoscalar and vector currents at zero momentum transfer, denoted as $M_{n,q}^{X,\rm th}$ ($X = P, V$), are calculated up to the $\mathcal{O}(\alpha_s^3)$ order. These corrections exhibit significant renormalization scheme and scale dependence, which introduces large theoretical uncertainties and deteriorates the precision of heavy quark mass determinations. In this work, we eliminate the renormalization scheme and scale ambiguities in the perturbative part of $M_{n,q}^{X,\rm th}$ by adopting the Principle of Maximum Conformality (PMC) within the characteristic operator (CO) approach. The CO approach, a novel extension of the standard PMC procedure, simultaneously determines the effective coupling $\alpha_s(Q_*)$ and the effective quark mass $\overline{m}_q(Q_*)$. It systematically absorbs the nonconformal $\{\beta_i\}$-terms and $\{\gamma_i\}$-terms via the renormalization group equations, yielding a strictly scheme- and scale-independent conformal perturbative series. Based on the improved PMC conformal series, we further provide reliable estimates for the unknown $\mathrm{N^4LO}$ contributions using the Pad\'e approximation method. {The final predicted heavy quark masses in the $\overline{\mathrm{MS}}$ scheme read: $\overline{m}_c(\overline{m}_c)=1278.8\pm 3.9~\text{MeV}$, extracted from the second moment of the charmed pseudoscalar correlator $M_{2,c}^{P}$; and $\overline{m}_b(\overline{m}_b) = 4161.0\pm 9.0~\text{MeV}$, extracted from the second moment of the bottom vector correlator $M_{2,b}^{V}$. The quoted uncertainties include those from input parameters such as $\Delta\langle \alpha_s G^2 \rangle$, $\Delta\alpha_s(m_Z)$, and $\Delta m_q$, the uncertainties originating from experimental measurements and lattice QCD calculations, as well as the uncertainties comes from unknown N$^4$LO ($\Delta_{\rm UHO}$). Both results agree well with the PDG world averages with deviations smaller than $2\sigma$.}}

\begin{document}

\maketitle

\flushbottom

\section{Introduction}
The quark masses and the strong coupling constant $\alpha_s$ are very important input parameters of the Standard Model. Their precise determinations are used for testing the validity of the Standard Model, as well as for indirectly searching for new physics. Due to the confinement of QCD, the quark masses and $\alpha_s$ are not directly measurable, and their values are extracted through other hadronic properties, making their values scheme-dependent. Quarks are conventionally separated into heavy quarks ($c, b, t$) and light quarks ($u, d, s$) based on their masses, with a boundary approximately at 1 GeV. In the low energy region, close to the threshold of spontaneous chiral symmetry breaking, one applies the chiral perturbation theory~\cite{Weinberg:1978kz, Gasser:1983yg, Pich:1995bw} to ascertain the mass ratio of light quarks~\cite{Weinberg:1977hb, Andersen:2023ivj}.  And the Lattice QCD provides a systematic method to determine the quark masses and $\alpha_s$ near the asymptotic scale, cf., the recent reviews~\cite{DelDebbio:2021ryq, FLAG:2021npn}. In the high energy region, the nonperturbative effects including chiral symmetry breaking become smaller, and the dominant radiative corrections in the theoretical calculations is computed within the framework of perturbative QCD (pQCD). Due to the asymptotic freedom inherent in QCD~\cite{Gross:1973id, Politzer:1973fx}, the strong interaction between quarks and gluons weakens, resulting in a sufficiently small $\alpha_s$ value to justify a perturbative expansion of physical observables. In doing so, one usually extracts the heavy quark masses and $\alpha_s$-value from experimental data by matching them with high-precision pQCD predictions.

One of the most important methods for determining heavy quark masses $m_q$ (with $q=c,b$) is the analysis of the low-energy moments of heavy-quark current correlators. For the vector channel ($V$-channel), the moments $M^{V}_{n,q}$ (with $n\geq 1$) can be compared with the weighted integrals of the experimentally measured $R_{q\bar{q}}(s)$-ratio~\cite{Novikov:1977dq} to extract the charm and bottom quark masses~\cite{Shifman:1978bx, Reinders:1984sr, Kuhn:2001dm, Kuhn:2007vp}. For the pseudoscalar channel ($P$-channel), the lattice simulations provide a competitive approach for determining the charm quark mass and the strong coupling $\alpha_s$ via the moments $M^{P}_{n,c}$ (with $n\geq 0$)~\cite{HPQCD:2008kxl, McNeile:2010ji, Nakayama:2016atf, Maezawa:2016vgv, Petreczky:2019ozv, Petreczky:2020tky}. 

Theoretically, the moments for $V$-channel and $P$-channel are related to the correlation function of heavy quark vector or pseudoscalar currents at zero momentum transfer ($s=0$)~\cite{Shifman:1978bx,Shifman:1978by}, i.e.,
\begin{eqnarray}
M^{V}_{n,q}&=&\frac{12\pi^2Q^2_q}{n!}\frac{{ d}^n}{{ d}s^{n}}\Pi_V(s)|_{s=0},
\label{thV}\\
M^{P}_{n,q}&=&\frac{12\pi^2Q^2_q}{n!}\frac{{ d}^n}{{ d}s^{n}}\left[\frac{\Pi_P(s)-\Pi_P(0)-s\Pi^{\prime}_P(s)}{s^2}\right]\bigg|_{s=0},\label{thP}
\end{eqnarray}
where $Q_q$ is the heavy quark electric charge and $\sqrt{s}=\sqrt{p^2}$ is the center-of-mass energy of the $e^+e^-$ system. The $\Pi_{V}$-function is satisfying
\begin{eqnarray}
(g_{\mu\nu}s-p_\mu p_\nu)\Pi_V(s) &=&-i\int dx e^{ip\cdot x}\left<0|T\{j_\mu(x)j_\nu(0)\}|0\right>,
\end{eqnarray}
where $j_\mu(x)=\bar{q}(x)\gamma^\mu q(x)$, and the $\Pi_{P}$-function has a similar formalization of
\begin{eqnarray}
\Pi_P(s) &=&i\int dx e^{ip\cdot x}\left<0|T\{j_P(x)j_P(0)\}|0\right>,
\end{eqnarray}
where $j_P(x)=2im_q\bar{q}(x)\gamma^5 q(x)$. For small $n$ of the moments $M^{X}_{n,q}$ ($X$ is for $V$ or $P$), the typical scale $m_q/n>>\Lambda_{\rm QCD}$ (the QCD asymptotic scale), so the theoretical moments $M^{X,\rm th}_{n,q}$ can be evaluated under the pQCD theory. In the calculations of Operator Product Expansion (OPE), the theoretical moments $M^{X,\rm th}_{n,q}$ are separated into two parts: perturbative term $M^{X,\rm pert}_{n,q}$ and non-perturbative term $M^{X, \rm n.p.}_{n,q}$, i.e., 
\begin{eqnarray}
M^{X,\rm th}_{n,q}=M^{X,\rm pert}_{n,q}+M^{X,\rm n.p.}_{n,q}.
\end{eqnarray}
Under the $\overline{\rm MS}$-scheme, the $\mathcal{O}(\alpha_s)$ correction to the moments $M^{X,\rm pert}_{n,q}$ has been computed for arbitrary $n$~\cite{Kallen:1955fb}. Refs.~\cite{Chetyrkin:1995ii, Chetyrkin:1996cf, Boughezal:2006uu, Czakon:2007qi, Maier:2007yn} has completed the $\mathcal{O}(\alpha^2_s)$ correction for high $n$ values. Furthermore, the $\mathcal{O}(\alpha^3_s)$ correction has been obtained for $n=0$ and $n$=1~\cite{Chetyrkin:2006xg, Boughezal:2006px}, $n$=2~\cite{Maier:2008he}, $n$=3~\cite{Hoang:2008qy, Kiyo:2009gb, Maier:2009fz, Maier:2017ypu}, as well as for higher $n$ values~\cite{Hoang:2008qy, Kiyo:2009gb, Greynat:2010kx, Boito:2021wbj}. Recently, the theoretical moments have been calaulated up to $\mathcal{O}(\alpha^4_s)$-level in the large $\beta_0$ limit~\cite{Boito:2021wbj}. 

It is well known that the pQCD prediction becomes unreliable in the region where the characteristic momentum flow approaches $\Lambda_{\rm QCD}$. Theoretical predictions for charm and bottom quark masses in this nonperturbative region are therefore typically provided by non-relativistic sum rules~\cite{Signer:2008da, Signer:2007dw}. In our analysis of the moments $M^{X,\rm th}_{n,q}$, these considerations dictate the selection of valid $n$ values to ensure the stability of the perturbative expansion. For instance, with $\overline{m}_c(\overline{m}_c)/n\simeq318$ MeV for the charm quark, which close to $\Lambda_{\rm QCD}$, we restrict the analysis to $M^{X,\rm th}_{n,c}$ with $n\leq3$ and $M^{X,\rm th}_{n,b}$ with $n\leq4$.

In pQCD calculations of $M^{X,\rm pert}_{n,q}$ under $\overline{\rm MS}$-scheme, the renormalization scale $\mu_r$ is introduced to handle the ultraviolet (UV) divergences. Specifically, the scale $\mu_r$ is used to cancel large logarithms arising from loop-momentum integrations. In an expansion of running quark mass $\overline{m}_q(\mu_m)$ and strong coupling $\alpha_s(\mu_r)$, an additional mass renormalization scale $\mu_m$ is introduced. Although the dependence on these artificial scales would cancel exactly if the perturbative series were calculated to all orders, such dependence persists in practice due to the truncation of the series at a finite order. Consequently, the renormalization scale dependence remains in the truncated pQCD series, manifesting in powers of logarithmic terms $\ln\mu^2_r/Q^2$ that are proportional to the divergent renormalon terms $n!\alpha^n_s\beta^n_0$~\cite{Gross:1974jv, Lautrup:1977hs, Beneke:1998ui}. The conventional method for estimating this theoretical uncertainty is to set $\mu_r=\mu_m=Q$, where $Q$ represents the typical momentum flow of the process, and then vary this scale within the range $[Q/\xi,\xi Q]$ ($\xi$ =2,3,4, or etc.). For the physical processes with a large typical scale $Q$, the scale dependence would be suppressed by powers of small $\alpha_s(Q)$. In contrast, for processes involving charm and bottom quarks, the coupling $\alpha_s$ evaluated at scales $\overline{m}_{c,b}$ is relatively large, which amplifies the scale sensitivity; therefore, narrower and process-specific ranges are typically adopted~\cite{Dehnadi:2015fra, Dehnadi:2011gc, AlamKhan:2023kgs}. 

According to the principle of renormalization group invariance (RGI)~\cite{Petermann:1953wpa, Peterman:1978tb, Callan:1970yg, Symanzik:1970rt}, a physical observable must be free of the renormalization scale. The arbitrary choice of $\mu_r$ and $\mu_m$ thus not only violates RGI but also constitutes a major source of theoretical uncertainty in pQCD calculation. This persistent scale dependence in low-order predictions originates from a mismatch between perturbative coefficients and the $\alpha_s$-values at each order. The following discussion focuses on the method to eliminate this unphysical scale dependence and obtain reliable estimates of the quark-mass parameters.

The principle of maximum conformality approach (PMC) is suggested to eliminate such renormalization scale ambiguity at fixed order~\cite{Brodsky:2011ta, Mojaza:2012mf, Brodsky:2012rj, Brodsky:2013vpa}. Its foundation lies in the renormalization group equation (RGE), which indicates that the scale dependence of the coupling $\alpha_s$ is governed by the QCD $\beta$-function. Therefore, the PMC procedure identifies and observes all RGE-related $\{\beta_i\}$-terms into the determination of the PMC scales, which successfully avoiding arbitrary scale choices. The resulting scheme-ivariant conformal series is independent of the initial renormalization scale and satisfies RGI. Moreover, the absorption of the $\{\beta_i\}$-terms also eliminates the associated factorially divergent renormalon contributions, leading to the PMC conformal series with improved convergence. It has been demonstrated that PMC predictions are also independent of the choice of renormalization scheme~\cite{Wu:2014iba, Wu:2015rga, Wu:2019mky}. 

In the PMC standard procedure, it is important to correctly absorb the nonconformal $\{\beta_i\}$-terms and thereby determine the effective coupling $\alpha_s$ at PMC scales~\cite{Yan:2023hra}. The PMC multi-scale setting approach (PMCm)~\cite{Brodsky:2011ta, Mojaza:2012mf, Brodsky:2011ig, Brodsky:2012rj, Brodsky:2013vpa} achevies this by assigning a distinct PMC scale to the coupling at each order, ensuring that different types of $\{\beta_i\}$-terms are properly resummed to achieve a scale-invariant prediction. However, the determination of the PMC scales exists residual scale uncertainty, as each effective coupling is fixed by absorbing the same type of $\{\beta_i\}$-term appearing at different orders, which is consequently influenced by the unknown higher-order (UHO) contributions. More explicitly, the last term in the perturbative expansion of each PMC scale is unknown, this ambiguity is referred as \textit{the first kind of residual scale dependence}~\cite{Zheng:2013uja}. And the PMC scale at the last order can not be strictly fixed since the one higher-order $\{\beta_i\}$-term is uncalculated, which is called \textit{the second kind of residual scale dependence}~\cite{Zheng:2013uja}. The two kinds of residual scale dependence caused by the UHO contributions generally can be suppressed by the powers of small $\alpha_s$-value, as the multi PMC scales are of perturbative. But there may be sizable residual scale uncertainty appearing in the charm region. To suppress such residual scale dependence, we suggest applying the PMC single-scale (PMCs) approach~\cite{Shen:2017pdu, Wu:2018cmb}. The PMCs suggests absorbing all the types of RGE-related $\{\beta_i\}$-terms into an global effective coupling $\alpha_s(Q_*)$ with a single PMC scale $Q_*$. The PMCs method not only maintains the scheme and scale invariance but also eliminate the \textit{the second kind of residual scale dependence}. The PMCs approach has been successfully applied to determine the value of $\alpha_s$ at various characteristic scales~\cite{Shen:2023qgz, Wang:2021tak, Yu:2021yvw}. 

With the help of QCD degeneracy relations among different orders~\cite{Bi:2015wea}, the PMC method translates the $n_f$-terms of the pQCD series into RGE-related $\{\beta_i\}$-terms. There are mixed $n_f$-terms in a general perturbative expansion of coupling $\alpha_s$ and quark mass $\overline{m}_q$. Since the PMC predictions are scheme independent, one transforms the quark mass under the $\overline{\rm MS}$-scheme into the on-shell scheme to avoid the misuse of $n_f$-terms~\cite{Huang:2020rtx, Yu:2018hgw, Wang:2013akk}. However, the extraction of quark masses is typically performed using physical processes calculated within the $\overline{\rm MS}$-scheme, such as the heavy quark related processes. The scale-dependent behavior of the coupling $\alpha_s$ and quark mass $\overline{m}_q$ are satisfying the RGE with QCD $\beta(\alpha_s)$-function and the mass anomalous dimension $\gamma_m(\alpha_s)$-function, respectively. Thus the PMC proposes the new degeneracy relations~\cite{Ma:2024xeq, Huang:2022rij} and employed it to extract the top-quark and bottom-quark masses under both the on-shell and $\overline{\text{MS}}$ schemes, which providing more precise and scale-independent results. 

Recently, an improved PMC procedure has indicated that the effective coupling $\alpha_s$ and $\overline{\rm MS}$ quark mass can be determined under the same scheme, which is the PMC characteristic operator (CO) approach~\cite{Yan:2024oyb}. More specifically, the CO provides the scale transformations for composite quantities of the form $\overline{m}^{n_\gamma}_q\alpha^{n_\beta}_s$. In doing so, all nonconformal contributions can be consistently resummed into a global effective coupling $\alpha_s(Q_*)$ and an effective quark mass $\overline{m}_q(Q_*)$ with PMC scale $Q_*$. The PMC scale is uniquely determined by systematically absorbing the $\{\beta_i\}$-terms and $\{\gamma_i\}$-terms at each perturbative order under the $\overline{\rm MS}$ scheme~\cite{Huang:2022rij, Yan:2024oyb}. As a result, the perturbative coefficients become conformal, and the resulting series satisfies the RGI at each fixed order. And the PMC scale-invariant series are also valuable for estimating the UHO term~\cite{Du:2018dma, Shen:2022nyr}.

The remainder of this paper is organized as follows. Sec.\ref{sec2} presents all the calculation techniques for the moments $M^{X,\rm th}_{n,q}$ with small $n$. We apply the PMC approach to the pQCD part $M^{X,\rm pert}_{n,q}$ and explain the non-perturbative part $M^{X,\rm n.p.}_{n,q}$. Sec.\ref{sec3} provides the numerical results and a detailed discussion on the extraction of the quark masses $\overline{m}_{q}(\overline{m}_{q})$. Finally, our conclusions are summarized in Sec.\ref{sec4}. 

\section{Calculation technology}
\label{sec2}

In the following, we first analyze the pQCD contributions to the moments $M^{X,\rm th}_{n,q}$ using the standard PMC formulas, and subsequently introduce an estimation of the nonperturbative contributions to these moments. 

\subsection{Perturbative series of the moments $M^{X,\rm th}_{n,q}$}

Under the $\overline{{\rm MS}}$-scheme, a general expression of the perturbative part $M^{X,\rm pert}_{n,q}$ up to $\mathcal{O}(a^3_s)$ can be written as
\begin{eqnarray}
M^{X,\rm pert}_{n,q}=\frac{9}{4}Q^2_q\left[2\overline{m}_q(\mu_m)\right]^{n_\gamma}a^{n_\beta}_s(\mu_r)\left[\sum^{3}_{i=0}r^X_{i}a^{i}_{s}(\mu_r)\right],
\label{Mnq}
\end{eqnarray}
where $a_s=\alpha_s/\pi$, and the indicators of $\overline{m}_q$ and $a_s$ are $n_\gamma=-2n$ and $n_\beta=0$. The perturbative coefficients $r^{X}_{i}$ for $M^{X,\rm pert}_{n,q}$ ($n=1,2,3,4$) within the ${\rm \overline{MS}}$-scheme have been presented in Refs.~\cite{Maier:2009fz, Maier:2017ypu}; further details are given in Appendix~\ref{appA}. The renormalization scale $\mu_r$ and $\mu_m$ are associated with the running coupling $\alpha_s(\mu_r)$ and the ${\rm \overline{MS}}$ running quark mass $\overline{m}_q(\mu_m)$, respectively.

To apply the PMC CO approach, we first rewrite the Eq.(\ref{Mnq}) by converting the $n_f$-series into $\{\beta_i\}-$ and $\{\gamma_i\}$-series. The perturbative coefficients $r_i$ are 
\begin{eqnarray}
r^X_0&=&r^X_0,\\
r^X_1&=&r^X_{1,0},\\
r^X_2&=&r^X_{2,0}+d_0^{[-2n,0;1,1]}r^X_{2,1},\\
r^X_3&=&r^X_{3,0}+d_1^{[-2n,0;1,1]} r^X_{2,1}
+d_0^{[-2n,0;1,2]} r^X_{3,1}+{1\over2!}d_0^{[-2n,0;2,1]} r^X_{3,2}, 
\end{eqnarray}	
where the coefficients $d_i^{[n_\gamma,n_\beta;k,l]}$ are functions of $\{\beta_i\}$ and $\{\gamma_i\}$~\cite{Yan:2024oyb}. The required coefficients for $M^{X,\rm pert}_{n,q}$ up to $\rm N^3LO$ are
\begin{eqnarray}
	d_i^{[n_\gamma,n_\beta;1,l]}&=&n_\gamma\gamma_i+(n_\beta+l)\beta_i,\\
	d_0^{[n_\gamma,n_\beta;2,l]}&=&\big[n_\gamma\gamma_0+(n_\beta+l+1)\beta_0\big]\big[n_\gamma\gamma_0+(n_\beta+l)\beta_0\big].
\end{eqnarray}

The coefficients $r^X_{i, {j\neq0}}$ are functions of the logarithm $\ln\mu_r^2/ \overline{m}^2_q(\mu_r)$, where $\mu_r$ denotes the renormalization scale~\footnote{If setting $\mu_r=\overline{m}_q(\overline{m}_q)$, all such logarithmic terms vanish, thereby yielding a renormalon-free and more convergent pQCD series. This explains why the conventional choice of $\mu_r=\overline{m}_q(\overline{m}_q)$ is usually regarded as the optimal scale.}. Moreover, those coefficients can be reformulated in the following form,
\begin{eqnarray}
r^X_{i,j} = \sum_{k=0}^{j} C_j^k \hat{r}^X_{i-k,j-k}\ln^k\frac{\mu_r^2}{\overline{m}^2_q(\mu_r)} ,
\label{rij}
\end{eqnarray}
where the combination coefficients $C_j^k={j!}/{k!(j-k)!}$, and the coefficients $\hat{r}^X_{i,j}=r^X_{i,j}|_{\mu_r=\overline{m}_q(\overline{m}_q)}$. The conformal coefficients $\hat{r}^X_{i,0}=r^X_{i,0}$ are scale-independent. For convenience, we present the conversion relationship between the perturbative coefficients $\hat{r}^X_{i,j}$ and the $n_f$-dependent coefficients $\mathcal{C}^X_{i,j}$ in Appendix \ref{appB}.

As RGI requires that a physical observe should be scale-invariant, i.e.,
\begin{eqnarray}
	\frac{\partial}{\partial\ln\mu^2}M^{X,\rm pert}_{n,q}=0,
\end{eqnarray}
which indicates
\begin{eqnarray}
\sum^{\infty}_{i=0}a^i_s\left[\frac{\partial}{\partial\ln\mu^2}+\gamma_m\overline{m}_{q}\frac{\partial}{\partial\overline{m}_q}\right]r_i=
-\sum^{\infty}_{i=0}r_i\left[n_\gamma\gamma_m+n_\beta\frac{\beta}{a_s}+\beta\frac{\partial}{\partial a_s}\right]a^{i}_s.
\label{RGI2}
\end{eqnarray}
The $\beta$-function and $\gamma_m$-function govern the scale $\mu$-dependence of the coupling $\alpha_s$ and quark mass $\overline{m}_q$, respectively, through the follows RGEs:
\begin{eqnarray}
\beta(\alpha_s)&=&{\partial \alpha_s\over\partial\ln\mu^2}=-\sum^j_{i=0}{\beta_i}{\alpha^{i+2}_s},\label{RGE1}\\
\gamma_m(\alpha_s)\overline{m}_{q}&=&{\partial \overline{m}_{q}\over\partial\ln\mu^2}=-\overline{m}_{q}\sum^j_{i=0}{\gamma_i}\alpha^{i+1}_s,
\label{RGE2}	
\end{eqnarray}
where the $\{\beta_i\}$-and-$\{\gamma_i\}$-functions have been known up to $i=4$ under $\overline{\rm MS}$-scheme~\cite{Chetyrkin:1997dh, Vermaseren:1997fq, Czakon:2004bu, Chetyrkin:2004mf, Baikov:2014qja, Baikov:2016tgj, Herzog:2017ohr}. In Eq.(\ref{RGI2}), the operator applied to the perturbative coefficients $r_i$ on the left is matched by the operator applied to the coupling $a_s$ on the right side. Based on this correspondence, the PMC CO method defines a characteristic operator $\hat{D}_{n_\gamma,n_\beta}$~\cite{Yan:2024oyb}. By acting upon the coupling $a^i_s$, this operator systematically determines the scale dependence of the coefficients $r_i$ through the $\hat{D}_{n_\gamma,n_\beta}[a^i_s]$. Formally, the PMC CO $\hat{D}_{n_\gamma,n_\beta}$ is defined as 
\begin{eqnarray}
	\hat{D}_{n_\gamma,n_\beta}=n_\gamma\gamma_m(\alpha_s)+n_\beta\frac{\beta(\alpha_s)}{\alpha_s}+\beta(\alpha_s)\frac{\partial}{\partial\alpha_s}.
\end{eqnarray}
The CO $\hat{D}_{n_\gamma,n_\beta}$ provides a general description of the scale evolution of the combination between coupling $\alpha_s$ and quark mass $\overline{m}_q$, 
\begin{eqnarray}
{\partial^k [\overline{m}^{n_\gamma}_{q}(\mu)\alpha^{n_\beta}_s(\mu)]\over\partial(\ln\mu^2)^k}&=&\overline{m}^{n_\gamma}_{q}(\mu)\alpha^{n_\beta}_s(\mu)\hat{D}^k_{n_\gamma,n_\beta}[1],
\end{eqnarray}
where the operator acts on 1, and
\begin{eqnarray}
\hat{D}^k_{n_\gamma,n_\beta}[1]=\sum^k_{i=0}C^i_k\hat{D}^k_{n_\gamma,0}[1]\hat{D}^k_{0,n_\beta}[1].
\end{eqnarray}
The cases $n_\gamma=0$ and $n_\beta=0$ for the operator $\hat{D}^k_{n_\gamma,n_\beta}$ correspond to the scale evolution of the coupling $\alpha_s$ and quark mass $\overline{m}_q$, respectively. 
The relation between the operator $\hat{D}^k_{n_\gamma,n_\beta}$ and the coefficients $d_i^{[n_\gamma,n_\beta;k,l]}$ is
\begin{eqnarray}
	\hat{D}^k_{n_\gamma,n_\beta}[a^l_s]=(-1)^k\sum^{\infty}_{i=0}d_i^{[n_\gamma,n_\beta;k,l]}a_s^{l+k+i}.
\end{eqnarray}
Based on the above CO equation, Eq.(\ref{Mnq}) can be rewrite as the following form:
\begin{eqnarray}
M^{X,\rm pert}_{n,q}&=&{9\over4}{Q^2_q\over[2\overline{m}_q(\mu_r)]^{2n}}\Big\{r^X_0+\sum^{3}_{i=1} r^X_{i,0}a^i_s(\mu_r) \nonumber\\
&+&\sum^{3}_{i=2}\sum^{i-1}_{j=1}\frac{(-1)^j}{j!}\sum^{j}_{k=0}C^{k}_{j}\hat{r}_{i-k,j-k}\ln^k\frac{\mu^2_r}{\overline{m}^2_q(\mu_r)}\hat{D}^{j}_{-2n,0}\left[a^{i-j}_s(\mu_r)\right]\Big\}.
\label{MnXco}
\end{eqnarray}

Under the PMC CO method, all the RGE-involved non-conformal $\{\beta_i\}$-terms and $\{\gamma_i\}$-terms eliminate by adopting an overall effective coupling constant $\alpha_s(Q_*)$ and quark mass $\overline{m}_q(Q_*)$. In other words,
\begin{eqnarray} \sum^{3}_{i=2}\sum^{i-1}_{j=1}\frac{(-1)^j}{j!}\sum^{j}_{k=0}&&C^{k}_{j}\hat{r}_{i-k,j-k}\ln^k\frac{Q^2_*}{\overline{m}^2_q(Q_*)}\times\hat{D}^{j}_{-2n,0}\left[a^{i-j}_s(Q_*)\right]=0.
\label{Qstar}
\end{eqnarray}
This yields a pQCD series with the following scheme-independent form
\begin{eqnarray}
M^{X,\rm pert}_{n,q}|_{\rm PMC}&=&{9\over4}{Q^2_q\over[2\overline{m}_q(Q_*)]^{2n}}\Big\{r^X_0+r^X_{1,0}a_s(Q_*)+r^X_{2,0}a_s^2(Q_*)+r^X_{3,0}a_s^3(Q_*)\Big\},
\label{MXnpmc}
\end{eqnarray}
where $Q_*$ is the PMC scale. Specifically, by applying the scale displacement relations to shift the scale from $\mu_r$ to $Q_*$ in Eq.(\ref{MnXco}), the coefficients of the nonconformal $\hat{r}^X_{i,j}$-terms are utilized to determine the magnitude of PMC scale $Q_*$. The scale evolution of the couplings and running quark masses between two distinct scales are presented in Appendix \ref{appC}. Numerical solution of $Q_*$ in Eq.(\ref{Qstar}) fixes the effective coupling constant $\alpha_s(Q_*)$ and effective quark mass $\overline{m}_q(Q_*)$, thereby avoiding \textit{the second kind of residual scale dependence} inherent in asymptotic approximations. Consequently, the resulting conformal series of coefficients $r^{X}_{i,0}$, $a_s(Q_*)$, and $\overline{m}_q(Q_*)$, becomes independent of the choice of $\mu_r$. As a byproduct, due to the elimination of divergent renormalon terms, the resultant PMC conformal series in Eq.~(\ref{MXnpmc}) exhibits the natural convergence of the perturbative series. In Sections \ref{sec3.1} and \ref{sec3.2}, we present the numerical results of the theoretical moments $M^{X,\rm th}_{n,q}$ before and after applying the PMC CO method.

\subsection{The non-perturbative terms $M^{X,\rm n.p.}_{n,q}$}

\begin{table}[htb]
\begin{center}
\begin{tabular}{c c c c c c c c c c c c}
\toprule
  & \multicolumn{4}{c}{Charm case} & \multicolumn{4}{c}{Bottom case} \\
\cmidrule(lr){2-5} \cmidrule(lr){6-9}
 $n$ & $a^{P,0}_{n,c}$ & $a^{P,1}_{n,c}$ & $a^{V,0}_{n,c}$ & $a^{V,1}_{n,c}$ & $a^{P,0}_{n,b}$ & $a^{P,1}_{n,b}$ & $a^{V,0}_{n,b}$ & $a^{V,1}_{n,b}$\\ 
\midrule
$0$ & 14.037 & $74.145$ & - & -   &3.509 &19.516 & - & -  \\ 
$1$ & 8.021 & 39.144 & $-16.042$ & $-143.364$  &2.005 &10.346 & $-4.011$ & $-36.960$   \\ 
$2$ & 0 & $-36.384$ & $-26.737$ & $-272.186$  &0 &$-9.096$ & $-6.684$ & $-69.912$   \\ 
$3$ & $-9.722$ & $-152.670$ & $-38.890$ & $-439.82$  &$-2.431$ &$-38.846$ & $-9.722$ & $-112.669$  \\ 
$4$ & - & - & - & -  & - & - & $-13.088$ & $-165.326$  \\ 
\bottomrule
\end{tabular}
\end{center}
\caption{Gluon condensate coefficients $a^{X,0}_{n,q}$ and $a^{X,1}_{n,q}$ for the nonperturbative contribution $M^{X,\mathrm{n.p.}}_{n,q}$, where $q=(c \;{\rm or}\; b)$, and $X=(P\; {\rm or}\;  V)$. Only the coefficients relevant to the present calculations are listed.}
\label{anX}
\end{table}

The non-perturbative part of the moments $M^{X,{\rm n.p.}}_{n,q}$ receives contributions from the gluon condensate and higher-order power-suppressed condensates, in the form
\begin{eqnarray}
M^{X,\rm n.p.}_{n,q}=\frac{\left<\alpha_sG^2\right>}{(2M_q)^{2n+4}} \left[a^{X,0}_{n,q}+a^{X,1}_{n,q}{\alpha_s(\mu_r)\over\pi}\right],	
\end{eqnarray}
where the gluon condensate $\left<\alpha_sG^2\right>$ is a RGI quantity. We adopt $\left<\alpha_sG^2\right>=0.006\pm0.012 {\rm GeV}^4$~\cite{Ioffe:2005ym} in the following discussions. The coefficients $a^{X,0}_{n,q}$ and $a^{X,1}_{n,q}$ are known up to NLO level~\cite{Narison:1983kn, Broadhurst:1994qj}, and all the calculated results are summarized in Table~\ref{anX}. As discussed in the introduction, in Table~\ref{anX}, we adopt $n\geq 0$ for $X=P$ and $n\geq 1$ for $X=V$; additionally, $n\leq 3$ for the charm case and $n\leq 4$ for the bottom case. As a subtle point, the perturbative coefficients $r^X_i$ for the bottom pseudoscalar moments at $n=4$ remain unknown. Therefore, we only consider the moments $M^{P,\mathrm{pert}}_{n\leq 3,b}$. 

The pole mass $M_q$ is related to the $\overline{\rm MS}$-scheme mass, and the relationship has been calculated up to four loop level~\cite{Marquard:2015qpa, Marquard:2016dcn}. To assist with the present nonperturbative coefficients $a^{X,i}_{n,q}$, we employ the one-loop relation between the on-shell mass $M_q$ and $\overline{\rm MS}$-scheme mass,
\begin{eqnarray}
M_q=\overline{m}_q(\mu_m)\left[1+{\alpha_s(\mu_r)\over\pi}\left({4\over3}-\ln{\overline{m}^2_q(\mu_m)\over\mu^2_r}\right)\right],
\end{eqnarray}
where the $\overline{\rm MS}$-scheme charm quark mass $\overline{m}_q(\mu_m)$ with mass-related renormalization scale $\mu_m$. 

\section{Numerical results}
\label{sec3}

The solution for the running behavior of $\alpha_s$ derived from the RGE is expressed as an expansion in powers of the logarithm $l_\Lambda=\ln{\mu^2/\Lambda^2_{\rm QCD}}$, with the asymptotic scale $\Lambda_{\rm QCD}$. As reported in Ref.~\cite{Chetyrkin:1997sg}, the expression of $a_s(\mu)$ under $\overline{\rm MS}$-scheme up to the four-loop is
\begin{eqnarray}
a^{\overline{\rm MS}}_s(\mu)&=&{1\over\beta_0l_\Lambda}\bigg\{1-{\beta_1\over\beta^2_0}{\ln{l_\Lambda}\over {l_\Lambda}}+{1\over\beta^4_0{l^2_\Lambda}} 
\times\left[{\beta^2_1}(\ln^2l_\Lambda
-\ln{l_\Lambda}\right.\nonumber\\&&\left. 
-1)+\beta_0{\beta^{\overline{\rm MS}}_2} \right] 
-{1\over\beta^6_0l^3_\Lambda} \left[{{\beta^{3}_1}\over2}(2\ln^3{l_\Lambda}
-{5}\ln^2{l_\Lambda}
\right.\nonumber\\&&\left. -4\ln{l_\Lambda}
+1)
+3{\beta_0\beta_1\beta^{\overline{\rm MS}}_2}\ln{l_\Lambda}-{\beta^2_0\beta^{\overline{\rm MS}}_3\over2}\right] \bigg\},
\label{alphas}
\end{eqnarray}
where the first two $\beta$-function coefficients are $\beta_0 = {(11 - 2n_f/3)}/{4}$ and $\beta_1 = {(102 - 38n_f/3)}/{16}$,
with $n_f$ denoting the number of active quark flavors. The higher-order coefficients $\{\beta^{\rm \overline{MS}}_{i\geq2}\}$, which are scheme-dependent, can be found in Refs.~\cite{Czakon:2004bu, Chetyrkin:2004mf}. 

As a fundamental parameter, $\Lambda_{\text{QCD}}$ plays a pivotal role in defining the cutoff for divergent integrals encountered in pQCD calculations. Typically, the value of $\Lambda_{\text{QCD}}$ is determined by fixing $\alpha_s$ at a reference scale---such as the mass of the $Z$ boson, $m_Z = 91.1876$ GeV---within a specified renormalization scheme. Using the world-average value from the Particle Data Group, $\alpha_s^{\overline{\text{MS}}}(m_Z) = 0.1180 \pm 0.0009$~\cite{ParticleDataGroup:2024cfk}, we obtain $\Lambda_{\overline{\text{MS}}}\big|_{n_f=3} = 344 \pm 14 \ \text{MeV}$, $\Lambda_{\overline{\text{MS}}}\big|_{n_f=4} = 293 \pm 13 \ \text{MeV}$, and $\Lambda_{\overline{\text{MS}}}\big|_{n_f=5} = 208^{+11}_{-10} \ \text{MeV}$. As for the running quark mass, we adopt~\cite{ParticleDataGroup:2024cfk}:
\begin{eqnarray}
\overline{m}_c(\overline{m}_c)&=&1273.0\pm4.6~{\rm MeV},  \label{PDGmcmass} \\
\overline{m}_b(\overline{m}_b)&=&4183\pm7~{\rm MeV}.   \label{PDGmbmass}
\end{eqnarray}
The electric charges of heavy quarks are $Q_c = {2}/{3}$ for the charm quark and $Q_b = -{1}/{3}$ for the bottom quark, respectively.

\subsection{Perturbative contributions to $M^{X,\mathrm{th}}_{n,c}$ up to $\mathrm{N^3LO}$ QCD corrections}  
\label{sec3.1}

The PMC scale for the perturbative part of the mements $M^{X,\text{th}}_{n,q}$ is determined by using all the RGE-involved non-conformal terms in their perturbative series, which can be obtained by solving Eq.~(\ref{Qstar}). The PMC scales for the perturbative part of the moments $M^{X,\text{pert}}_{n,c}$ are 
\begin{eqnarray}
Q^{P,0}_{*,c} &=& 1.4333(^{-0.0171}_{+0.0163})_{\Delta \alpha_s}(\pm0.0067)_{\Delta m_c}~{\rm GeV}, \\
Q^{P,1}_{*,c} &=& 1.3345(\pm0.0018)_{\Delta \alpha_s}(\pm0.0047)_{\Delta m_c}~{\rm GeV}, \\
Q^{P,2}_{*,c} &=& 1.3912(^{+0.0005}_{-0.0007})_{\Delta \alpha_s}(\pm0.0050)_{\Delta m_c}~{\rm GeV}, \\
Q^{P,3}_{*,c} &=& 1.5408(^{+0.0021}_{-0.0023})_{\Delta \alpha_s}(\pm0.0054)_{\Delta m_c}~{\rm GeV}, \\
Q^{V,1}_{*,c} &=& 1.3412(\pm0.0018)_{\Delta \alpha_s}(\pm0.0047)_{\Delta m_c}~{\rm GeV}, \\
Q^{V,2}_{*,c} &=& 1.4203(^{+0.0011}_{-0.0012})_{\Delta \alpha_s}(\pm0.0050)_{\Delta m_c}~{\rm GeV}, \\
Q^{V,3}_{*,c} &=& 1.6416(\pm0.0055)_{\Delta \alpha_s}(\pm0.0054)_{\Delta m_c}~{\rm GeV}.
\end{eqnarray}
Here the first error originates from $\Delta\alpha_s(m_Z) = \pm 0.0009$, and the second error arises from $\Delta\overline{m}_c(\overline{m}_c) = \pm 4.6\,\mathrm{MeV}$. The PMC scale $Q_{*}$ characterizes the underlying momentum flow of each moment, and determines the effective values of the running coupling $\alpha_s(Q_*)$ and the running quark mass $\overline{m}_q(Q_*)$ for the relevant quantities. 

The corresponding PMC results for the moments $M^{X,\mathrm{th}}_{n,c}|_{\mathrm{PMC}}$ are as follows:
\begin{eqnarray}
M^{P,\rm th}_{0,c}|_{\rm PMC} &=& 1.7357(^{+0.0137}_{-0.0128})_{\Delta \alpha_s}(\mp0.0012)_{\Delta m_c}(\pm0.0040)_{\Delta{\rm n.p.}}, \\
M^{P,\rm th}_{1,c}|_{\rm PMC} &=& 1.4353(^{+0.0277}_{-0.0256})_{\Delta \alpha_s}(^{-0.0127}_{+0.0129})_{\Delta m_c}(\pm0.0024)_{\Delta{\rm n.p.}},\\
M^{P,\rm th}_{2,c}|_{\rm PMC} &=& 1.3805(^{+0.0332}_{-0.0305})_{\Delta \alpha_s}(^{-0.0226}_{+0.0231})_{\Delta m_c}(\mp0.0011)_{\Delta{\rm n.p.}},\\
M^{P,\rm th}_{3,c}|_{\rm PMC} &=& 1.3482(^{+0.0276}_{-0.0258})_{\Delta \alpha_s}(^{-0.0311}_{+0.0320})_{\Delta m_c}(^{-0.0106}_{+0.0102})_{\Delta{\rm n.p.}},\\
M^{V,\rm th}_{1,c}|_{\rm PMC} &=& 2.1765(^{+0.0166}_{-0.0159})_{\Delta \alpha_s}(^{-0.0170}_{+0.0173})_{\Delta m_c}(^{-0.0066}_{+0.0064})_{\Delta{\rm n.p.}},\\
M^{V,\rm th}_{2,c}|_{\rm PMC} &=& 1.4548(\pm0.0094)_{\Delta \alpha_s}(^{-0.0216}_{+0.0220})_{\Delta m_c}(^{-0.0154}_{+0.0148})_{\Delta{\rm n.p.}},\\
M^{V,\rm th}_{3,c}|_{\rm PMC} &=& 1.0775(^{-0.0209}_{+0.0164})_{\Delta \alpha_s}(^{-0.0212}_{+0.0217})_{\Delta m_c}(^{-0.0403}_{+0.0386})_{\Delta{\rm n.p.}},
\end{eqnarray}
where the uncertainty associated with $\Delta\mathrm{n.p.}$ is estimated from the error of the nonperturbative gluon condensate term $\Delta\langle \alpha_s G^2 \rangle = \pm 0.012\,\mathrm{GeV}^4$, and $M^{X,{\rm th}}_{n,c}$ are in units of $10^{-n}{\rm GeV}^{-2n}$. When considering the uncertainty from one error source, the other input parameters are set to be their central values. For comparison, the uncertainties in $\Delta\alpha_s(m_Z)$, $\Delta m_q$, and $\Delta\mathrm{n.p.}$ for the results under the conventional scale-setting approach with $\mu_{r,m}=\overline{m}_c(\overline{m}_c)$ are also provided:

\begin{align}
M^{P,\rm th}_{0,c}|_{\rm Conv.}&=1.7370(_{-0.0580}^{+0.0127})_{\Delta\mu_r}(_{+0.1037}^{-0.0385})_{\Delta\mu_m}(^{+0.0135}_{-0.0127})_{\Delta \alpha_s}(\pm0.0012)_{\Delta m_c}(\pm0.0036)_{\Delta\rm n.p.}\\
M^{P,\rm th}_{1,c}|_{\rm Conv.}&=1.4466(_{-0.1657}^{+0.0533})_{\Delta\mu_r}(_{-1.9420}^{-0.1084})_{\Delta\mu_m}(^{+0.0293}_{-0.0270})_{\Delta \alpha_s}(^{-0.0129}_{+0.0130})_{\Delta m_c}(\pm0.0022)_{\Delta{\rm n.p.}},\\
M^{P,\rm th}_{2,c}|_{\rm Conv.}&=1.4124(_{-0.2264}^{+0.1084})_{\Delta\mu_r}(_{-4.1483}^{-0.2199})_{\Delta\mu_m}(^{+0.0370}_{-0.0338})_{\Delta \alpha_s}(^{-0.0233}_{+0.0239})_{\Delta m_c}(\mp0.0009)_{\Delta{\rm n.p.}},\\
M^{P,\rm th}_{3,c}|_{\rm Conv.}&=1.4499(_{-0.2585}^{+0.1409})_{\Delta\mu_r}(_{-5.2345}^{-0.3441})_{\Delta\mu_m}(^{+0.0408}_{-0.0371})_{\Delta \alpha_s}(^{-0.0344}_{+0.0354})_{\Delta m_c}(\mp0.0063)_{\Delta{\rm n.p.}},\\
M^{V,\rm th}_{1,c}|_{\rm Conv.}&=2.1983(_{-0.0899}^{+0.0052})_{\Delta\mu_r}(_{-3.5448}^{-0.0958})_{\Delta\mu_m}(^{+0.0198}_{-0.0186})_{\Delta \alpha_s}(^{-0.0175}_{+0.0177})_{\Delta m_c}(\mp0.0059)_{\Delta{\rm n.p.}},\\
M^{V,\rm th}_{2,c}|_{\rm Conv.}&=1.5165(_{-0.0881}^{+0.0036})_{\Delta\mu_r}(_{-5.1771}^{-0.1503})_{\Delta\mu_m}(^{+0.0174}_{-0.0163})_{\Delta \alpha_s}(^{-0.0231}_{+0.0236})_{\Delta m_c}(\mp0.0118)_{\Delta{\rm n.p.}},\\
M^{V,\rm th}_{3,c}|_{\rm Conv.}&=1.3093(_{-0.0686}^{-0.0042})_{\Delta\mu_r}(_{-4.1764}^{-0.2004})_{\Delta\mu_m}(^{+0.0132}_{-0.0124})_{\Delta \alpha_s}(^{-0.0290}_{+0.0298})_{\Delta m_c}(\mp0.0205)_{\Delta{\rm n.p.}}.
\end{align}
Similarly, when evaluating the uncertainty from a single error source, all other input parameters are fixed to their central values; in particular, the central values of $\mu_{r,m}$ are taken as $\overline{m}_c(\overline{m}_c)$, and $\Delta\mu_{r,m}$ is estimated by varying the scale within the range of [1,4] GeV. 

It is worth noting that after eliminating scale dependence in the moments $M^{X,\mathrm{th}}_{n,c}$, the PMC predictions show that for small $n$, the theoretical moments $M^{X,\mathrm{th}}_{n,c}$ are more sensitive to variations in $\Delta\alpha_s(m_Z)$, whereas the uncertainty from $\Delta m_q$ becomes more pronounced as $n$ increases. For instance, for the pseudoscalar moment $M^{P,\mathrm{th}}_{0,c}|_{\mathrm{PMC}}$, the uncertainty caused by $\Delta\alpha_s$ is about $\left(^{+0.8\%}_{-0.7\%}\right)$, which becomes $\left(\pm 0.1\%\right)$ for $\Delta m_q$; while for the moment $M^{P,\mathrm{th}}_{1,c}|_{\mathrm{PMC}}$, the $\Delta\alpha_s$ error changes to be about $\left(^{+1.9\%}_{-1.8\%}\right)$ and the $\Delta m_q$ error changes to $\left(\mp 0.9\%\right)$. Since the PMC removes scale ambiguities, it yields more reliable physical predictions that reflect the intrinsic dependence of the moments on the fundamental parameters $\alpha_s$ and $\overline{m}_c$. 

\begin{figure}[htbp]
\centering
\makebox[\textwidth][c]{
\begin{minipage}{1.06\textwidth}
\centering
\includegraphics[width=0.49\textwidth,trim=35 10 35 20,clip]{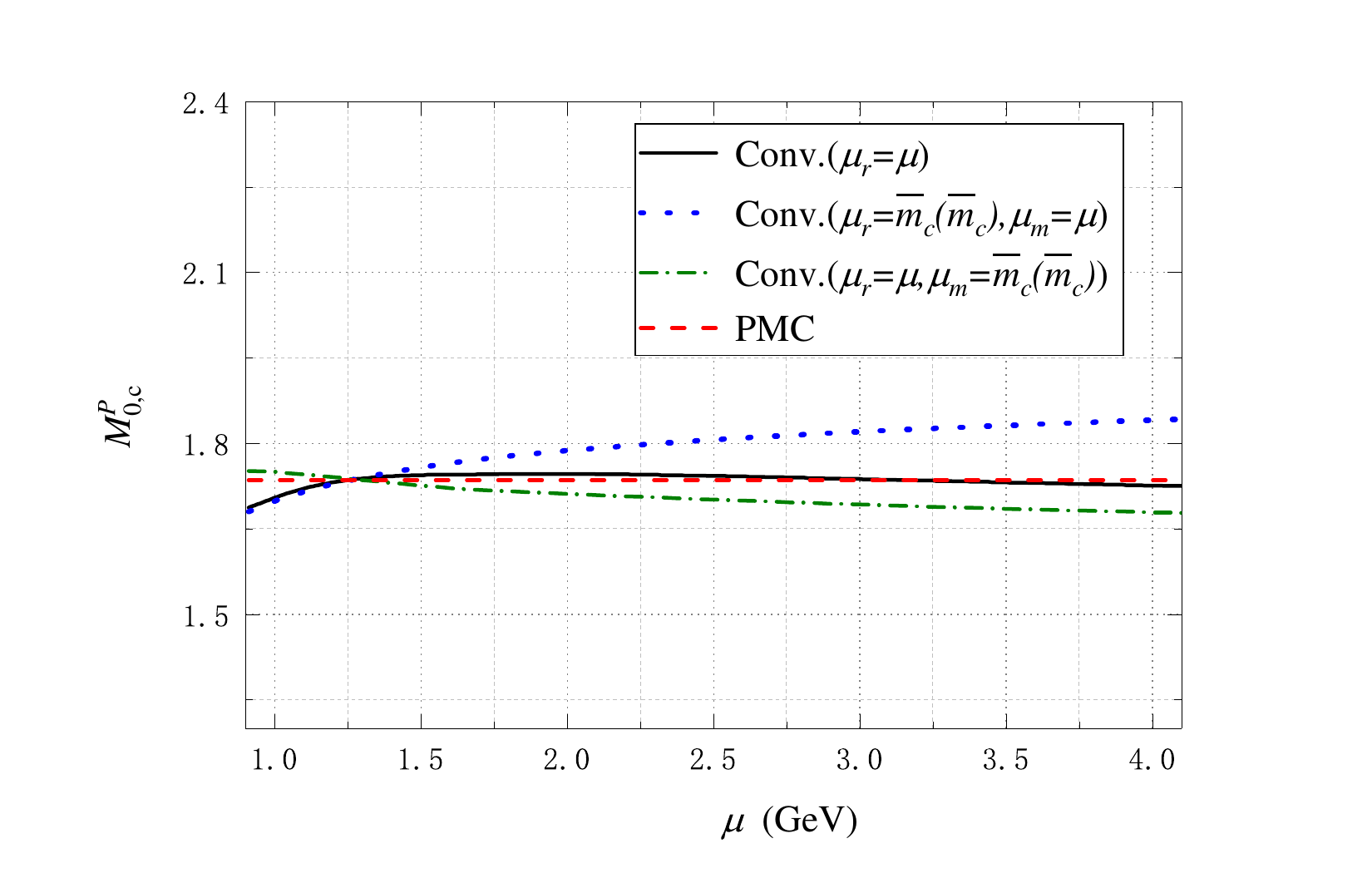}
\hfill
\includegraphics[width=0.49\textwidth,trim=35 10 35 20,clip]{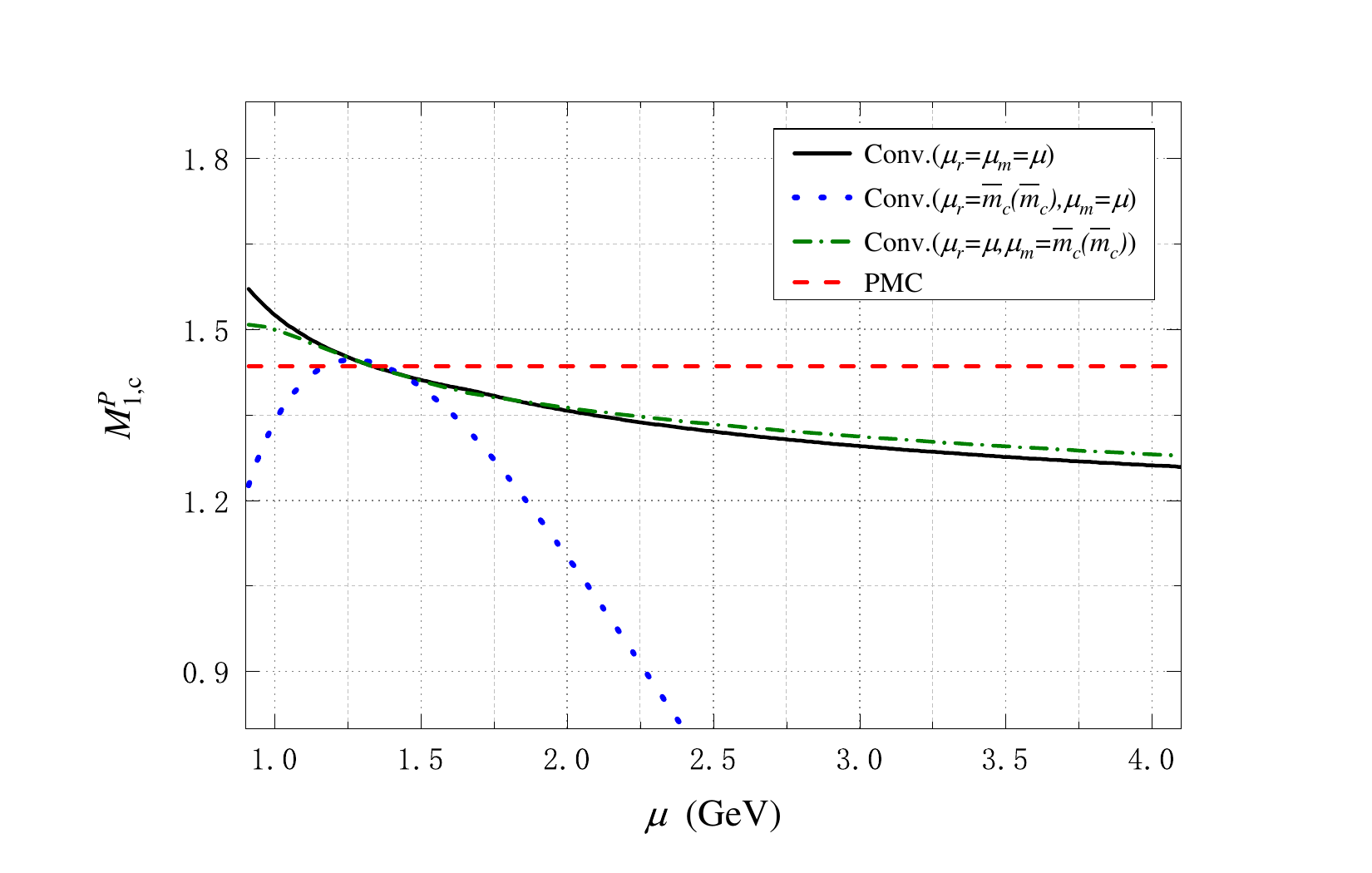}

\vspace{0.35cm}

\includegraphics[width=0.49\textwidth,trim=35 10 35 20,clip]{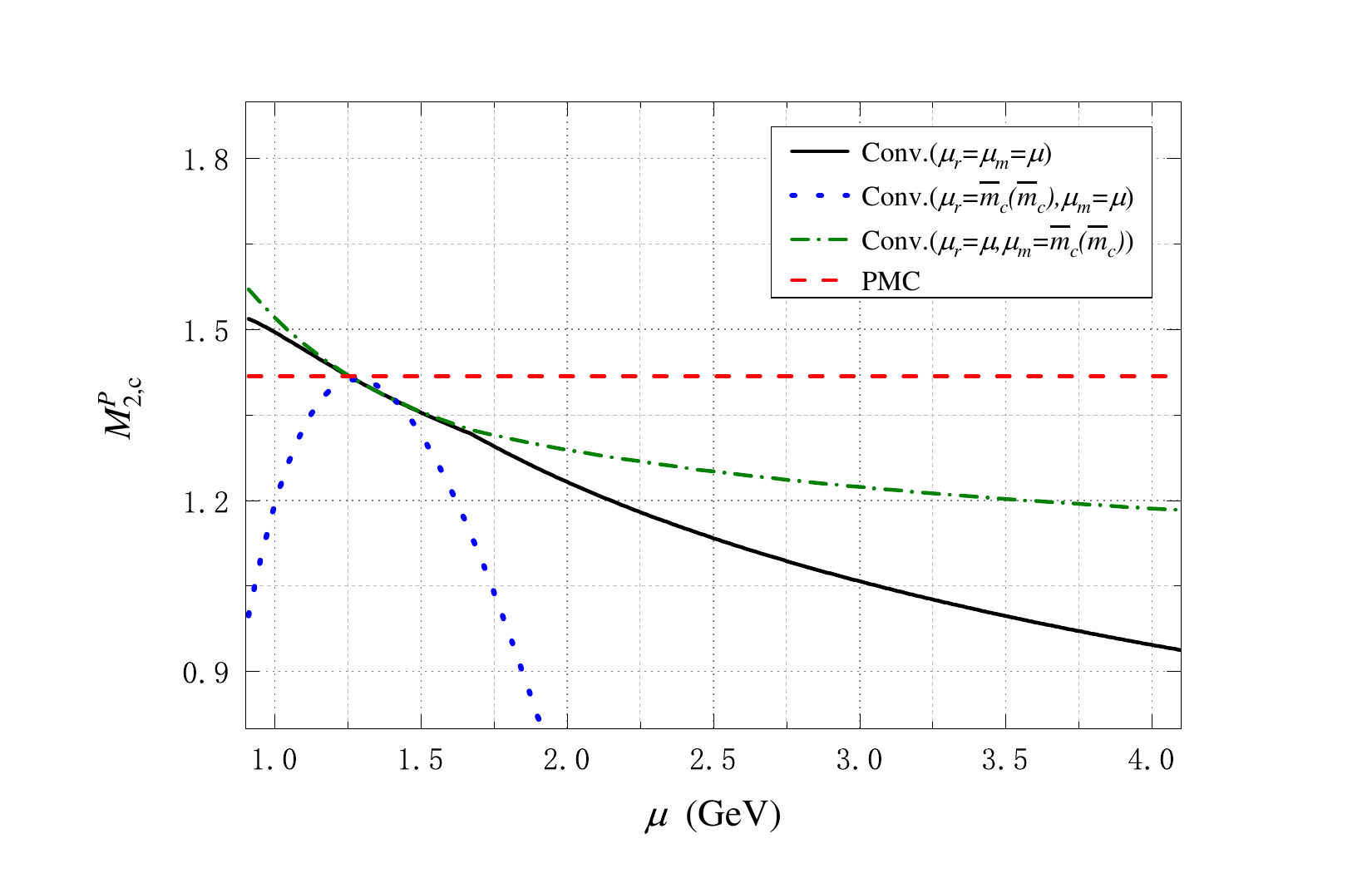}
\hfill
\includegraphics[width=0.49\textwidth,trim=35 10 35 20,clip]{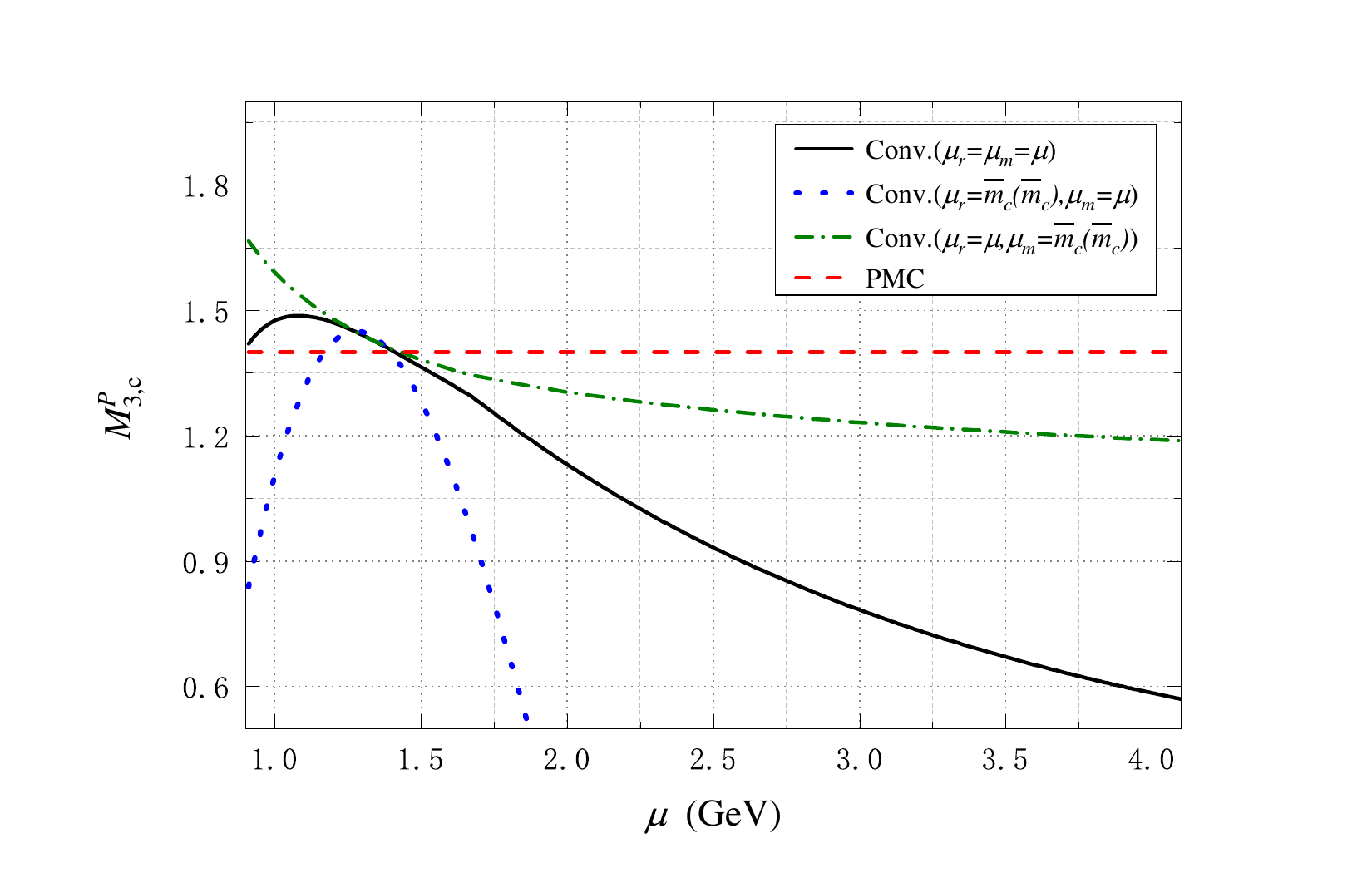}
\end{minipage}
}        
\caption{The moments $M^{P,{\rm th}}_{n,c}$ (in units of $10^{-n}{\rm GeV}^{-2n}$) versus the scale $\mu$ in the range $\mu\in[1,4]$ GeV, $n=(0,1,2,3)$. The solid, dotted and dash-dotted lines represent results under conventional scale-setting approach by taking [$\mu_r=\mu_m=\mu$], [$\mu_r=\overline{m}_c(\overline{m}_c)$ and $\mu_m=\mu$], [$\mu_r=\mu$ and $\mu_m=\overline{m}_c(\overline{m}_c)$], respectively. The dashed line represents the scale-invariant PMC prediction.}     \label{mnPc}
\end{figure}

\begin{figure}[htbp]
\centering
\makebox[\textwidth][c]{
\begin{minipage}{1.06\textwidth}
\centering
\includegraphics[width=0.49\textwidth,trim=35 10 35 20,clip]{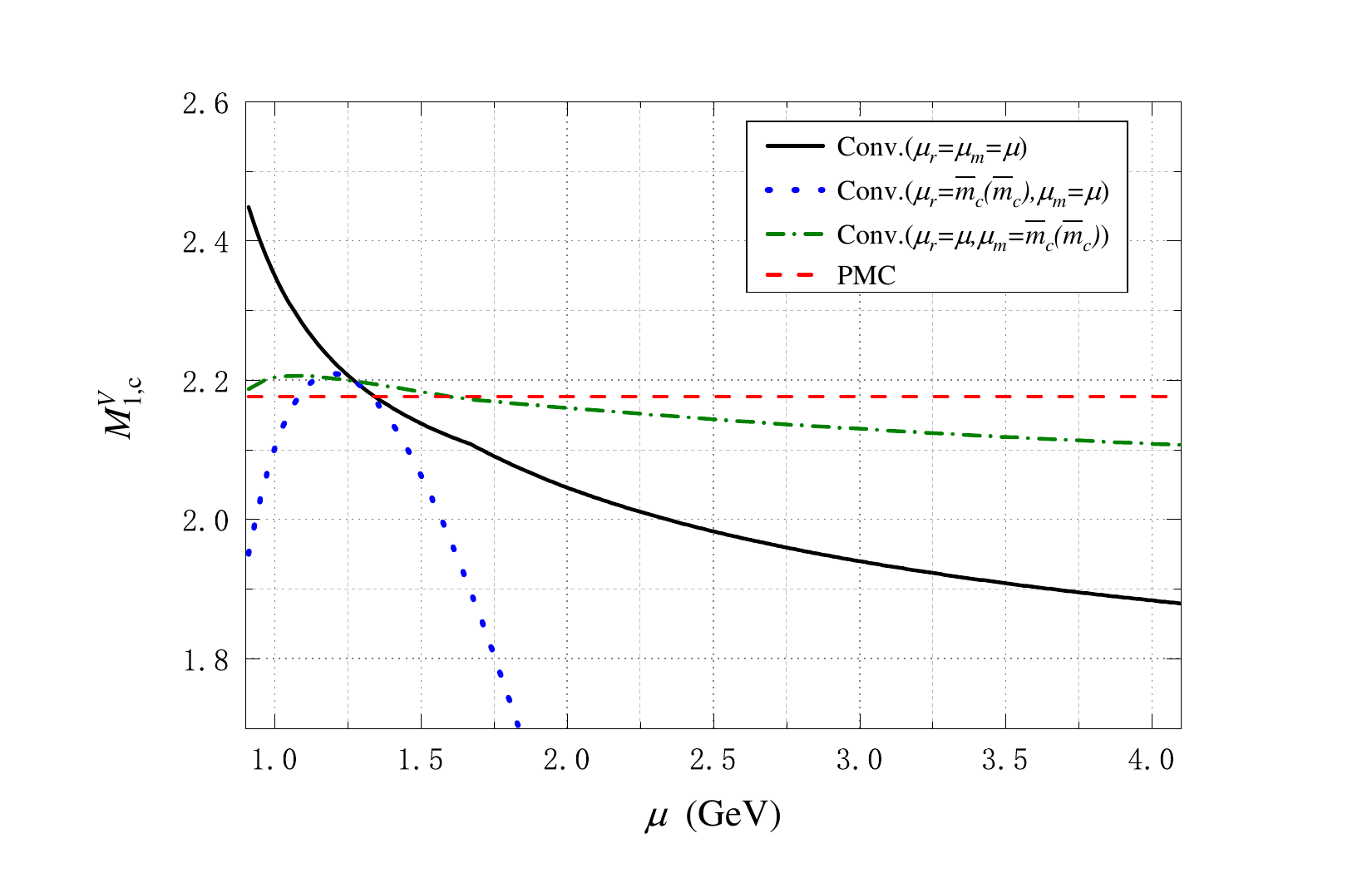}
\hfill
\includegraphics[width=0.49\textwidth,trim=35 10 35 20,clip]{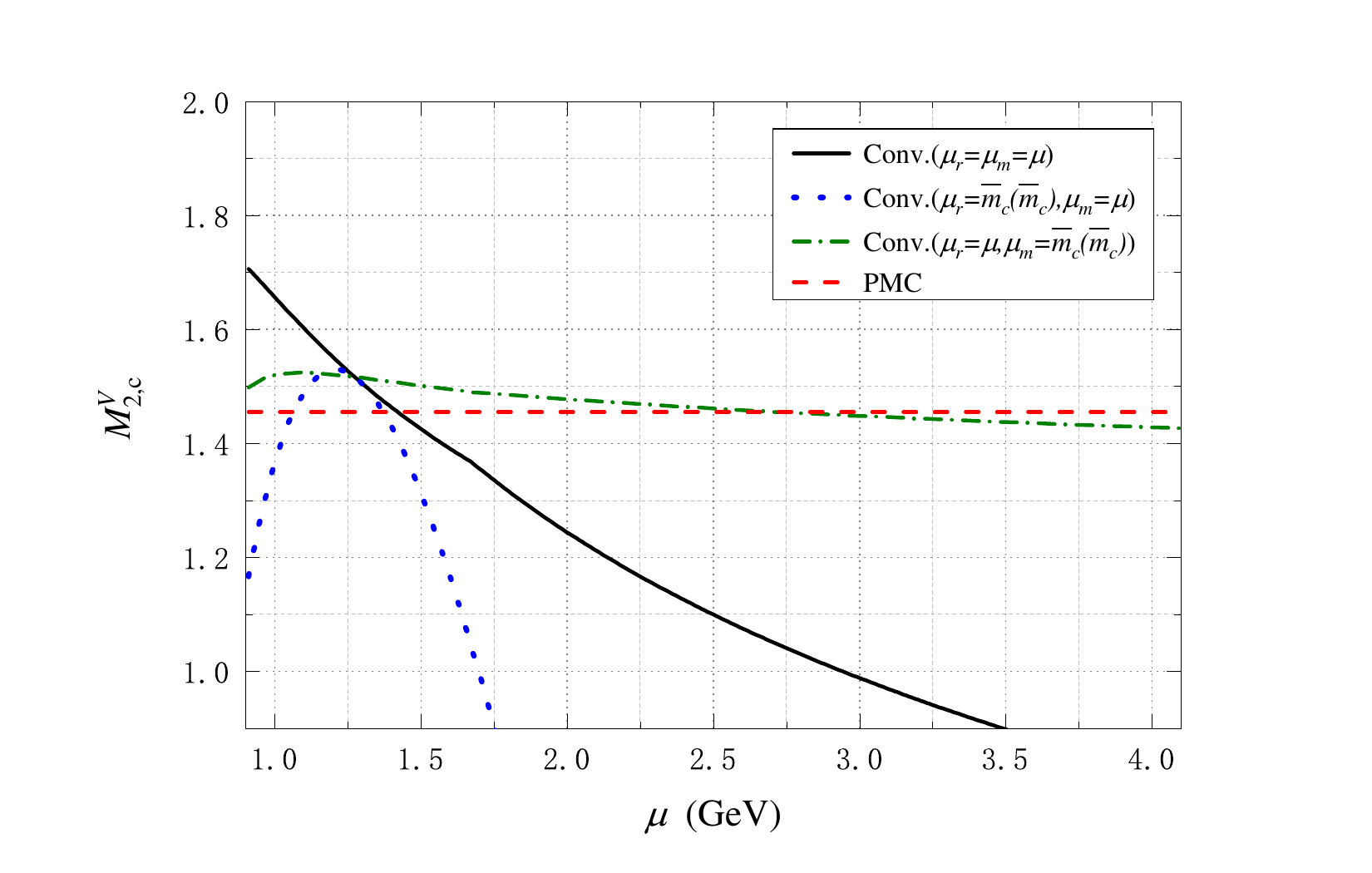}

\vspace{0.35cm}

\includegraphics[width=0.49\textwidth,trim=35 10 35 20,clip]{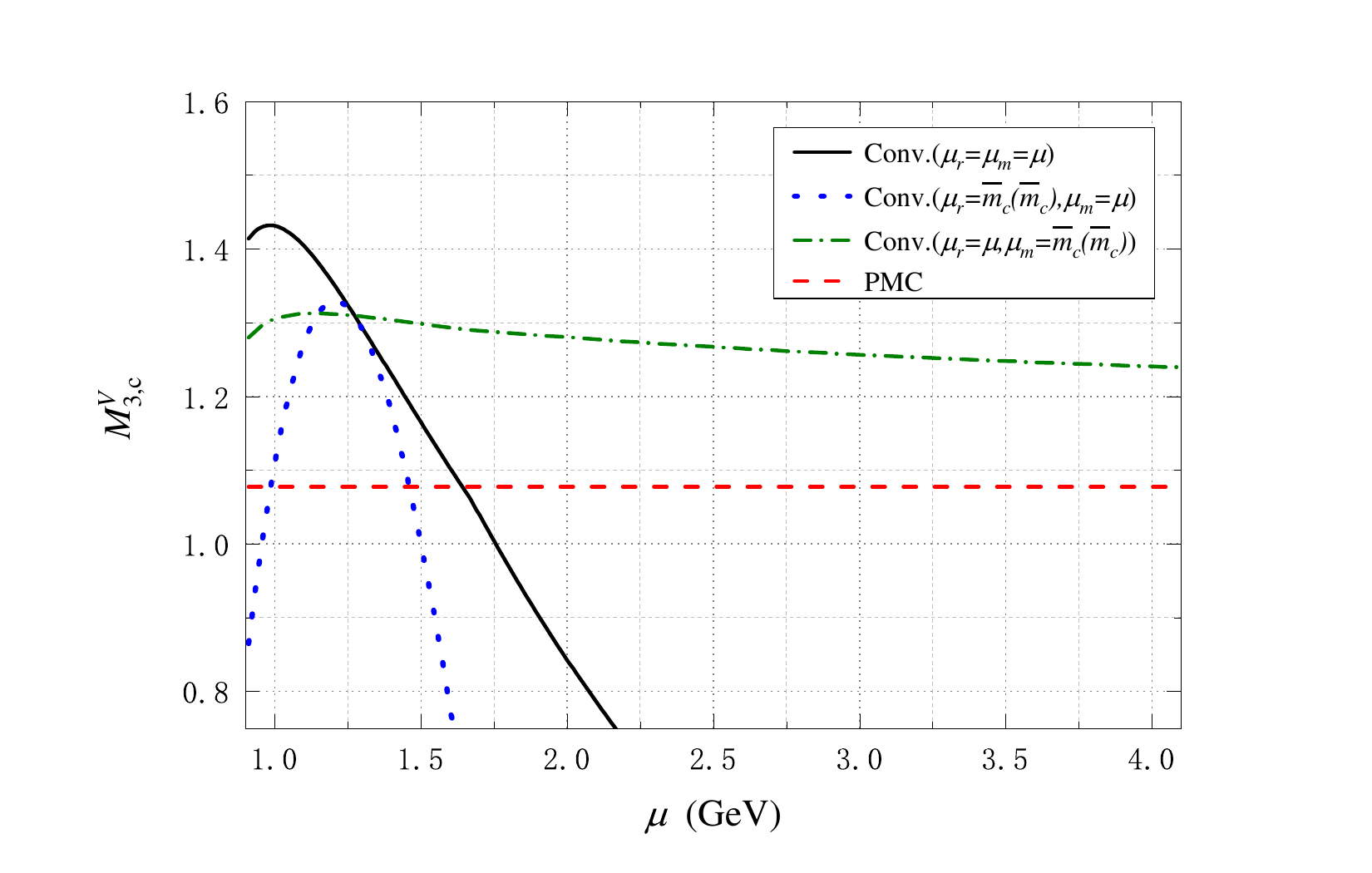}
\end{minipage}
}
\caption{The moments $M^{V,{\rm th}}_{n,c}$ (in units of $10^{-n}{\rm GeV}^{-2n}$) versus the scale $\mu$ in the range $\mu\in[1,4]$ GeV, $n=(1,2,3)$. The solid, dotted and dash-dotted lines represent results under conventional scale-setting approach by taking [$\mu_r=\mu_m=\mu$], [$\mu_r=\overline{m}_c(\overline{m}_c)$ and $\mu_m=\mu$], [$\mu_r=\mu$ and $\mu_m=\overline{m}_c(\overline{m}_c)$], respectively. The dashed line represents the scale-invariant PMC prediction.}
    \label{mnVc}
\end{figure}

Figs.~\ref{mnPc} and \ref{mnVc} show the theoretical predictions to the moments $M^{X,\rm th}_{n,c}$ versus the scale $\mu$ under both conventional (Conv.) and PMC scale-setting approaches, and $X=P$ or $V$, respectively. Under the conventional scale-setting approach, we examine three typical choices for the renormalization scale: (i) $\mu_r=\mu_m=\mu$, (ii) $\mu_r=\overline{m}_c(\overline{m}_c)$ with $\mu_m=\mu$, (iii) $\mu_r=\mu$ with $\mu_m=\overline{m}_c(\overline{m}_c)$. The scale $\mu$ is varied within [1,4] GeV for the charmonium case. The conventional predictions for $M^{X,\rm th}_{n,c}|_{\rm Conv.}$ show a pronounced sensitivity to these arbitrary scale choices. This sensitivity is particularly acute in the charm sector due to the relatively large value of the coupling constant $\alpha_s(\overline{m}_c)$ at the charm quark mass scale. In contrast, the PMC predictions are manifestly stable and scale-invariant: by applying the PMC method, all the RGE-related non-conformal terms are incorporated into the determination of $a_s(Q_*)$ and $\overline{m}_q(Q_*)$, resulting in a scale-independent conformal perturbative series. Numerical results for the resultant series obtained via the PMC CO approach for the moments $M^{\mathrm{X},\mathrm{pert}}_{n,c}$ are presented in Appendix~\ref{appA}. 

\begin{table}[htbp]
\centering
\small
\setlength{\tabcolsep}{0.5pt}
\renewcommand{\arraystretch}{1.5}
\makebox[\textwidth][c]{
\begin{tabular}{l c c c c c c c}
\toprule
&~Conv.~  &~LO~ &~NLO~  &~$\rm N^2LO$~ &~$\rm N^3LO$~  &~n.p.~ &~total~\\
\midrule
&$M^{P,\rm th}_{0,c}$   &$1.333$  &$0.403_{-0.175}^{+0.106}\pm0$ &$0.002_{+0.078}^{-0.082}$$_{+0.074}^{-0.034}$ &$-0.003_{+0.039}^{-0.011}$$_{+0.029}^{-0.004}$ &$(4.5_{+0.3}^{-0.2}$$_{+1.2}^{-1.0})\times10^{-5}$ &$1.735_{-0.058}^{+0.013}$$_{+0.103}^{-0.038}$ \\

&$M^{P,\rm th}_{1,c}$   &$0.823\pm{0}_{+0.805}^{-0.220}$  &$0.412_{-0.179}^{+0.109}$$_{-0.849}^{+0.013}$ &$0.187_{-0.046}^{+0.026}$$_{-0.906}^{+0.045}$ &$0.024_{+0.059}^{-0.082}$$_{-0.992}^{+0.054}$ &$(110.9_{+26.3}^{-13.7}$$_{+45.7}^{-35.4}))\times10^{-5}$ &$1.447_{-0.166}^{+0.053}$$_{-1.942}^{-0.108}$ \\

&$M^{P,\rm th}_{2,c}$   &$0.725\pm{0}_{+2.115}^{-0.336}$  &$0.373_{-0.162}^{+0.099}$$_{-3.282}^{-0.013}$ &$0.239_{-0.089}^{+0.065}$$_{-1.737}^{+0.046}$ &$0.075_{+0.025}^{-0.056}$$_{-1.244}^{+0.083}$ &$(-44.8^{+3.1}_{+1.8}$$^{+18.0}_{-26.2})\times10^{-5}$ &$1.412_{-0.226}^{+0.108}$$_{-4.148}^{-0.220}$ \\

&$M^{P,\rm th}_{3,c}$   &$0.746\pm{0}_{+5.034}^{-0.453}$  &$0.339_{-0.148}^{+0.090}$$_{-11.054}^{-0.025}$ &$0.262_{-0.112}^{+0.086}$$_{-0.060}^{+0.033}$ &$0.106_{+0.002}^{-0.036}$$_{+0.848}^{+0.099}$  &$(-314.6^{+61.8}_{-117.4}$$^{+148.8}_{-244.7})\times10^{-5}$ &$1.450_{-0.259}^{+0.141}$$_{-5.234}^{-0.344}$\\  
\addlinespace[1em]
&$M^{V,\rm th}_{1,c}$   &$1.646\pm{0}_{+1.611}^{-0.440}$  &$0.510_{-0.222}^{+0.135}$$_{-2.006}^{+0.111}$ &$0.064_{+0.056}^{-0.068}$$_{-1.752}^{+0.144}$ &$-0.019_{+0.076}^{-0.062}$$_{-1.397}^{+0.088}$ &$(-293.2^{+27.0}_{-41.2}$$^{+93.6}_{-120.9})\times10^{-5}$ &$2.198_{-0.090}^{+0.005}$$_{-3.545}^{-0.096}$\\

&$M^{V,\rm th}_{2,c}$   &$1.088\pm{0}_{+3.172}^{-0.504}$  &$0.342_{-0.149}^{+0.090}$$_{-5.558}^{+0.081}$  &$0.110_{-0.008}^{-0.005}$$_{-2.172}^{+0.146}$ &$-0.018_{+0.071}^{-0.082}$$_{-0.616}^{+0.125}$ &$-589.3_{-163.8}^{+89.8}$$_{-344.5}^{+236.3}\times10^{-5}$  &$1.516_{-0.088}^{+0.004}$$_{-5.177}^{-0.150}$ \\

&$M^{V,\rm th}_{3,c}$   &$0.995\pm{0}_{+6.712}^{-0.604}$  &$0.246_{-0.107}^{+0.065}$$_{-16.120}^{+0.091}$ &$0.101_{-0.020}^{+0.009}$$_{+1.609}^{+0.153}$ &$-0.023_{+0.062}^{-0.080}$$_{+3.631}^{+0.155}$ &$(-1022.5_{-447.2}^{+214.8}$$_{-795.4}^{+483.6})\times10^{-5}$  &$1.309_{-0.069}^{-0.004}$$_{-4.176}^{-0.200}$ \\  
\bottomrule
\end{tabular}}
\caption{The perturbative contributions up to N$^3$LO-level for charmonium moments in the P- and V-channels under conventional scale-setting approach (in units of $10^{-n}\,\mathrm{GeV}^{-2n}$). The central values correspond to $\mu_r=\overline{m}_c(\overline{m}_c)$ and $\mu_m=\overline{m}_c(\overline{m}_c)$. The first uncertainty arises from $\mu_r\in[1,4]\,\mathrm{GeV}$ with $\mu_m=\overline{m}_c(\overline{m}_c)$ fixed, and the second uncertainty from $\mu_m\in[1,4]\,\mathrm{GeV}$ with $\mu_r=\overline{m}_c(\overline{m}_c)$ fixed. Nonperturbative contributions are labeled ``n.p.".}
\label{tab1c}
\end{table}

\begin{table}
\centering
\begin{tabular}{l c c c c c c c}
\toprule
 &~PMC~ &~LO~ &~NLO~  &~$\rm N^2LO$~ &~$\rm N^3LO$~  &~n.p.~ &~total~\\
\midrule
&$M^{P,\rm th}_{0,c}$   &1.333  &0.368 &0.109 &$-0.076$ &$5.0\times10^{-5}$ &1.734\\ 

&$M^{P,\rm th}_{1,c}$   &0.863  &0.416 &0.186 &$-0.031$ &$119.2\times10^{-5}$ &1.435\\

&$M^{P,\rm th}_{2,c}$   &0.864  &0.415 &0.121 &$-0.018$ &$-51.7\times10^{-5}$ &1.381\\

&$M^{P,\rm th}_{3,c}$  &1.269  &0.500 &$-0.194$ &$-0.222$ &$-496.8\times10^{-5}$  &1.348\\  
\addlinespace[1em]
 &$M^{V,\rm th}_{1,c}$   &1.735  &0.516 &0.071 &$-0.142$ &$-315.3\times10^{-5}$ &2.177\\

&$M^{V,\rm th}_{2,c}$   &1.347  &0.388 &$-0.102$ &$-0.171$ &$-728.7\times10^{-5}$  &1.455\\

&$M^{V,\rm th}_{3,c}$   &1.969  &0.407 &$-0.713$ &$-0.567$ &$-1884.0\times10^{-5}$  &1.081\\  
\bottomrule
\end{tabular}
\caption{The scale-invariant PMC predictions for the charmonium moments in the P- and V-channels (in units of $10^{-n}{\rm GeV}^{-2n}$), in which the scales $\mu_{r}$ and $\mu_{m}$ are varied independently within the range of $[1,4]\,\mathrm{GeV}$.}
\label{tab1c2}
\end{table}

Results for the perturbative and nonperturbative contributions (``n.p.") to the moments $M^{X}_{n,c}$ under conventional and PMC scale-setting approaches are listed in Tables~\ref{tab1c} and~\ref{tab1c2}, respectively. Table~\ref{tab1c} shows that for the total moments $M^{X,\mathrm{th}}_{n,c}|_{\rm Conv.}$ under the conventional scale-setting approach, their $\mu_r$-uncertainty is notably smaller than the $\mu_m$-uncertainty. As an example, for $M^{P,\mathrm{th}}_{1,c}|_{\rm Conv.}$, its $\mu_r$-uncertainty is $\left(^{-11.5\%}_{+3.7\%}\right)$ of the central value, while its $\mu_m$-uncertainty reaches $\left(^{-134.2\%}_{-7.5\%}\right)$. This relatively smaller $\mu_r$-error stems from the cancellation of large $\mu_r$-dependent variations across perturbative orders, although the dependence within individual orders remains significant. Specifically, for $M^{P,\mathrm{th}}_{1,c}|_{\rm Conv.}$, the $\mu_r$-uncertainty is $\left(^{-43.4\%}_{+26.5\%}\right)$ for the NLO term, $\left(^{-13.9\%}_{+24.6\%}\right)$ for the $\mathrm{N^2LO}$ term, and $\left(^{-341.7\%}_{+245.8\%}\right)$ for the $\mathrm{N^3LO}$ term. When the renormalization scale is set to the quark mass, $\mu_{r,m}=\overline{m}_q$, the conventional predictions exhibit better convergence for certain moments, such as $M^{V,\mathrm{th}}_{3,c}$. This apparent improvement, however, stems from an accidental cancellation between large conformal terms and divergent non-conformal terms at the $\mathrm{N^3LO}$ order, which does not occur for $M^{P,\mathrm{th}}_{0,c}$. Furthermore, a comparison of moments with different $n$ reveals that the scale dependence is significantly amplified for higher $n$.

Table~\ref{tab1c2} shows that after applying the PMC, the conventional scale dependence at each order of the moments $M^{X,\mathrm{th}}_{n,c}$ is eliminated. Thus, the perturbative behavior of the PMC series can be regarded as an intrinsic property. Thanks to the removal of divergent renormalon contributions, the PMC method generally improves the convergence of the pQCD expansion. Nevertheless, it has been found that for certain charm-related processes, such as the ratio $R = \Gamma_{\eta_c\to \mathrm{LH}} / \Gamma_{\eta_c\to\gamma\gamma}$, the conformal terms remain moderately suppressed. And its resulting PMC series converges more slowly than anticipated, since the power suppression induced by $\alpha_s(Q_*\sim{\cal O}(m_c))\sim 0.3$ is less effective than in typical QCD scenarios~\cite{Yu:2019mce}. A similar behavior occurs in the current study. As illustrated in Table~\ref{tab1c2}, the magnitudes of the $\mathrm{N^3LO}$ terms for the V-channel moments $M^{V,\mathrm{th}}_{n,q}$ are comparable to their $\mathrm{N^2LO}$ counterparts, which weakens the convergence of the perturbative series. Therefore, a systematic treatment to estimate missing higher-order corrections is essential for obtaining reliable pQCD predictions, which we will address in the subsequent analysis.

\subsection{Perturbative contributions to $M^{X,\mathrm{th}}_{n,b}$ up to $\mathrm{N^3LO}$ QCD corrections}  
\label{sec3.2}

For the bottom quark case, the determined PMC scales are as follows
\begin{align}
Q^{P,0}_{*,b} &= 5.6380(^{-0.0201}_{+0.0199})_{\Delta \alpha_s}(\pm0.0103)_{\Delta m_b}~{\rm GeV}, \\
Q^{P,1}_{*,b} &= 4.2971(\pm0.0024)_{\Delta \alpha_s}(^{+0.0388}_{+0.0245})_{\Delta m_b}~{\rm GeV},\\
Q^{P,2}_{*,b} &= 4.5069(\pm0.0033)_{\Delta \alpha_s}(^{+0.0305}_{-0.0156})_{\Delta m_b}~{\rm GeV}, \\
Q^{P,3}_{*,b} &= 4.8869(\pm0.0077)_{\Delta \alpha_s}(^{+0.0330}_{+0.0169})_{\Delta m_b}~{\rm GeV},\\
Q^{V,1}_{*,b} &= 4.3161(\pm0.0027)_{\Delta \alpha_s}(^{+0.0462}_{-0.0320})_{\Delta m_b}~{\rm GeV}, \\
Q^{V,2}_{*,b} &= 4.5762(\pm0.0044)_{\Delta \alpha_s}(^{+0.0316}_{+0.0166})_{\Delta m_b}~{\rm GeV},\\
Q^{V,3}_{*,b} &= 5.0797(^{+0.0118}_{-0.0117})_{\Delta \alpha_s}(^{+0.0128}_{-0.0037})_{\Delta m_b}~{\rm GeV}, \\
Q^{V,4}_{*,b} &= 6.0875(^{+0.0367}_{-0.0357})_{\Delta \alpha_s}(^{+0.2204}_{-0.2388})_{\Delta m_b}~{\rm GeV},
\end{align}
where the first error originates from $\Delta\alpha_s(m_Z) = \pm 0.0009$, and the second error arises from $\Delta\overline{m}_b(\overline{m}_b) = \pm 7\,\mathrm{MeV}$. Accounting for the uncertainties associated with $\Delta\alpha_s(m_Z)$, $\Delta m_b$, and $\Delta\mathrm{n.p.}$, the corresponding PMC results for the moments $M^{X,\mathrm{th}}_{n,b}$ are:
\begin{align}
M^{P,\rm th}_{0,b}|_{\rm PMC}&=3.8958(^{+0.0087}_{-0.0086})_{\Delta \alpha_s}(\mp0.0003)_{\Delta m_b}(\pm0.0001)_{\Delta{\rm n.p.}},\\
M^{P,\rm th}_{1,b}|_{\rm PMC}&=2.5808(^{+0.0129}_{-0.0126})_{\Delta \alpha_s}(^{-0.0032}_{+0.0151})_{\Delta m_b}(\mp0.0001)_{\Delta{\rm n.p.}},\\
M^{P,\rm th}_{2,b}|_{\rm PMC}&=2.1864(^{+0.0129}_{-0.0126})_{\Delta \alpha_s}(^{-0.0071}_{+0.0233})_{\Delta m_b}(\mp0.0000)_{\Delta{\rm n.p.}},\\
M^{P,\rm th}_{3,b}|_{\rm PMC}&=2.0418(^{+0.0112}_{-0.0110})_{\Delta \alpha_s}(^{-0.0103}_{+0.0325})_{\Delta m_b}(\mp0.0002)_{\Delta{\rm n.p.}},\\
M^{V,\rm th}_{1,b}|_{\rm PMC}&=4.5107(^{+0.0112}_{-0.0111})_{\Delta \alpha_s}(^{-0.0005}_{+0.0307})_{\Delta m_b}(\mp0.0001)_{\Delta{\rm n.p.}},\\
M^{V,\rm th}_{2,b}|_{\rm PMC}&=2.8071(^{+0.0078}_{-0.0077})_{\Delta \alpha_s}(^{-0.0074}_{+0.0310})_{\Delta m_b}(\mp0.0003)_{\Delta{\rm n.p.}},\\
M^{V,\rm th}_{3,b}|_{\rm PMC}&=2.2705(^{+0.0029}_{-0.0030})_{\Delta \alpha_s}(^{-0.0209}_{+0.0258})_{\Delta m_b}(\mp0.0006)_{\Delta{\rm n.p.}},\\
M^{V,\rm th}_{4,b}|_{\rm PMC}&=1.9704(^{-0.0065}_{+0.0060})_{\Delta \alpha_s}(^{-0.1559}_{-0.1069})_{\Delta m_b}(\mp0.0011)_{\Delta{\rm n.p.}}.	
\end{align}
where $M^{X,{\rm th}}_{n,b}$ are in units of $10^{-(2n+1)}{\rm GeV}^{-2n}$. The uncertainty associated with $\Delta\mathrm{n.p.}$, or equivalently the nonperturbative gluon condensate term $\Delta\langle \alpha_s G^2 \rangle = \pm 0.012\,\mathrm{GeV}^4$ is negligible. For comparison, the uncertainties in $\Delta\alpha_s(m_Z)$, $\Delta m_q$, and $\Delta\mathrm{n.p.}$ for the results under the conventional scale-setting approach with $\mu_{r,m}=\overline{m}_b(\overline{m}_b)$ are also provided:
\begin{align}
M^{P,\rm th}_{0,b}|_{\rm Conv.}&=3.8958(_{-0.0104}^{+0.0083})_{\Delta\mu_r}(_{+0.0257}^{-0.0313})_{\Delta\mu_m}(^{+0.0087}_{-0.0086})_{\Delta \alpha_s}(\mp0.0003)_{\Delta m_b}(\pm0.0001)_{\Delta\rm n.p.}\\
M^{P,\rm th}_{1,b}|_{\rm Conv.}&=2.5829(_{-0.0356}^{+0.0056})_{\Delta\mu_r}(_{-0.4052}^{-0.2381})_{\Delta\mu_m}(^{+0.0130}_{-0.0127})_{\Delta \alpha_s}(\mp0.0091)_{\Delta m_b}(\pm0.0001)_{\Delta{\rm n.p.}},\\
M^{P,\rm th}_{2,b}|_{\rm Conv.}&=2.1942(_{-0.0477}^{+0.0272})_{\Delta\mu_r}(_{-0.7088}^{-0.4038})_{\Delta\mu_m}(^{+0.0135}_{-0.0131})_{\Delta \alpha_s}(^{-0.0151}_{+0.0152})_{\Delta m_b}(\mp0.0000)_{\Delta{\rm n.p.}},\\
M^{P,\rm th}_{3,b}|_{\rm Conv.}&=2.0662(_{-0.0525}^{+0.0392})_{\Delta\mu_r}(_{-0.9947}^{-0.5680})_{\Delta\mu_m}(^{+0.0131}_{-0.0127})_{\Delta \alpha_s}(^{-0.0211}_{+0.0214})_{\Delta m_b}(\mp0.0001)_{\Delta{\rm n.p.}},\\
M^{V,\rm th}_{1,b}|_{\rm Conv.}&=4.5149(_{-0.0177}^{-0.0023})_{\Delta\mu_r}(_{-0.7635}^{-0.3802})_{\Delta\mu_m}(^{+0.0116}_{-0.0114})_{\Delta \alpha_s}(^{-0.0155}_{+0.0156})_{\Delta m_b}(\mp0.0001)_{\Delta{\rm n.p.}},\\
M^{V,\rm th}_{2,b}|_{\rm Conv.}&=2.8207(_{-0.0186}^{-0.0082})_{\Delta\mu_r}(_{-0.9755}^{-0.4742})_{\Delta\mu_m}(^{+0.0088}_{-0.0086})_{\Delta \alpha_s}(^{-0.0191}_{+0.0193})_{\Delta m_b}(\mp0.0003)_{\Delta{\rm n.p.}},\\
M^{V,\rm th}_{3,b}|_{\rm Conv.}&=2.3224(_{-0.0155}^{-0.0090})_{\Delta\mu_r}(_{-1.1671}^{-0.5824})_{\Delta\mu_m}(^{+0.0065}_{-0.0064})_{\Delta \alpha_s}(^{-0.0234}_{+0.0237})_{\Delta m_b}(\mp0.0005)_{\Delta{\rm n.p.}},\\
M^{V,\rm th}_{4,b}|_{\rm Conv.}&=2.1366(_{-0.0100}^{-0.0142})_{\Delta\mu_r}(_{-1.2435}^{-0.7040})_{\Delta\mu_m}(^{+0.0042}_{-0.0041})_{\Delta \alpha_s}(^{-0.0285}_{+0.0290})_{\Delta m_b}(\mp0.0008)_{\Delta{\rm n.p.}},	
\end{align}
where $\Delta\mu_{r,m}$ is estimated by varying the scale within the range of [2,10] GeV.

\begin{figure}[htbp]
\centering
\makebox[\textwidth][c]{
\begin{minipage}{1.06\textwidth}
\centering
\includegraphics[width=0.49\textwidth,trim=35 10 35 20,clip]{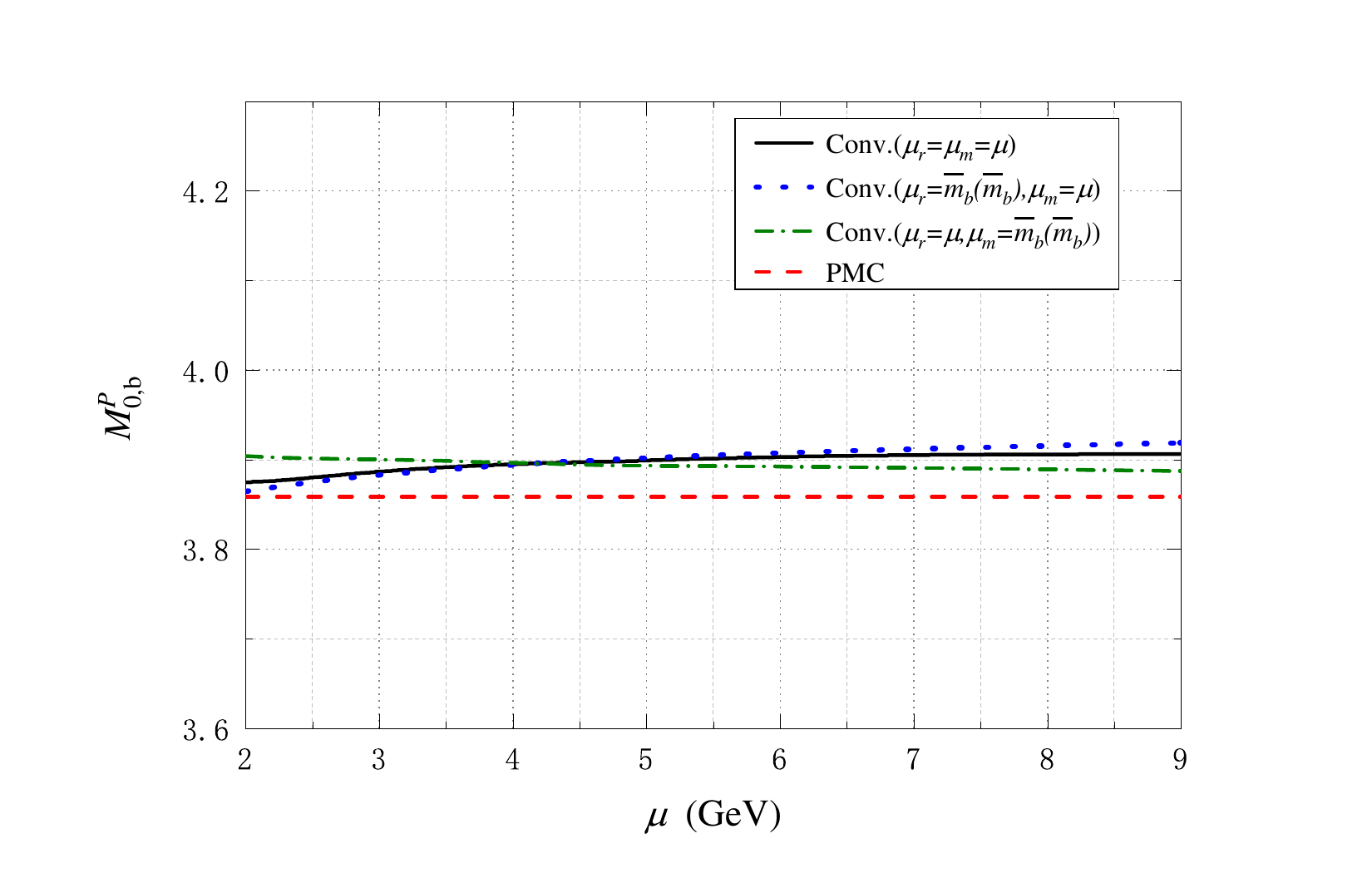}
\hfill
\includegraphics[width=0.49\textwidth,trim=35 10 35 20,clip]{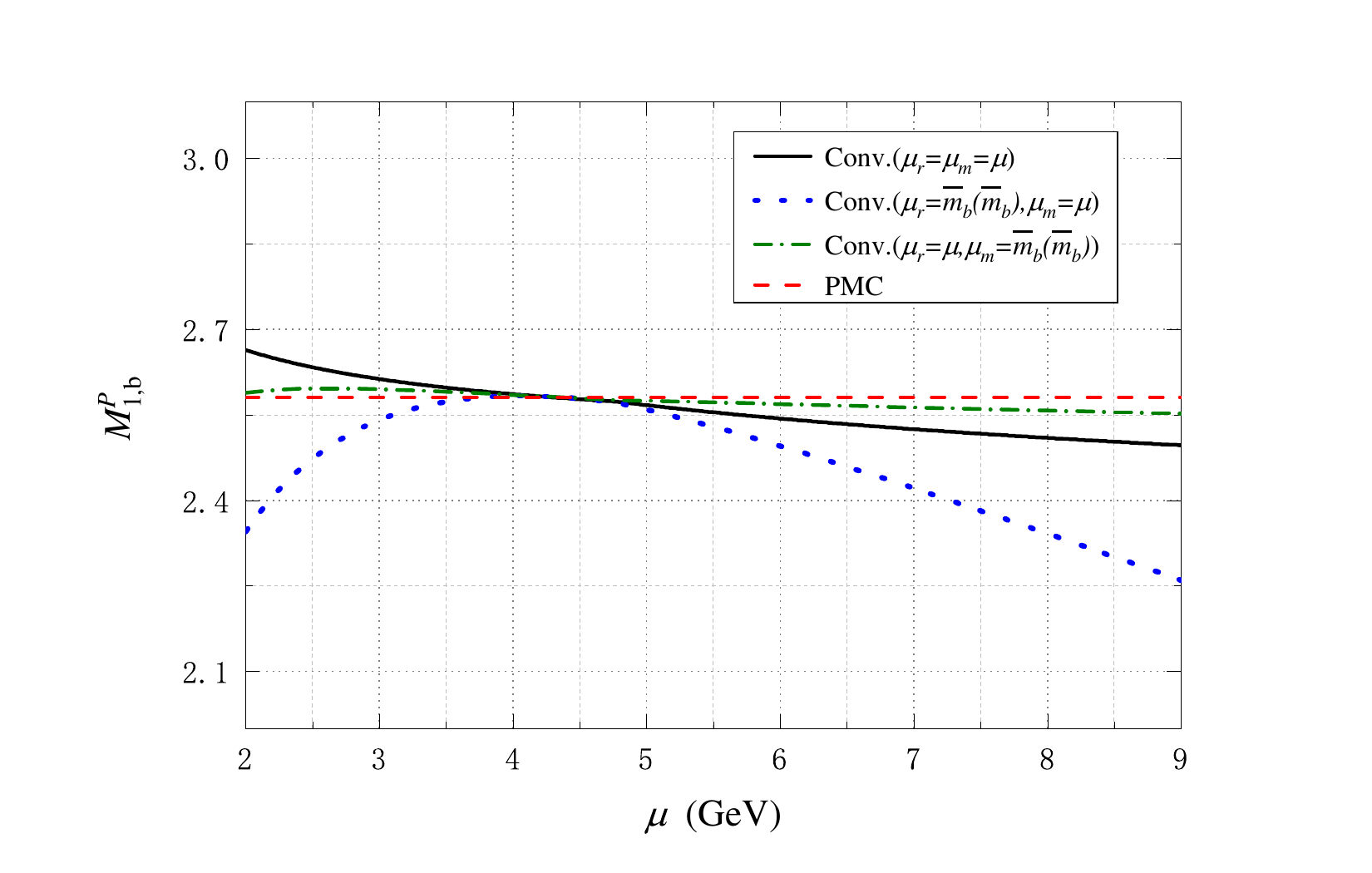}

\vspace{0.35cm}

\includegraphics[width=0.49\textwidth,trim=35 10 35 20,clip]{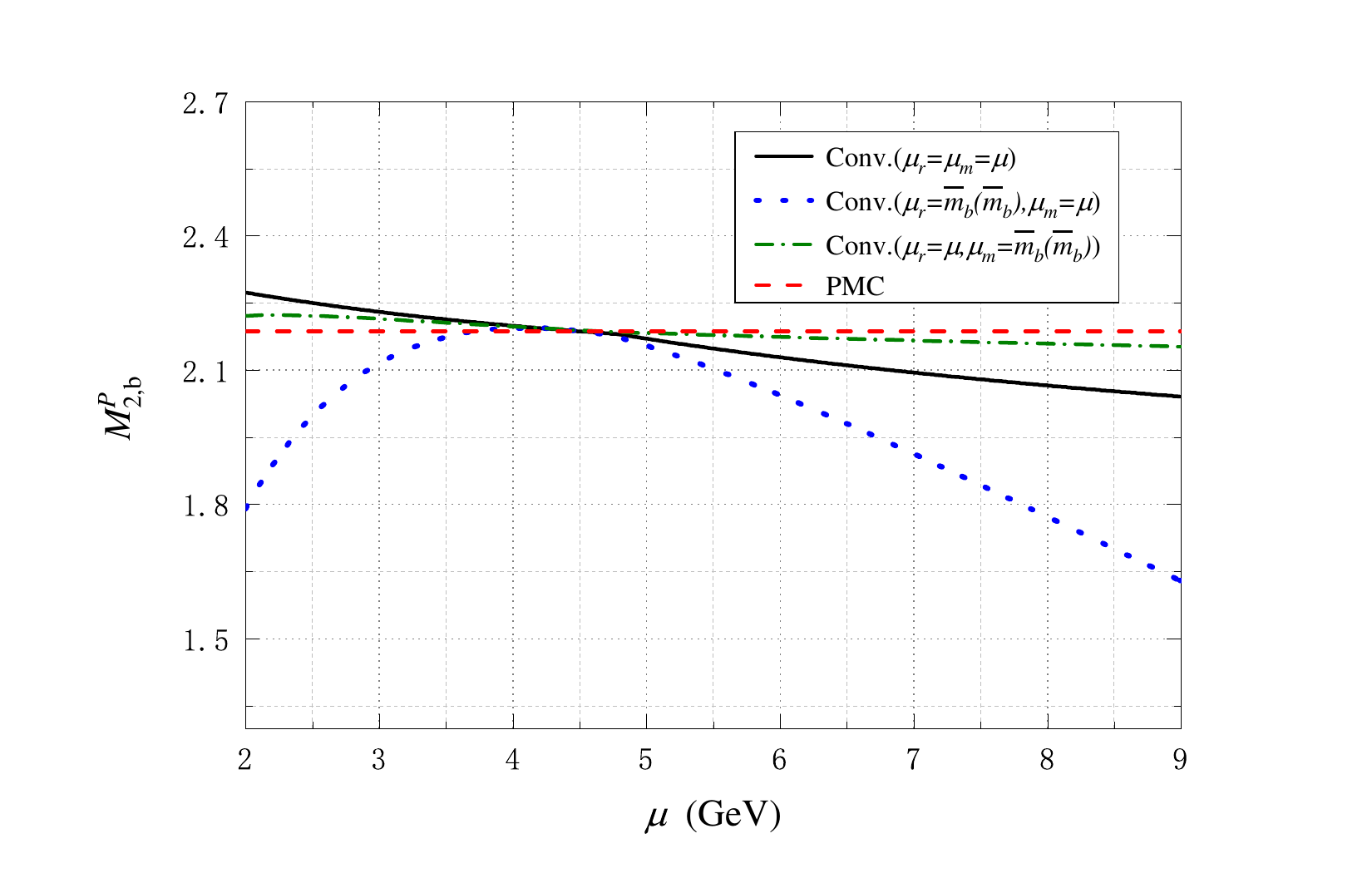}
\hfill
\includegraphics[width=0.49\textwidth,trim=35 10 35 20,clip]{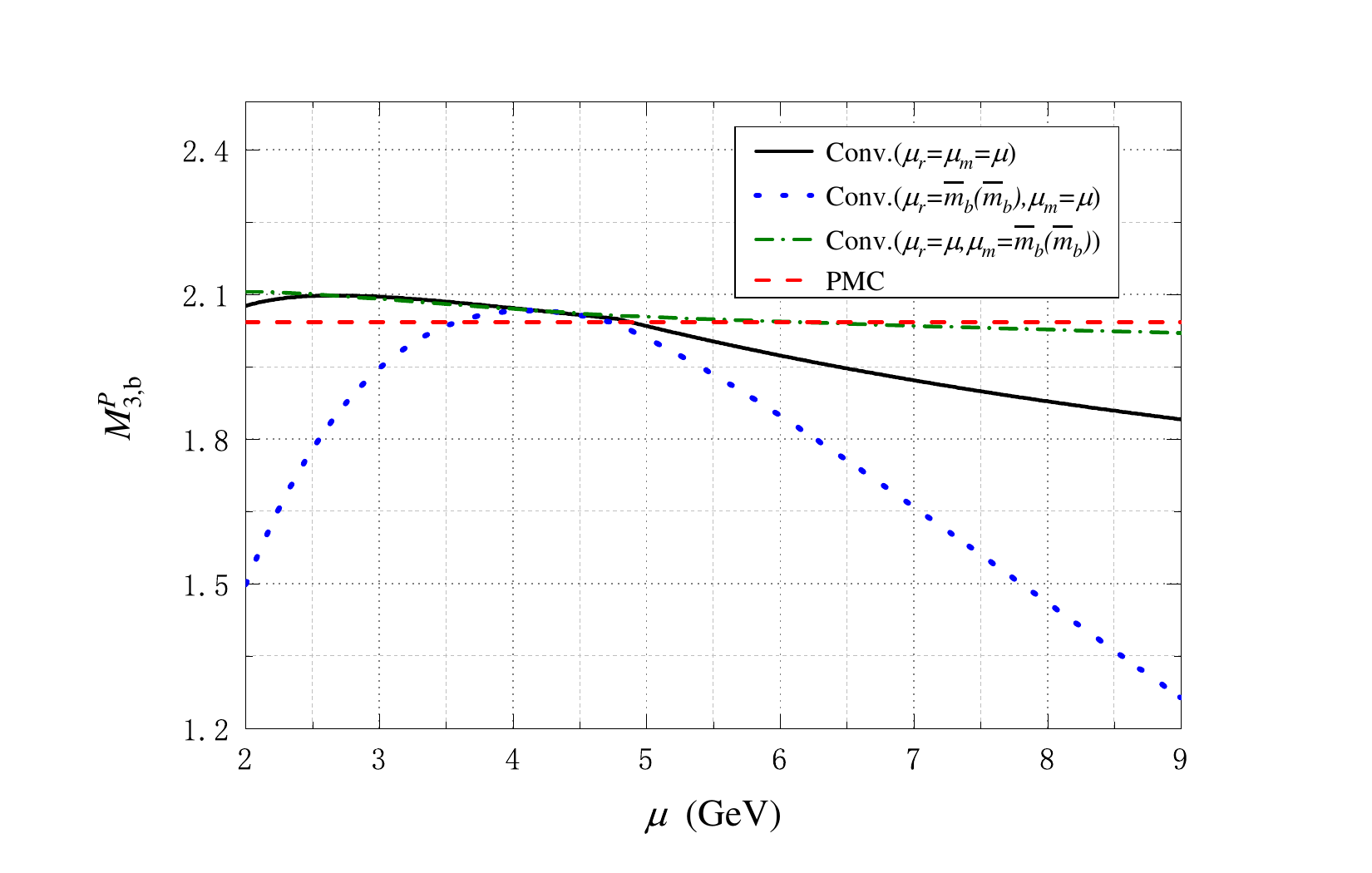}
\end{minipage}
}
\caption{The moments $M^{P,{\rm th}}_{n,b}$ (in units of $10^{-(2n+1)}{\rm GeV}^{-2n}$) versus the scale $\mu$ in the range $\mu\in[2,10]$ GeV, $n=(0,1,2,3)$. The solid, dotted and dash-dotted lines represent results under conventional scale-setting approach by taking [$\mu_r=\mu_m=\mu$], [$\mu_r=\overline{m}_b(\overline{m}_b)$ and $\mu_m=\mu$], [$\mu_r=\mu$ and $\mu_m=\overline{m}_b(\overline{m}_b)$], respectively. The dashed line represents the scale-invariant PMC prediction.}
    \label{mnPb}
\end{figure}

\begin{figure}[htbp]
\centering
\makebox[\textwidth][c]{
\begin{minipage}{1.06\textwidth}
\centering
\includegraphics[width=0.49\textwidth,trim=35 10 35 20,clip]{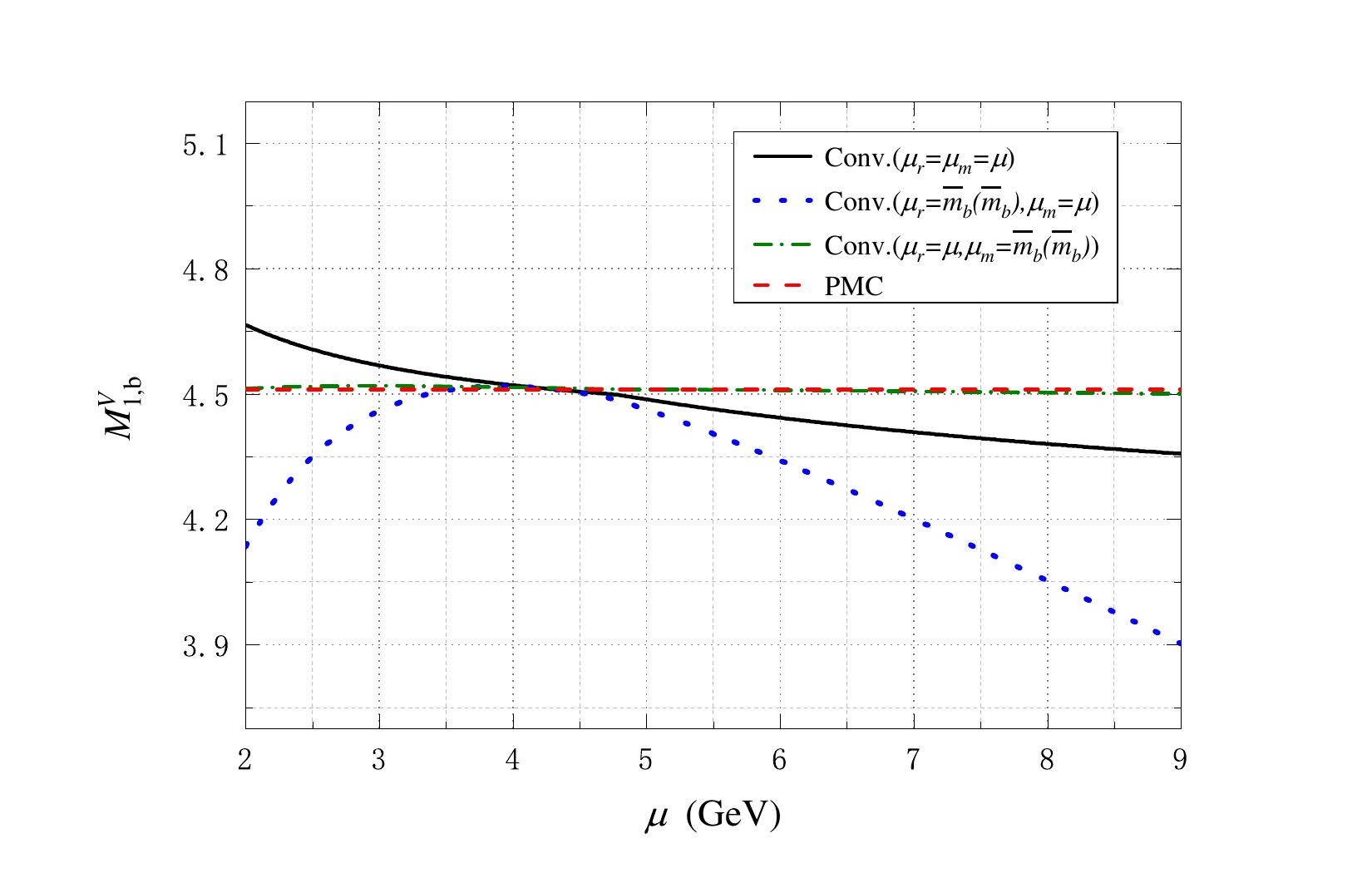}
\hfill
\includegraphics[width=0.49\textwidth,trim=35 10 35 20,clip]{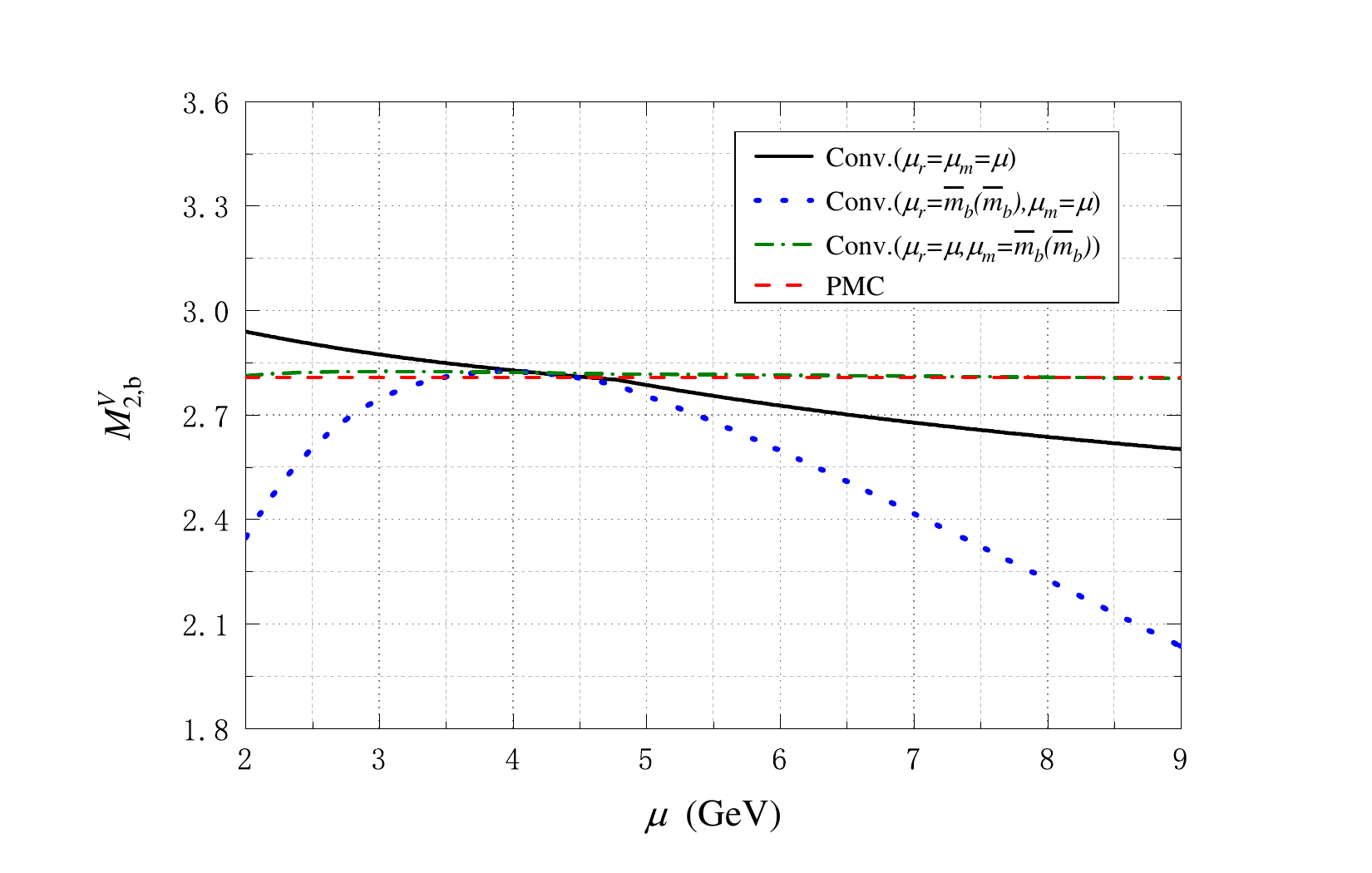}

\vspace{0.35cm}

\includegraphics[width=0.49\textwidth,trim=35 10 35 20,clip]{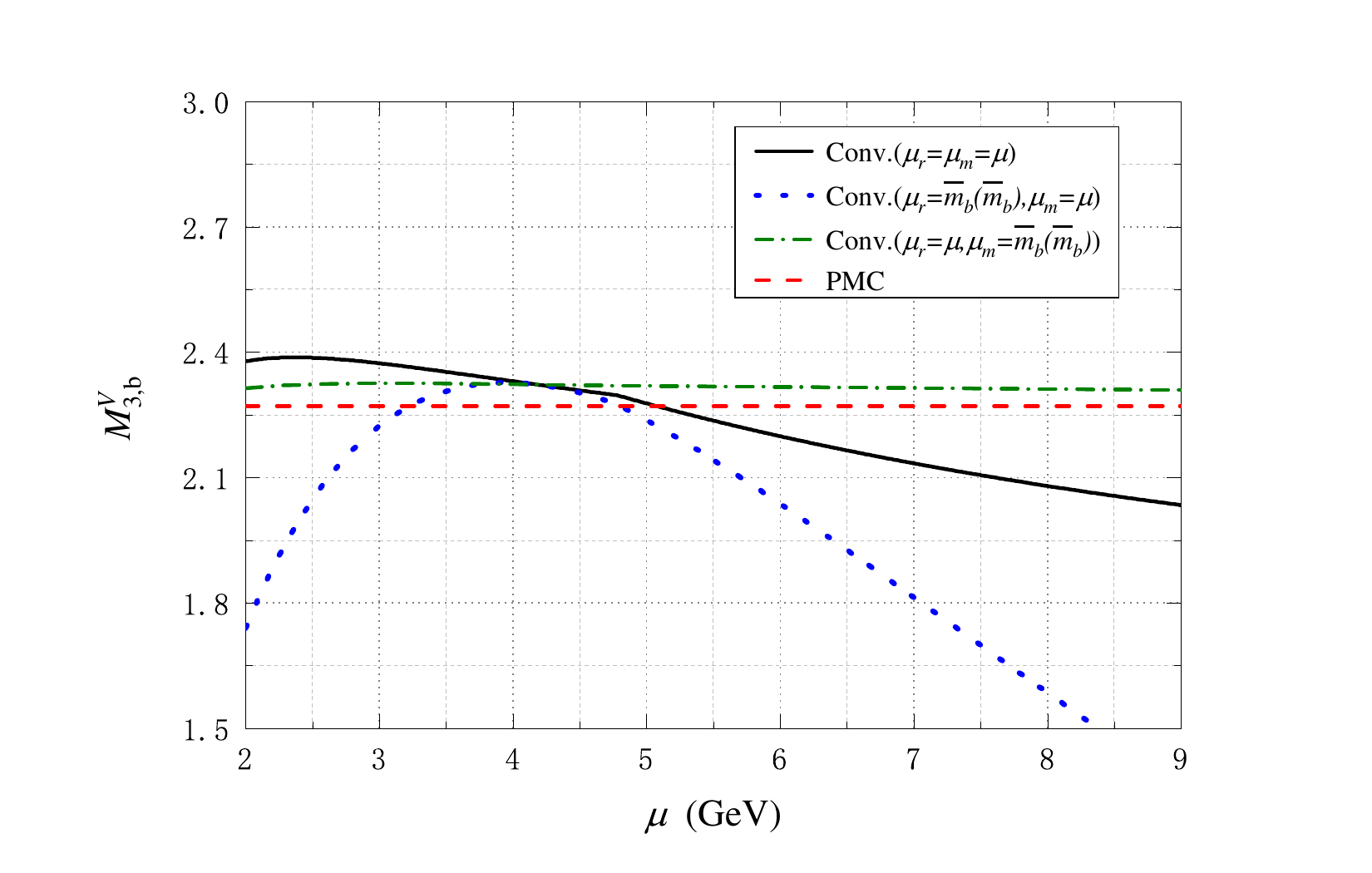}
\hfill
\includegraphics[width=0.49\textwidth,trim=35 10 35 20,clip]{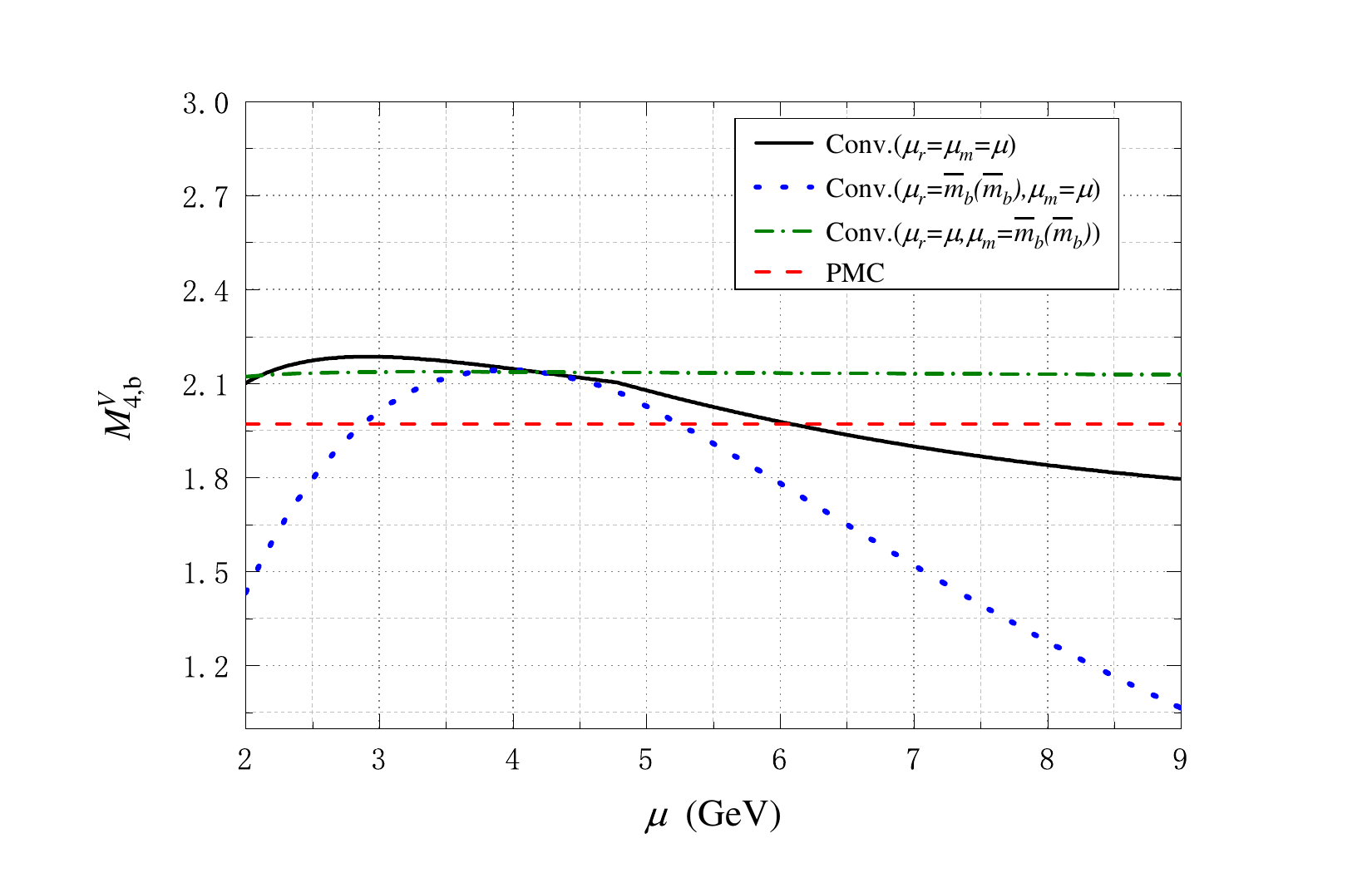}
\end{minipage}
}
\caption{The moments $M^{V,{\rm th}}_{n,b}$ (in units of $10^{-(2n+1)}{\rm GeV}^{-2n}$) versus the scale $\mu$ in the range $\mu\in[2,10]$ GeV, $n=(1,2,3,4)$. The solid, dotted and dash-dotted lines represent results under conventional scale-setting approach by taking [$\mu_r=\mu_m=\mu$], [$\mu_r=\overline{m}_b(\overline{m}_b)$ and $\mu_m=\mu$], [$\mu_r=\mu$ and $\mu_m=\overline{m}_b(\overline{m}_b)$], respectively. The dashed line represents the scale-invariant PMC prediction.}
    \label{mnVb}
\end{figure}

\begin{table}
\centering
\small
\setlength{\tabcolsep}{0.5pt}
\renewcommand{\arraystretch}{1.5}
\makebox[\textwidth][c]{
\begin{tabular}{c c c c c c c c c c c c c c c}
\toprule
 &~Conv.~  &~LO~ &~NLO~  &~$\rm N^2LO$~ &~$\rm N^3LO$~  &~n.p.~  &~total~\\
\midrule
&$M^{P,\rm th}_{0,b}$   &3.333  &$0.560^{+0.196}_{-0.118}\pm0$  &$0.009^{-0.200}_{+0.080}$$^{-0.029}_{+0.022}$  &$-0.006^{+0.012}_{+0.028}$$^{-0.002}_{+0.004}$  &$(4.2^{-0.1}_{+0.0}$$^{-0.9}_{+0.3})\times10^{-5}$ &$3.896^{+0.008}_{-0.010}$$^{-0.031}_{+0.026}$ \\
&$M^{P,\rm th}_{1,b}$   &$1.905\pm0^{-0.599}_{+0.630}$  &$0.531^{+0.186}_{-0.112}$$^{+0.181}_{-0.564}$  &$0.139^{-0.082}_{+0.027}$$^{+0.124}_{-0.315}$  &$0.008^{-0.098}_{+0.049}$$^{+0.056}_{-0.156}$  &$(2.8\mp0.2^{-0.8}_{+0.3})\times10^{-5}$ &$2.583^{+0.006}_{-0.036}$$^{-0.238}_{-0.405}$ \\
&$M^{P,\rm th}_{2,b}$   &$1.555\pm0^{-0.825}_{+1.200}$  &$0.445^{+0.155}_{-0.094}$$^{+0.154}_{-1.266}$  &$0.166^{-0.029}_{+0.004}$$^{+0.165}_{-0.464}$  &$0.028^{-0.099}_{+0.042}$$^{+0.102}_{-0.179}$  &$(-0.8\mp0.0^{+0.3}_{-0.1})\times10^{-5}$ &$2.194^{+0.027}_{-0.048}$$^{-0.404}_{-0.709}$ \\
&$M^{P,\rm th}_{3,b}$   &$1.481\pm0^{-1.005}_{+2.011}$  &$0.374^{+0.130}_{-0.079}$$^{+0.128}_{-2.552}$  &$0.172^{+0.003}_{-0.008}$$^{+0.174}_{-0.366}$  &$0.039^{-0.094}_{+0.035}$$^{+0.135}_{-0.088}$  &$(-7.5^{+0.9}_{-0.5}$$^{+3.3}_{-1.5})\times10^{-5}$ &$2.066^{+0.039}_{-0.052}$$^{-0.568}_{-0.995}$ \\
\addlinespace[1em]
&$M^{V,\rm th}_{1,b}$   &$3.810\pm{0}_{+1.260}^{-1.199}$  &$0.657_{-0.139}^{+0.229}$$_{-1.263}^{+0.490}$ &$0.058_{+0.076}^{-0.195}$$_{-0.567}^{+0.258}$ &$-0.010_{+0.045}^{-0.036}$$_{-0.194}^{+0.071}$ &$(-6.7_{-0.2}^{+0.3}$$_{-0.8}^{+2.0})\times10^{-5}$ &$4.515_{-0.018}^{-0.002}$$_{-0.764}^{-0.380}$ \\
&$M^{V,\rm th}_{2,b}$   &$2.333\pm{0}_{+1.799}^{-1.237}$  &$0.408_{-0.086}^{+0.142}$$_{-2.099}^{+0.369}$  &$0.085_{+0.029}^{-0.081}$$_{-0.570}^{+0.276}$ &$-0.005_{+0.038}^{-0.069}$$_{-0.105}^{+0.118}$ &$(-14.1^{+1.4}_{-0.8}$$^{+5.3}_{-2.2})\times10^{-5}$ &$2.821_{-0.019}^{-0.008}$$_{-0.975}^{-0.474}$\\
&$M^{V,\rm th}_{3,b}$   &$1.975\pm{0}_{+2.681}^{-1.339}$  &$0.272_{-0.058}^{+0.095}$$_{-3.709}^{+0.324}$ &$0.078_{+0.011}^{-0.037}$$_{-0.247}^{+0.280}$ &$-0.003_{+0.031}^{-0.067}$$_{+0.108}^{+0.153}$ &$(-25.5^{+3.6}_{-2.3}$$^{+11.4}_{-5.0})\times10^{-5}$ &$2.322_{-0.016}^{-0.009}$$_{-1.167}^{-0.582}$\\ 
&$M^{V,\rm th}_{4,b}$   &$1.924\pm{0}_{+4.113}^{-1.499}$  &$0.152_{-0.032}^{+0.053}$$_{-6.726}^{+0.334}$ &$0.066_{-0.002}^{-0.002}$$_{+0.950}^{+0.281}$ &$-0.005_{+0.024}^{-0.065}$$_{+0.420}^{+0.180}$ &$(-42.5^{+7.9}_{-5.3}$$^{+21.6}_{-10.2})\times10^{-5}$ &$2.137_{-0.010}^{-0.014}$$_{-1.243}^{-0.704}$\\
\bottomrule
\end{tabular}}
\caption{The perturbative contributions up to N$^3$LO-level for bottomonium moments in the P- and V-channels under conventional scale-setting approach (in units of $10^{-(2n+1)}\,\mathrm{GeV}^{-2n}$). The central values correspond to $\mu_r=\overline{m}_b(\overline{m}_b)$ and $\mu_m=\overline{m}_b(\overline{m}_b)$. The first uncertainty arises from $\mu_r\in[2,10]\,\mathrm{GeV}$ with $\mu_m=\overline{m}_b(\overline{m}_b)$ fixed, and the second uncertainty from $\mu_m\in[2,10]\,\mathrm{GeV}$ with $\mu_r=\overline{m}_b(\overline{m}_b)$ fixed. Nonperturbative contributions are labeled ``n.p.".}
\label{tab1b}
\end{table}

\begin{table}
\centering
\begin{tabular}{c c c c c c c c c c c c}
\toprule
 &~PMC~  &~LO~ &~NLO~  &~$\rm N^2LO$~ &~$\rm N^3LO$~   &~n.p.~ &~total~\\
\midrule
&$M^{P,\rm th}_{0,b}$   &3.333  &0.511 &0.085 &$-0.033$ &$+4.4\times10^{-5}$ &3.896\\
&$M^{P,\rm th}_{1,b}$   &1.927  &0.532 &0.135 &$-0.013$ &$+2.8\times10^{-5}$ &2.581\\
&$M^{P,\rm th}_{2,b}$   &1.654  &0.461 &0.078 &$-0.007$ &$-0.8\times10^{-5}$ &2.186\\
&$M^{P,\rm th}_{3,b}$   &1.785  &0.428 &$-0.101$ &$-0.070$ &$-8.2\times10^{-5}$ &2.042\\
\addlinespace[1em]
&$M^{V,\rm th}_{1,b}$   &3.860  &0.658 &0.052 &$-0.059$ &$-6.8\times10^{-5}$ &4.511\\
&$M^{V,\rm th}_{2,b}$   &2.512  &0.426 &$-0.066$ &$-0.065$ &$-14.7\times10^{-5}$ &2.807\\
&$M^{V,\rm th}_{3,b}$  &2.479  &0.321 &$-0.353$ &$-0.176$ &$-28.7\times10^{-5}$ &2.271\\ 
&$M^{V,\rm th}_{4,b}$   &3.321  &0.235 &$-1.063$ &$-0.522$ &$-55.3\times10^{-5}$ &1.970\\
\bottomrule
\end{tabular}
\caption{The scale-invariant PMC predictions for the bottomonium moments in the P- and V-channels (in units of $10^{-(2n+1)}{\rm GeV}^{-2n}$), in which the scales $\mu_{r}$ and $\mu_{m}$ are varied independently within the range of $[2,10]\,\mathrm{GeV}$.}
\label{tab1b2}
\end{table}

Figs.~\ref{mnPb} and \ref{mnVb} display the theoretical predictions to the moments $M^{X,\rm pert}_{n,b}$ as a function of the scale $\mu$ under both conventional and PMC scale-setting approaches, respectively. Under conventional scale-setting approach, we examine three typical choices for the renormalization scale: (i) $\mu_r=\mu_m=\mu$, (ii) $\mu_r=\overline{m}_b(\overline{m}_b)$ with $\mu_m=\mu$, (iii) $\mu_r=\mu$ with $\mu_m=\overline{m}_b(\overline{m}_b)$. For the bottomonium case, the scale $\mu$ is varied within [2,10] GeV. Although the dependence on $\mu_{r,m}$ is suppressed by the $\alpha_s$ in the bottomonium energy region, the conventional predictions for $M^{X,\rm th}_{n,b}|_{\rm Conv.}$ still exhibit a significant reliance on the choice of the renormalization scale $\mu_{r,m}$, especially for higher $n$. In contrast to the scale-dependent conventional predictions, the PMC predictions are demonstrably scale-independent for all $n$. Results for the perturbative and nonperturbative contributions (``n.p.") to the moments $M^{X}_{n,b}$ under conventional and PMC scale-setting approaches are listed in Tables~\ref{tab1b} and~\ref{tab1b2}, respectively. A better convergent behavior compared with the charmonium case is observed, owing to stronger $\alpha_s$-power suppression, where $\alpha_s(Q_*\sim \mathcal{O}(m_b))\sim 0.1$.

\subsection{Estimated $\text{N}^4\text{LO}$ contribution using the Pad\'{e} approximation approach}  
\label{sec3.3}

For the perturbative series of the moments $M_{n,q}^{X,\rm th}$, achieving reliable theoretical predictions requires not only the elimination of scale dependence but also an accurate estimation of the UHO terms. Under the conventional scale-setting method, scale uncertainties (e.g., $\mu_r$- and $\mu_m$-uncertainties here), as shown in Tables~\ref{tab1c} and \ref{tab1b}, are generally regarded as indicators of UHO contributions. However, this treatment of UHO contributions is not precise, since it only provides a rough estimation of non-conformal UHO terms and fails to account for the contributions of the scale-invariant conformal UHO terms. Since a complete $\rm N^4LO$ calculation is not expected to be available in the near future, these UHO terms must be estimated carefully. It has been noted that conventional perturbative series -- being inherently scheme- and scale-dependent -- are incapable of providing a reliable estimate for such UHO contributions. In this work, we apply the PMC series together with the Pad\'{e} approximation approach (PAA)~\cite{Basdevant:1972fe, Samuel:1992qg, Samuel:1995jc} to estimate the unknown $\rm N^4LO$ terms, i.e., the $r^{X}_{4,0}a_s^4(Q_*)$ term, which yields $\Delta M^{X,\rm pert}_{n,q}\big|^{\rm N^4LO}_{\rm PMC+PAA}$. 

The PAA method is a type of resummation technique that employs a fractional generating function. The generating function builds a bridge between the series and the functions using the formal $\alpha_s$-power series, which allows us to use functional analysis to investigate the properties of the fixed-order series. Its accuracy can be systematically improved by incorporating additional known higher-order loop terms, which constrains the form of the fractional generating function, enables a more reliable description of the perturbative series, and suppresses uncertainties from UHO terms -- thereby facilitating better estimates of UHO contributions. Given different sets of loop terms, various types of generating functions can be constructed.

\begin{table}
\center
\small
\setlength{\tabcolsep}{2pt}
\renewcommand{\arraystretch}{1.5}
\makebox[\textwidth][c]{
\begin{tabular}{c c c c c c c c c c c c c}
\toprule
  ~~~   &~$\Delta M^{P,\rm pert}_{n,c}|^{\rm N^4LO}_{\rm PMC+PAA}$~    &~$\Delta M^{V,\rm pert}_{n,c}|^{\rm N^4LO}_{\rm PMC+PAA}$~     &~$\Delta M^{P,\rm pert}_{n,b}|^{\rm N^4LO}_{\rm PMC+PAA}$~   &~$\Delta M^{V,\rm pert}_{n,b}|^{\rm N^4LO}_{\rm PMC+PAA}$~   \\
\midrule
{$n=0$}  
&$-0.0552^{-0.0068}_{+0.0059}|_{[0/2]}$  &-   &$-0.0132\mp0.0007|_{[0/2]}$  &-\\
&$+0.0536^{+0.0066}_{-0.0058}|_{[1/1]}$  &-   &$+0.0128\pm0.0007|_{[1/1]}$    &-\\

{$n=1$}  
&$-0.0645^{-0.0087}_{+0.0074}|_{[0/2]}$  &$-0.0403^{-0.0054}_{+0.0046}|_{[0/2]}$   &$-0.0154\mp0.0010|_{[0/2]}$  &$-0.0097\mp0.0006|_{[0/2]}$\\\
&$+0.0052^{+0.0007}_{-0.0006}|_{[1/1]}$  &$+0.2835^{+0.0378}_{-0.0326}|_{[1/1]}$   &$+0.0012\pm0.0001|_{[1/1]}$    &$+0.0682^{+0.0043}_{-0.0040}|_{[1/1]}$\\

{$n=2$}  
&$-0.0213^{-0.0027}_{+0.0024}|_{[0/2]}$  &$+0.0971^{+0.0121}_{-0.0105}|_{[0/2]}$   &$-0.0046\mp0.0003|_{[0/2]}$  &$+0.0215^{+0.0013}_{-0.0012}|_{[0/2]}$\\\
&$+0.0030^{+0.0004}_{-0.0003}|_{[1/1]}$  &$-0.2861^{-0.0357}_{+0.0311}|_{[1/1]}$   &$+0.0007\pm0.0000|_{[1/1]}$    &$-0.0634^{-0.0039}_{+0.0037}|_{[1/1]}$\\
{$n=3$}  
&$+0.2014^{+0.0231}_{-0.0203}|_{[0/2]}$  &$+4.1743^{+0.4495}_{-0.3994}|_{[0/2]}$   &$+0.0389^{+0.0023}_{-0.0022}|_{[0/2]}$  &$+0.8133^{+0.0482}_{-0.0455}|_{[0/2]}$\\\
&$-0.2539^{-0.0291}_{+0.0256}|_{[1/1]}$  &$-0.4497^{-0.0484}_{+0.0430}|_{[1/1]}$   &$-0.0490^{-0.0029}_{+0.0028}|_{[1/1]}$    &$-0.0876^{-0.0052}_{+0.0049}|_{[1/1]}$\\

{$n=4$}  
&-  &-   &-  &$+26.488^{+1.4808}_{-1.4032}|_{[0/2]}$\\\
&-  &-   &-    &$-0.2568^{-0.0144}_{+0.0136}|_{[1/1]}$\\
\bottomrule
\end{tabular}}
\caption{Two typical PAA estimations for the $\rm N^4LO$-terms of the moments $M^{X,\rm pert}_{n,q}$ using the PMC series. The predictions include the $\Delta\alpha_s(m_z)$-uncertainty for $\Delta M^{X,\rm pert}_{n,q}|^{\rm N^4LO}_{\rm PMC+PAA}$.}
\label{paa2}
\end{table}

Following the PAA, there are several $[N/M]$ types for the generating function, which consist the systematic error of the method itself. For definiteness, the presently considered perturbative part of the moments $M^{X}_{n,q}$ up to $\mathcal{O}(a^3_s)$, two typical types that are frequently adopted for the generating functions, namely the $[0/2]$ type and the diagonal type ($[1/1]$ type)~\cite{Du:2018dma}, in which the $[0/2]$ type is more consistent with the Generalized Crewther Relations~\cite{Shen:2016dnq}. Following the standard procedure for determining the N$^4$LO-level generating function of the moments $M^{X,\rm pert}_{n,q}$, for the $[0/2]$ type we have
\begin{eqnarray}
r^{X}_{4,0}=&& \frac{r^{X}_{2,0}\left(2r^{X}_{1,0}r^{X}_{3,0}-r^{X,2}_{2,0}\right)}{r^{X,2}_{1,0}},
\label{PAA02}
\end{eqnarray}
and for the [1/1] type~\cite{Du:2018dma}, we have
\begin{eqnarray}
r^{X}_{4,0}= \frac{r^{X,2}_{3,0}}{r^{X}_{2,0}}.
\label{PAA11}
\end{eqnarray}
 
Table~\ref{paa2} shows that the $\mathrm{N^4LO}$ terms exhibit relatively large magnitudes and are non-negligible within a slowly converging perturbative series; such contributions are further enhanced for higher moments. For the perturbative part of the heavy quarkonium moments $M^{X}_{n,q}$, the $[0/2]$ type generally performs better, following the convergent behavior of the given series. However, the diagonal $[1/1]$ type sometimes behaves better due to the oscillatory and divergent behavior of the series and accidentally large cancellations among different orders~\cite{Wu:2019mky, Gardi:1996iq, Cvetic:1997qm}. 

A comparison of the theoretical moments $M^{X,\mathrm{th}}_{n,q}|_{\mathrm{PMC}}$ obtained from the PMC approach with the results using experimental data and lattice QCD simulations will be given in Subsection~\ref{sec.3.3}.

\subsection{Experimental and lattice results for the moments $M^{V,\mathrm{exp}}_{n,q}$ and $M^{P,\mathrm{lat}}_{n,q}$} \label{sec3.4}

\begin{table*}
\center
\begin{tabular}{ c c  c  c c c c c c}
\toprule
   &Charm case~    &~$n=1$~     &~$n=2$~   &~$n=3$~    &~$n=4$~\\
\midrule
 &~\cite{Dehnadi:2011gc} $M^{V,\rm exp}_{n,c}$  &$2.121\pm0.036$  &$1.478\pm0.028$  &$1.302\pm0.027$   &$1.243\pm0.028$                   \\
&~\cite{Chetyrkin:2017lif}  $M^{V,\rm exp}_{n,c}$    &$2.154\pm{0.023}$  &$1.490\pm{0.017}$  &$1.308\pm{0.016}$ &$1.248\pm{0.016}$\\
\bottomrule
\end{tabular}
\caption{The vector moments $M^{V,\rm exp}_{n,c}$ for the charm quark (in units of $10^{-n}{\rm GeV}^{-2n}$), $n=1,2,3,4$.}
\label{expMnVc}
\end{table*}

\begin{table*}
\center
\begin{tabular}{c c c c c c c c c}
\toprule
 & ~Bottom case~~   &~$n=1$~     &~$n=2$~   &~$n=3$~    &~$n=4$~\\
\midrule
&\cite{Kuhn:2007vp}  $M^{V,\rm exp}_{n,b}$  &$4.601\pm0.043$  &$2.881\pm0.037$   &$2.370\pm0.034$    &$2.178\pm0.032$\\
&\cite{Dehnadi:2015fra}  $M^{V,\rm exp}_{n,b}$ &$4.526\pm0.112$   &$2.834\pm0.052$   &$2.338\pm0.036$   &$2.154\pm0.030$\\
&\cite{Chetyrkin:2009fv} $M^{V,\rm exp}_{n,b}$ &$4.592\pm0.031$   &$2.872\pm0.028$   &$2.362\pm0.026$   &$2.170\pm0.026$  \\
\bottomrule
\end{tabular}
\caption{The vector moments $M^{V,\rm exp}_{n,b}$ for the bottom quark (in units of $10^{-(2n+1)}{\rm GeV}^{-2n}$), $n=1,2,3,4$.}
\label{expMnVb}
\end{table*}

The moments for the vector channel $M^V_{n,q}$ can be obtained from experimental measurements of the ratio $R_{e^+e^-\to q\bar{q}+X}(s)$, which is defined as
\begin{eqnarray}
R_{e^+e^-\rightarrow q\bar{q}+X}(s)=\frac{\sigma_{e^+e^-\rightarrow q\bar{q}+X}(s)}{\sigma_{e^+e^-\rightarrow\mu^+\mu^-+X}(s)}.
\end{eqnarray}
Specifically, they are defined as weighted integrals of the normalized hadronic cross section:
\begin{eqnarray}
M^V_{n,q}&=&\int\frac{ds}{s^{n+1}}R_{e^+e^-\rightarrow q\bar{q}+X}(s), \label{mnexp}
\end{eqnarray}
where $q=c,b$ and $n\geq1$. A substantial amount of experimental measurements for the ratio $R_{e^+e^- \to c\bar{c} + X}(s)$ is available over a region from 2 GeV to 10.538 GeV, including BES~\cite{BES:1999wbx, BES:2001ckj, BES:2004hbv, BES:2006dso, Ablikim:2006mb, BES:2009ejh}, CrystalBall~\cite{Osterheld:1986hw, Edwards:1990pc}, CLEO~\cite{CLEO:1997eca, CLEO:1984vfn, CLEO:2007suf, CLEO:2008ojp}, MD1~\cite{Blinov:1993fw}, PLUTO~\cite{Criegee:1981qx}, MARK I~\cite{Siegrist:1976br, Rapidis:1977cv, Siegrist:1981zp}, and MARK II~\cite{Abrams:1979cx}. By incorporating these experimental data together with a theoretical analysis~\cite{Gorishnii:1988bc, Gorishnii:1991se} above 10.538 GeV as an input, one determines the experimental moments $M^{V,\text{exp}}_{n,c}$~\cite{Dehnadi:2011gc, Chetyrkin:2017lif} by Eq.~\eqref{mnexp}. In this paper, we do not discuss the explicit calculations of these moments, but instead employ the latest results from Refs.~\cite{Kuhn:2007vp, Chetyrkin:2009fv, Dehnadi:2011gc, Dehnadi:2015fra, Chetyrkin:2017lif}. The resulting vector channel experimental moments $M^{V,\text{exp}}_{n,q}$ for the charmonium and bottomonium are summarized in Tables~\ref{expMnVc} and~\ref{expMnVb}.

\begin{table}
\center
\small
\setlength{\tabcolsep}{2pt}
\makebox[\textwidth][c]{
\begin{tabular}{c c c c c c c c}
\toprule
 & ~Charm case~~   &~\cite{Maezawa:2016vgv}~  &~\cite{Petreczky:2019ozv}~ &~\cite{HPQCD:2008kxl}~   &~\cite{McNeile:2010ji}~   &~\cite{Petreczky:2020tky}~  \\
\midrule
&$M^{P,\rm lat}_{0,c}$    &-                  &$1.705\pm0.005$ &$1.708\pm0.007$   &$1.708\pm0.005$   &$1.7037\pm0.0027$  \\
&$M^{P,\rm lat}_{1,c}$    &$1.385\pm0.007$  &$1.386\pm0.005$   &$1.404\pm0.019$   &$1.395\pm0.005$   &$1.387\pm0.004$  \\
&$M^{P,\rm lat}_{2,c}$    &$1.345\pm0.032$    &$1.349\pm0.012$ &$1.359\pm0.041$   &$1.365\pm0.012$  &$1.344\pm0.010$  \\
&$M^{P,\rm lat}_{3,c}$    &$1.406\pm0.048$   &$1.461\pm0.050$  &$1.425\pm0.059$   &$1.415\pm0.010$   &$1.395\pm0.022$  \\
\bottomrule
\end{tabular}}
\caption{The pseudoscalar moments $M^{P,\rm lat}_{n,c}$ for the charm quark are presented in units of $10^{-n}{\rm GeV}^{-2n}$ for $n=0,1,2,3$.}
\label{expMnPc}
\end{table}

At present, the experimental analysis on the pseudoscalar moments $M^{P,\text{exp}}_{n,c}$ are not available. So we adopt the lattice simulations of $M^{P,\text{lat}}_{n,c}$~\cite{HPQCD:2008kxl, McNeile:2010ji, Nakayama:2016atf, Maezawa:2016vgv, Petreczky:2019ozv, Petreczky:2020tky} to do our following comparison with the PMC predictions. To derive the regular moments, we use the relation between the reduced moments $\mathcal{R}_k$ and the moments $M^{P,\text{lat}}_{n,c}$,
\begin{equation}
	M^{P,\text{lat}}_{n,c} = r^X_0 \left( \frac{\mathcal{R}_{2n+4}}{m_{\eta_c}} \right),
\end{equation}
where ``lat" is short notation for the result derived using the Lattice QCD calculation, the coefficients $r^{X}_{0}$ are taken from the pQCD series in Eq.~\eqref{Mnq}, and the charmonium mass of $\eta_c$ is set to $m_{\eta_c} = 2983.9 \pm 0.4$ MeV, following Refs.~\cite{HPQCD:2008kxl, McNeile:2010ji}. Utilizing recent Lattice QCD results for the reduced moments $\mathcal{R}_k$ from Refs.~\cite{HPQCD:2008kxl, McNeile:2010ji, Maezawa:2016vgv, Petreczky:2019ozv, Petreczky:2020tky}, we present the calculated charm moments in the pseudoscalar channel $M^{P,\text{lat}}_{n,c}$ in Table~\ref{expMnPc}. It should be noted that corresponding lattice simulation results for the bottom quark case $M^{P,\text{lat}}_{n,b}$ are currently unavailable.

\subsection{Extraction of $\overline{m}_{q}(\overline{m}_{q})$ via the moments $M^{X}_{n,q}$}
\label{sec.3.3}

Theoretical expressions for these moments, $M^{X,\text{th}}_{n\neq0,q}$, take the form $(\overline{m}_q)^{-2n}$, making them highly sensitive to the quark mass values. Therefore, by comparing the theoretical predictions of these moments with the corresponding experimental data for $M^{V,\text{exp}}_{n,c}$~\cite{Dehnadi:2011gc,Chetyrkin:2017lif}, $M^{V,\text{exp}}_{n,b}$~\cite{Kuhn:2007vp,Chetyrkin:2009fv,Dehnadi:2015fra}, and lattice QCD results for $M^{P,\text{lat}}_{n,c}$~\cite{HPQCD:2008kxl,McNeile:2010ji,Maezawa:2016vgv,Petreczky:2019ozv,Petreczky:2020tky}, the charm and bottom quark masses can be determined inversely with high accuracy. In this subsection, we perform such a determination of the $\overline{m}_{c,b}(\overline{m}_{c,b})$ value by inversely using the aforementioned experimental and lattice QCD moments $M^{X}_{n,q}$.

\begin{table}[htb]
\centering
\setlength{\tabcolsep}{1.5pt}
\makebox[\textwidth][c]{
\begin{tabular}{cc r c c c c c r c c c c}
\toprule
& & \multicolumn{6}{c}{Conv.} & \multicolumn{5}{c}{PMC} \\
\cmidrule(lr){3-8}\cmidrule(lr){9-13}
& Moments & $\overline m_c(\overline m_c)$ & $\Delta_{\mu_r,m}$ & $\Delta_{\alpha_s}$ & $\Delta_{\rm n.p.}$ & $\Delta_{\rm lat}$ & ${\Delta_{\rm input}}$ & $\overline m_c(\overline m_c)$ & $\Delta_{\alpha_s}$ & $\Delta_{\rm n.p.}$ & $\Delta_{\rm lat}$ & ${\Delta_{\rm input}}$ \\
\midrule
\cite{Maezawa:2016vgv} & $M^P_{1,c}$ & 1295.5 & $^{+31.2}_{-73.7}$ & $^{+10.0}_{-9.6}$ & $\pm0.8$ & $^{-2.6}_{+2.7}$ & {$^{+32.9}_{-74.4}$} & 1291.6 & $^{+9.7}_{-9.3}$ & $\pm0.8$ & $\mp{2.7}$ & {$^{+10.1}_{-9.7}$} \\
   & $M^P_{2,c}$ & 1286.5 & $^{+16.6}_{-140.8}$ & $^{+7.0}_{-6.7}$ & {$\mp0.2$} & $^{-6.5}_{+6.7}$ & {$^{+19.2}_{-141.1}$} & 1280.3 & $^{+6.5}_{-6.2}$ & $\mp0.2$ & {$^{-6.6}_{+6.8}$} & {$^{+9.3}_{-9.2}$} \\
   & $M^P_{3,c}$ & 1278.9 & $^{+3.2}_{-144.2}$ & $^{+5.2}_{-5.0}$ & {$\mp0.8$} & $^{-6.4}_{+6.7}$ & {$^{+8.9}_{-144.4}$} & 1264.8 & $^{+4.0}_{-3.8}$ & {$\mp1.5$} & $^{-6.5}_{+6.8}$ & {$^{+7.8}_{-7.9}$} \\
\addlinespace
\cite{Petreczky:2019ozv} & $M^P_{1,c}$ & 1295.1 & $^{+31.2}_{-73.8}$ & $^{+10.0}_{-9.6}$ & $\pm0.8$ & {$\mp1.9$} & {$^{+32.8}_{-74.5}$} & 1291.2 & $^{+9.7}_{-9.3}$ & $\pm0.8$ & {$\mp1.9$} & {$^{+9.9}_{-9.5}$} \\
   & $M^P_{2,c}$ & 1285.7 & $^{+16.6}_{-141.1}$ & $^{+7.0}_{-6.7}$ & {$\mp0.2$} & {$\mp2.5$} & {$^{+18.2}_{-141.3}$} & 1279.5 & $^{+6.5}_{-6.2}$ & {$\mp0.2$} & {$\mp2.5$} & {$^{+7.0}_{-6.7}$} \\
   & $M^P_{3,c}$ & 1271.5 & $^{+3.3}_{-155.6}$ & {$^{+5.3}_{-5.0}$} & {$\mp0.8$} & $^{-6.4}_{+6.7}$ & {$^{+8.9}_{-155.8}$} & 1257.3 & $^{+4.0}_{-3.8}$ & {$\mp1.5$} & $^{-6.5}_{+6.8}$ & {$^{+7.8}_{-7.9}$} \\
\addlinespace
\cite{HPQCD:2008kxl} & $M^P_{1,c}$ & 1288.4 & $^{+30.6}_{-73.9}$ & $^{+10.1}_{-9.7}$ & $\pm0.8$ & $^{-7.0}_{+7.1}$ & {$^{+33.0}_{-74.9}$} & 1284.5 & $^{+9.7}_{-9.3}$ & $\pm0.8$ & $^{-7.0}_{+7.1}$ & {$^{+12.0}_{-11.7}$} \\
   & $M^P_{2,c}$ & 1283.7 & $^{+16.5}_{-141.8}$ & $^{+7.0}_{-6.7}$ & {$\mp0.2$} & $^{-8.2}_{+8.6}$ & {$^{+19.7}_{-142.2}$} & 1277.4 & {$^{+6.6}_{-6.2}$} & {$\mp0.2$} & $^{-8.3}_{+8.6}$ & {$\pm10.6$} \\
   & $M^P_{3,c}$ & 1276.3 & $^{+3.2}_{-147.7}$ & $^{+5.3}_{-5.0}$ & {$\mp0.8$} & $^{-7.7}_{+8.1}$ & {$^{+9.9}_{-148.0}$} & 1262.2 & $^{+4.0}_{-3.8}$ & {$\mp1.5$} & $^{-7.9}_{+8.3}$ & {$^{+9.0}_{-9.3}$} \\
\addlinespace
\cite{McNeile:2010ji} & $M^P_{1,c}$ & 1291.8 & $^{+30.9}_{-73.8}$ & $^{+10.1}_{-9.6}$ & $\pm0.8$ & {$\mp1.9$} & {$^{+32.6}_{-74.5}$} & 1287.8 & $^{+9.7}_{-9.3}$ & $\pm0.8$ & {$\mp1.9$} & {$^{+9.9}_{-9.5}$} \\
   & $M^P_{2,c}$ & 1282.4 & $^{+16.5}_{-142.3}$ & $^{+7.0}_{-6.7}$ & {$\mp0.2$} & $^{-2.4}_{+2.5}$ & {$^{+18.1}_{-142.5}$} & 1276.2 & $^{+6.6}_{-6.2}$ & {$\mp0.2$} & $^{-2.4}_{+2.5}$ & {$^{+7.0}_{-6.7}$} \\
   & $M^P_{3,c}$ & 1277.7 & $^{+3.2}_{-145.8}$ & $^{+5.2}_{-5.0}$ & {$\mp0.8$} & {$\mp1.4$} & {$^{+6.3}_{-145.9}$} & 1263.5 & $^{+4.0}_{-3.8}$ & {$\mp1.5$} & {$\mp1.4$} & {$^{+4.5}_{-4.3}$} \\
\addlinespace
\cite{Petreczky:2020tky} & $M^P_{1,c}$ & 1294.8 & $^{+31.2}_{-73.8}$ & $^{+10.0}_{-9.6}$ & $\pm0.8$ & {$\mp1.5$} & {$^{+32.8}_{-74.4}$} & 1290.8 & $^{+9.7}_{-9.3}$ & $\pm0.8$ & {$\mp1.5$} & {$^{+9.8}_{-9.5}$} \\
   & $M^P_{2,c}$ & 1286.8 & $^{+16.6}_{-140.7}$ & $^{+7.0}_{-6.7}$ & {$\mp0.2$} & {$\mp2.1$} & {$^{+18.2}_{-140.9}$} & 1280.5 & $^{+6.5}_{-6.2}$ & {$\mp0.2$} & {$\mp2.1$} & {$^{+6.9}_{-6.5}$} \\
   & $M^P_{3,c}$ & 1280.4 & $^{+3.2}_{-142.4}$ & $^{+5.2}_{-5.0}$ & {$\mp0.8$} & $^{-3.0}_{+3.1}$ & {$^{+6.8}_{-142.5}$} & 1266.3 & $^{+4.0}_{-3.8}$ & {$\mp1.5$} & {$\mp3.1$} & {$^{+5.3}_{-5.1}$} \\
\bottomrule
\end{tabular}}
\caption{Determination of the charm quark mass $\overline{m}_c(\overline{m}_c)$ (in units of MeV) via the pseudoscalar ($P$)-channel charmonium moments, before and after applying the PMC method. {The uncertainties (``$\Delta_{\rm input}$") }include variations in the scales ($\mu_{r,m} \in [1,4]\,\text{GeV}$ for $M^{P}_{n\neq3,c}$ and $\mu_{r,m} \in [1,2.6]\,\text{GeV}$ for $M^{P}_{3,c}$), the coupling constant $\Delta\alpha_s(m_Z) = \pm 0.0009$, nonperturbative contributions (``$\Delta {\rm n.p.}$") arising from the gluon condensate term $\Delta\left\langle \alpha_sG^2\right\rangle= \pm 0.012\,\text{GeV}^4$, and the input lattice uncertainties for $M^{P,\rm lat}_{n,c}$ (``$\Delta\rm lat$"). The predictions using the Lattice QCD predictions~\cite{HPQCD:2008kxl, McNeile:2010ji, Maezawa:2016vgv, Petreczky:2019ozv, Petreczky:2020tky} together with their uncertainties are propagated from the errors of $M^{P,\text{lat}}_{n,c}$ listed in Table~\ref{expMnPc}.}
\label{mcmcp}
\end{table}

\begin{table}[htb]
\centering
\setlength{\tabcolsep}{1.5pt}
\makebox[\textwidth][c]{
\begin{tabular}{cc r c c c c c r c c c c}
\toprule
& & \multicolumn{6}{c}{Conv.} & \multicolumn{5}{c}{PMC} \\
\cmidrule(lr){3-8}\cmidrule(lr){9-13}
& Moments & $\overline m_c(\overline m_c)$ & $\Delta_{\mu_r,m}$ & $\Delta_{\alpha_s}$ & $\Delta_{\rm n.p.}$ & $\Delta_{\rm exp}$ & {$\Delta_{\rm input}$} & $\overline m_c(\overline m_c)$ & $\Delta_{\alpha_s}$ & $\Delta_{\rm n.p.}$ & $\Delta_{\rm exp}$ & {$\Delta_{\rm input}$} \\
\midrule
\cite{Dehnadi:2011gc} & $M^V_{1,c}$ & 1293.8 & $^{+44.5}_{-97.8}$ & $^{+5.1}_{-4.9}$ & {$\mp1.5$} & $^{-9.8}_{+10.1}$ & {$^{+45.9}_{-98.5}$} & 1288.2 & $^{+4.4}_{-4.3}$ & {$\mp1.6$} & $^{-9.9}_{+10.2}$ & {$^{+11.0}_{-11.2}$} \\
   & $M^V_{2,c}$ & 1280.7 & $^{+27.7}_{-153.4}$ & $^{+3.4}_{-3.2}$ & {$\mp2.3$} & $^{-5.6}_{+5.8}$ & {$^{+28.6}_{-153.6}$} & 1268.2 & $\pm2.0$ & {$\mp2.9$} & $^{-5.7}_{+5.9}$ & {$^{+6.7}_{-6.9}$} \\
   & $M^V_{3,c}$ & 1274.2 & $^{+18.1}_{-184.4}$ & $^{+2.1}_{-1.9}$ & {$\mp3.2$} & $^{-4.3}_{+4.3}$ & {$^{+19.0}_{-184.5}$} & 1230.6 & $^{+5.8}_{-4.1}$ & $^{-7.5}_{+6.9}$ & $^{-4.7}_{+4.8}$ & {$^{+10.6}_{-9.4}$} \\
\addlinespace
\cite{Chetyrkin:2017lif} & $M^V_{1,c}$ & 1284.8 & $^{+43.5}_{-98.5}$ & $^{+5.1}_{-4.9}$ & {$\mp1.5$} & $^{-6.2}_{+6.3}$ & {$^{+44.3}_{-98.8}$} & 1279.1 & $^{+4.4}_{-4.3}$ & $^{-1.7}_{+1.6}$ & $^{-6.2}_{+6.3}$ & {$\pm{7.8}$} \\
    & $M^V_{2,c}$ & 1278.3 & $^{+27.6}_{-156.9}$ & $^{+3.4}_{-3.2}$ & {$\mp2.3$} & $\mp3.4$ & {$^{+28.1}_{-157.0}$} & 1265.8 & $\pm2.0$ & $^{-3.0}_{+2.9}$ & $\mp3.5$ & {$\pm{5.0}$} \\
    & $M^V_{3,c}$ & 1273.2 & $^{+18.1}_{-181.8}$ & $^{+2.1}_{-1.9}$ & {$\mp3.2$} & $\mp2.5$ & {$^{+18.7}_{-181.9}$} & 1229.5 & $^{-5.8}_{+4.2}$ & {$^{-7.5}_{+6.9}$} & $\mp2.8$ & {$^{+9.9}_{-8.5}$} \\
\bottomrule
\end{tabular}}
\caption{The charm quark mass $\overline{m}_c(\overline{m}_c)$ (in unit: MeV) is determined through the charmonium moments for V-channel, both before and after applying the PMC method. {The uncertainties (``$\Delta_{\rm input}$") }encompass variations from the scales ($\mu_{r,m}\in[1,4]$ GeV for $M^{V}_{1,c}$, $\mu_{r,m}\in[1,2.5]$ GeV for $M^{V}_{2,c}$, and $\mu_{r,m}\in[1,1.8]$ GeV for $M^{V}_{3,c}$), the coupling constant $\Delta\alpha_s(m_Z)=\pm0.0009$, nonperturbative contributions (``$\Delta {\rm n.p.}$") arising from the gluon condensate term $\Delta\left<\alpha_sG^2\right>=\pm0.012~\rm GeV^4$, and the input experimental experimental uncertainties for $M^{V,\rm exp}_{n,c}$ (``$\Delta\rm exp$"). The experimental uncertainties are propagated from the experimental errors in $M^{V,\rm exp}_{n,c}$  listed in Table.~\ref{expMnVc}.}
\label{mcmcv}
\end{table}

\begin{table}[htb]
\centering
\setlength{\tabcolsep}{1.5pt}
\makebox[\textwidth][c]{
\begin{tabular}{cc r c c c c c r c c c c}
\toprule
& & \multicolumn{6}{c}{Conv.} & \multicolumn{5}{c}{PMC} \\
\cmidrule(lr){3-8}\cmidrule(lr){9-13}
& Moments & $\overline m_b(\overline m_b)$ & $\Delta_{\mu_r,m}$ & $\Delta_{\alpha_s}$ & $\Delta_{\rm n.p.}$ & $\Delta_{\rm exp}$ & {$\Delta_{\rm input}$} & $\overline m_b(\overline m_b)$ & $\Delta_{\alpha_s}$ & $\Delta_{\rm n.p.}$ & $\Delta_{\rm exp}$ & {$\Delta_{\rm input}$} \\
\midrule
\cite{Kuhn:2007vp} & $M^V_{1,b}$ & 4178.0 & $^{+67.9}_{-82.7}$ & $^{+5.2}_{-5.1}$ & {$\mp0.1$} & $^{-49.4}_{+51.2}$ & {$^{+84.1}_{-97.4}$} & 4176.1 & $^{+5.1}_{-5.0}$ & {$\mp0.1$} & $^{-49.4}_{+51.2}$ & {$^{+49.7}_{-51.4}$} \\
   & $M^V_{2,b}$ & 4178.2 & $^{+42.5}_{-97.3}$ & $^{+3.2}_{-3.1}$ & {$\mp0.1$} & $^{-18.6}_{+19.1}$ & {$^{+46.5}_{-99.2}$} & 4173.2 & $^{+2.9}_{-2.8}$ & {$\mp0.1$} & $^{-18.6}_{+19.1}$ & {$^{+18.8}_{-19.3}$} \\
   & $M^V_{3,b}$ & 4178.4 & $^{+16.2}_{-107.1}$ & $\pm1.9$ & {$\mp0.2$} & {$^{-10.5}_{+10.7}$} & {$^{+19.4}_{-107.7}$} & 4163.1 & $\pm0.9$ & {$\mp0.2$} & $^{+10.4}_{-10.5}$ & {$^{+10.4}_{-10.5}$} \\
   & $M^V_{4,b}$ & 4178.8 & $^{-8.3}_{-100.5}$ & $\pm1.0$ & {$\mp0.2$} & $^{-7.2}_{+7.3}$ & {$^{+11.0}_{-100.8}$} & 4137.1 & $^{-1.7}_{+1.6}$ & {$\mp0.3$} & $^{-7.1}_{+7.2}$ & {$^{+7.3}_{-7.4}$} \\
\addlinespace
\cite{Dehnadi:2015fra} & $M^V_{1,b}$ & 4144.7 & $^{+66.9}_{-82.7}$ & $^{+5.2}_{-5.1}$ & {$\mp0.1$} & $^{-18.7}_{+19.0}$ & {$^{+69.7}_{-85.0}$} & 4142.9 & $^{+5.1}_{-5.0}$ & {$\mp0.1$} & $^{-18.7}_{+19.0}$ & {$^{+19.4}_{-19.6}$} \\
   & $M^V_{2,b}$ & 4161.3 & $^{+42.3}_{-97.5}$ & $^{+3.2}_{-3.1}$ & {$\mp0.1$} & $^{-13.0}_{+13.2}$ & {$^{+44.4}_{-98.6}$} & 4156.3 & $^{+2.9}_{-2.8}$ & {$\mp0.1$} & $^{-13.0}_{+13.2}$ & {$^{+13.3}_{-13.5}$} \\
   & $M^V_{3,b}$ & 4169.0 & $^{+16.3}_{-107.2}$ & $\pm1.9$ & {$\mp0.2$} & $^{-9.8}_{+10.0}$ & {$^{+19.1}_{-107.7}$} & 4153.9 & $\pm0.9$ & {$\mp0.2$} & $^{-9.6}_{+9.8}$ & {$^{+9.6}_{-9.8}$} \\
   & $M^V_{4,b}$ & 4173.0 & $^{-8.2}_{-100.5}$ & $\pm1.0$ & {$\mp0.2$} & $^{-7.6}_{+7.7}$ & {$^{+11.2}_{-100.8}$} & 4131.5 & $^{-1.7}_{+1.6}$ & {$\mp0.3$} & $^{-7.4}_{+7.6}$ & {$^{+7.6}_{-7.8}$} \\
\addlinespace
\cite{Chetyrkin:2009fv} & $M^V_{1,b}$ & 4148.7 & $^{+67.1}_{-82.7}$ & $^{+5.2}_{-5.1}$ & {$\mp0.1$} & $^{-13.6}_{+13.7}$ & {$^{+68.7}_{-84.0}$} & 4146.8 & $^{+5.1}_{-5.0}$ & {$\mp0.1$} & $^{-13.6}_{+13.7}$ & {$^{+14.5}_{-14.6}$} \\
    & $M^V_{2,b}$ & 4164.5 & $^{+42.3}_{-97.4}$ & $^{+3.2}_{-3.1}$ & {$\mp0.1$} & $^{-9.9}_{+10.0}$ & {$^{+43.6}_{-98.0}$} & 4159.5 & $^{+2.9}_{-2.8}$ & {$\mp0.1$} & $^{-9.9}_{+10.0}$ & {$^{+10.3}_{-10.4}$} \\
    & $M^V_{3,b}$ & 4171.4 & $^{+16.2}_{-107.2}$ & $\pm1.9$ & {$\mp0.2$} & $^{-7.5}_{+7.6}$ & {$^{+18.0}_{-107.5}$} & 4156.2 & $\pm0.9$ & {$\mp0.2$} & $^{-7.4}_{+7.5}$ & {$^{+7.5}_{-7.6}$} \\
    & $M^V_{4,b}$ & 4174.9 & $^{-8.2}_{-100.5}$ & $\pm1.0$ & {$\mp0.2$} & $^{-6.2}_{+6.3}$ & {$^{+10.3}_{-100.7}$} & 4133.3 & $^{-1.7}_{+1.6}$ & {$\mp0.3$} & $^{-6.1}_{+6.2}$ & {$^{+6.3}_{-6.4}$} \\
\bottomrule
\end{tabular}}
\caption{The bottom quark mass $\overline{m}_b(\overline{m}_b)$ (in unit: MeV) is determined through the charmonium moments for V-channel, both before and after applying the PMC method. {The uncertainties (``$\Delta_{\rm input}$") include variations from the scales $\mu_{r,m}\in[2,10]$ GeV,} the error of the coupling constant $\Delta\alpha_s(m_Z)=\pm0.0009$, the nonperturbative contributions (``$\Delta {\rm n.p.}$") from the gluon condensate term $\Delta\left<\alpha_sG^2\right>=\pm0.012~\rm GeV^4$, and the input experimental uncertainties for $M^{V,\rm exp}_{n,b}$ (``$\Delta\rm exp$"). The experimental uncertainties are propagated from the experimental errors in $M^{V,\rm exp}_{n,b}$  listed in Table.~\ref{expMnVb}.}
\label{mbmbv}
\end{table}

Using the currently known perturbative contributions to $M^{X,\text{th}}_{n,q}$ up to $\text{N}^3\text{LO}$ QCD corrections, the charm quark mass in the $\overline{\text{MS}}$ scheme, determined from the pseudoscalar moments $M^{P}_{n,c}$ and vector moments $M^{V}_{n,c}$, is presented in Tables~\ref{mcmcp} and~\ref{mcmcv}, respectively. Similarly, the bottom quark mass $\overline{m}_{b}(\overline{m}_{b})$, extracted from the vector moments $M^{V}_{n,b}$, is listed in Table~\ref{mbmbv}. The input values from Lattice QCD and experimental data analyses for $M^{V,\text{exp}}_{n,q}$ and $M^{P,\text{lat}}_{n,c}$ are used to determine $\overline{m}_q(\overline{m}_q)$ with uncertainties labeled as ``$\Delta \text{exp}$" and ``$\Delta \text{lat}$", respectively. {The error ``$\Delta_{ \rm input}$" is obtained by combining the root-mean-square of correlated errors for the four input parameters.} These include: the experimental errors; the uncertainty in the input parameter $\Delta\alpha_s(m_Z) = \pm 0.0009$; the gluon condensate $\Delta\left\langle \alpha_s G^2 \right\rangle = \pm 0.012\,\text{GeV}^4$ for the nonperturbative part $M^{X,\text{n.p.}}_{n,q}$, labeled as ``$\Delta\text{n.p.}$"; and the renormalization scale uncertainty, denoted by ``$\Delta\mu_{r,m}$". The latter is evaluated by setting $\mu_r = \mu_m$ and varying it over $[1,4]\,\text{GeV}$ for the charm quark case and $[2,10]\,\text{GeV}$ for the bottom quark case. The conventional predictions are sensitive to the choice of renormalization scales. Thus, the extraction of quark masses is restricted to specific moments within certain ranges of $\mu_{r,m}$, such as $M^{V}_{2,c}|_{\text{Conv.}}$ within $\mu_{r,m}\in[1,2.5]\,\text{GeV}$, $M^{V}_{3,c}|_{\text{Conv.}}$ within $\mu_{r,m}\in[1,1.8]\,\text{GeV}$, and $M^{P}_{3,c}|_{\text{Conv.}}$ within $\mu_{r,m}\in[1,2.6]\,\text{GeV}$~\footnote{Principally, the scale can take any value in the perturbative region. In practice, these narrower scale ranges yield reasonable moments within the conventional scale-setting method, and even within these ranges, the resulting uncertainties remain sizable.}. As illustrated in Tables~\ref{mcmcp},~\ref{mcmcv}, and~\ref{mbmbv}, for moments with higher $n$, the scale uncertainty $\Delta\mu_{r,m}$ dominates the total uncertainty of the conventional predictions. In contrast, the PMC method successfully eliminates this sizable scale dependence. The nonperturbative corrections are determined using the running quark mass $\overline{m}_q(Q_*)$, which yields a slightly different $\Delta_{\text{n.p.}}$ compared to the conventional predictions. 

We begin by comparing the PMC and conventional predictions for the quark masses derived from the lower moments $(n=1)$ in both the $P$- and $V$-channels, respectively. The $\overline{\rm MS}$ quark masses for the charm and bottom quarks are determined by evaluating the weighted average of the results extracted from the moments $M^{X}_{1,q}$, as provided in Tables.~\ref{mcmcp},~\ref{mcmcv} and~\ref{mbmbv}. {At N$^3$LO, the conventional predictions yield
\begin{eqnarray}
	&\overline{m}_c(\overline{m}_c)|^{P,1}_{\rm Conv.}=1293.1\pm25.7{\rm MeV},\\
		&\overline{m}_c(\overline{m}_c)|^{V,1}_{\rm Conv.}=1289.3\pm54.2{\rm MeV},\\
		&\overline{m}_b(\overline{m}_b)|^{V,1}_{\rm Conv.}=4155.0\pm46.8{\rm MeV},
\end{eqnarray}
 with the uncertainty of $\sim\pm2.0\%$ for $M^{P}_{1,c}$ case, $\sim\pm4.2\%$ for $M^{V}_{1,c}$ case, and $\sim\pm1.1\%$ for $M^{V}_{1,b}$ case.  After applying the PMC method to eliminate the scale dependence, we have
\begin{eqnarray}	
&\overline{m}_c(\overline{m}_c)|^{P,1}_{\rm PMC}=1289.5\pm4.5{\rm MeV},\\
&\overline{m}_c(\overline{m}_c)|^{V,1}_{\rm PMC}=1282.1\pm6.4 {\rm MeV},\\
&\overline{m}_b(\overline{m}_b)|^{V,1}_{\rm PMC}=4147.0\pm11.4{\rm MeV},
\end{eqnarray}
whose error is small $\sim\pm0.3\%$ for $M^{P}_{1,c}$ case, $\sim\pm0.5\%$ for $M^{V}_{1,c}$ case, and $\sim\pm0.3\%$ for $M^{V}_{1,b}$ case.} The conventional results for the quark mass are consistent with the PMC predictions within uncertainties; however, the PMC predictions exhibit significantly smaller uncertainties. Furthermore, the weighted averages of the results determined via PMC predictions for other moments are also provided as follows:{
\begin{eqnarray}	
&\overline{m}_c(\overline{m}_c)|^{P,2}_{\rm PMC}=1278.9\pm3.4{\rm MeV},\\
&\overline{m}_c(\overline{m}_c)|^{P,3}_{\rm PMC}=1263.6\pm2.7{\rm MeV},\\
&\overline{m}_c(\overline{m}_c)|^{V,2}_{\rm PMC}=1266.6\pm4.0 {\rm MeV},\\
&\overline{m}_c(\overline{m}_c)|^{V,3}_{\rm PMC}=1230.0\pm6.8 {\rm MeV},\\
&\overline{m}_b(\overline{m}_b)|^{V,2}_{\rm PMC}=4160.6\pm7.5{\rm MeV},\\
&\overline{m}_b(\overline{m}_b)|^{V,3}_{\rm PMC}=4157.2\pm5.2{\rm MeV},\\
&\overline{m}_b(\overline{m}_b)|^{V,4}_{\rm PMC}=4134.0\pm4.1{\rm MeV}.
\end{eqnarray}}
These values are consistent with the PDG world averages for $\overline{m}_c(\overline{m}_c)$ and $\overline{m}_b(\overline{m}_b)$, as shown in Eqs.~\eqref{PDGmcmass} and \eqref{PDGmbmass}, within reasonable uncertainties. To quantify this agreement more explicitly, we evaluate the deviations between the PMC results and the PDG world averages in units of standard deviations ($\sigma$). The deviation between a theoretical prediction $T \pm \sigma_T$ and an experimental measurement $E \pm \sigma_E$ is generally defined as
\begin{displaymath}
n_\sigma = \frac{|T - E|}{\sqrt{\sigma_T^2 + \sigma_E^2}},
\end{displaymath}
where $\sigma_T$ and $\sigma_E$ are the respective uncertainties of the theoretical and experimental results. This expression follows the law of error propagation for independent uncertainties and corresponds to a standard normal significance test. For the charm quark mass, the deviations are: {$2.56\sigma$ for $\overline{m}_c^{P,1}$, $1.03\sigma$ for $\overline{m}_c^{P,2}$, $1.76\sigma$ for $\overline{m}_c^{P,3}$, $1.16\sigma$ for $\overline{m}_c^{V,1}$, $1.04\sigma$ for $\overline{m}_c^{V,2}$, and $5.24\sigma$ for $\overline{m}_c^{V,3}$, respectively. Correspondingly, for the bottom quark mass, the deviations are: $2.70\sigma$ for $\overline{m}_b^{V,1}$, $2.18\sigma$ for $\overline{m}_b^{V,2}$, $2.96\sigma$ for $\overline{m}_b^{V,3}$, and $6.05\sigma$ for $\overline{m}_b^{V,4}$, respectively.} These standard deviation values demonstrate that the extraction of quark masses via lower moments is more reliable.

{ 
\begin{table}[htbp]
\centering
\renewcommand{\arraystretch}{1.25}
\setlength{\tabcolsep}{7pt}
\resizebox{\textwidth}{!}{%
\begin{tabular}{c c c c c c c}
\toprule
Quark mass & Moment
& NLO ($n_\sigma$)
& NNLO ($n_\sigma$)
& N$^3$LO ($n_\sigma$)
& N$^3$LO with $\Delta_{\rm input+UHO}$($n_\sigma$) \\
\midrule

$\overline m_c(\overline m_c)$
& $M_{1,c}^P$
& $1209.4\pm3.2$ (11.39)
& $1285.2\pm4.4$ (1.92)
& $1289.5\pm4.5$ (2.56)
& $1289.2\pm11.3$ (1.33) \\

& $M_{2,c}^P$
& $1212.7\pm2.2$ (11.80)
& $1271.2\pm3.3$ (0.31)
& $1278.9\pm3.4$ (1.03)
& $1278.8\pm3.9$ (0.96) \\

& $M_{3,c}^P$
& $1220.1\pm1.9$ (10.64)
& $1266.4\pm2.9$ (1.21)
& $1263.6\pm2.7$ (1.76)
& $1262.9\pm14.2$ (0.68) \\

\addlinespace

$\overline m_c(\overline m_c)$
& $M_{1,c}^V$
& $1275.7\pm6.4$ (0.35)
& $1292.8\pm6.7$ (2.44)
& $1282.1\pm6.4$ (1.16)
& $1283.0\pm10.0$ (0.91) \\

& $M_{2,c}^V$
& $1260.2\pm3.9$ (2.12)
& $1281.6\pm4.3$ (1.36)
& $1266.6\pm4.0$ (1.04)
& $1267.0\pm15.4$ (0.37) \\

& $M_{3,c}^V$
& $1260.9\pm3.6$ (2.08)
& $1268.5\pm4.0$ (0.74)
& $1230.0\pm6.8$ (5.24)
& --- \\

\addlinespace

$\overline m_b(\overline m_b)$
& $M_{1,b}^V$
& $4127.1\pm11.3$ (4.21)
& $4153.4\pm11.4$ (2.21)
& $4147.0\pm11.4$ (2.70)
& $4149.7\pm23.0$ (1.39) \\

& $M_{2,b}^V$
& $4135.9\pm7.5$ (4.60)
& $4167.0\pm7.6$ (1.55)
& $4160.6\pm7.5$ (2.18)
& $4161.0\pm9.0$ (1.93) \\

& $M_{3,b}^V$
& $4149.8\pm5.3$ (3.78)
& $4170.5\pm5.2$ (1.42)
& $4157.2\pm5.2$ (2.96)
& $4157.7\pm16.1$ (1.44) \\

& $M_{4,b}^V$
& $4160.7\pm4.0$ (2.76)
& $4156.2\pm4.0$ (3.33)
& $4134.0\pm4.1$ (6.05)
& $4134.0\pm39.9$ (1.21) \\

\bottomrule
\end{tabular}}
\caption{ {Quark masses (MeV) extracted from different moments at N$^i$LO using PMC. Parenthesized numbers denote deviations from PDG averages in units of combined standard deviation ($n_\sigma$). The Uncertainties (``$\Delta_{\rm input}$") include $\Delta\alpha_s$, nonperturbative (``$\Delta\rm n.p.$"), and experimental/lattice inputs (``$\Delta\rm exp/lat$"). And the ``N$^3$LO+PAA'' column incorporates additional uncertainties (``$\Delta_{\rm UHO}$"), which comes from PAA estimates for unknown N$^4$LO.}}
\label{table13}
\end{table}

Table~\ref{table13} compares heavy quark mass extractions from different moments using pQCD series up to N$^i$LO. In the order-by-order comparison, the extracted masses shift substantially from NLO to NNLO, whereas the NNLO-to-N$^3$LO shifts are generally smaller. For example, in the PMC analysis of the $M^{P}_{2,c}$ channel, the charm quark mass shifts by $|1271.2-1212.7|=58.5~\mathrm{MeV}$ from NLO to NNLO, but only by $|1278.9-1271.2|=7.7~\mathrm{MeV}$ from NNLO to N$^3$LO; both shifts exceed the input uncertainties. This indicates that not every moment exhibits a well-converged perturbative sequence. Therefore, a reliable estimation of UHO contributions is essential to ensure the convergence of the series and the accuracy of the final results.

Within the PMC framework, UHO contributions are incorporated via PAA estimates of the $\pm r^X_{4,0}a^4_s$ terms in $M_{n,q}^{X,\text{th}}$, which introduces an additional higher-order uncertainty (``$\Delta_{\rm UHO}$"). Comparisons with experimental moments show that [0/2]-type PAA estimates are preferred for the charm quark and the $M_{2,b}^{V}$ moment, while [1/1]-type estimates are favored for the remaining bottom quark moments. For $M^V_{3,c}$, the PAA variation does not yield a finite mass solution; thus, no uncertainty is quoted for this case. After including the $\Delta_{\rm UHO}$-uncertainty, all extracted quark masses from different moments are compatible with the PDG averages within approximately $2\sigma$. This compatibility is largely driven by the PAA-induced mass shifts in specific cases. Nevertheless, the refined quark masses retain high theoretical precision for some moments. For instance, the refined charm quark mass via the $M_{2,c}^{P}$ moment is $\overline{m}_c(\overline{m}_c) \big|_{\rm PMC+PAA}^{\rm [0/2]} = 1278.5\pm 3.9~\text{MeV}$, with a relative uncertainty of $\sim\pm0.3\%$, and the refined bottom quark mass via the $M_{2,b}^{V}$ moment is $\overline{m}_b(\overline{m}_b) \big|_{\rm PMC+PAA}^{\rm [0/2]} = 4161.0\pm 9.0~\text{MeV},$with a relative uncertainty of $\sim\pm0.2\%$.

Based on the standard deviation analysis between the PMC predictions and the PDG world averages, we identify the charm quark mass extracted via the $M_{2,c}^{P}$ moment and the bottom quark mass extracted via the $M_{2,b}^{V}$ moment as our final results. These choices are motivated by their minimal deviations from the PDG averages and their small theoretical uncertainties.}

\section{Summary}
\label{sec4}

In this paper, we present a highly precise determination of the charm and bottom quark masses through a systematic analysis of the heavy quark moments $M^{X}_{n,q}$ derived from charmonium and bottomonium sum rules. Under the conventional scale-setting method, the pQCD corrections to $M^{X,\rm th}_{n,q}$ are still highly scale dependent for the present known up to N$^3$LO QCD corrections, whose error ranges are estimated by proper choice of scale. This however introduces a dominant uncertainty to the theoretical prediction. According to the RGI, physical predictions should be independent of such artificial scale dependence. Thus, we apply the PMC combined with the CO method to the pQCD contributions of the moments $M^{X}_{n,q}$ up to $\rm N^3LO$ level. Based on the CO redefined scale-displacement relations between $\overline{m}^{n_\gamma}_q \alpha^{n_\beta}_s$ and the QCD degeneracy relations, the PMC approach adopts the effective coupling $\alpha_s(Q_*)$ and running quark mass $\overline{m}_q(Q_*)$ to absorb all the nonconformal $\{\beta_i\}$-terms and $\{\gamma_i\}$-terms, thereby obtaining a renormalization scheme-and-scale independent prediction. A detailed comparison between the conventional and PMC results is provided in Tables~\ref{tab1c}--\ref{tab1b2}. The PMC conformal series is independent of the renormalization scale, reflecting the intrinsic nature of the pQCD series. For a slowly converging pQCD series, a precise estimation of the UHO contribution becomes essential; we therefore employ the PAA method to derive reliable predictions for the $\rm N^4LO$ term of $M^{X}_{n,q}$, as summarized in Table~\ref{paa2}.

In the extraction of the quark masses, since the pQCD contributions dominate when $\overline{m}_q/n\geq\Lambda_{\rm QCD}$ within the OPE framework, we restrict our study to the moments $M^{X}_{n,q}$ with small values of $n$. As important inputs, the experimental moments of the vector heavy-quark current correlators $M^{V,\rm exp}_{n,q}$ are determined by fitting the normalized quarkonium cross sections, as summarized in Tables~\ref{expMnVc} and~\ref{expMnVb}. The moments of the pseudoscalar charm-quark current correlator $M^{P,\rm lat}_{n,c}$, computed via lattice QCD, are provided in Table~\ref{expMnPc}. {By comparing theoretical predictions with lattice simulations for $M^{P}_{2,c}$ and experimental data for $M^{V}_{2,b}$, we obtain precise extractions of the quark masses: $\overline{m}_c(\overline{m}_c)|^{P,2}_{\rm PMC+PAA}=1278.8\pm 3.9$ MeV and $\overline{m}_b(\overline{m}_b)|^{V,2}_{\rm PMC+PAA}=4161.0\pm 9.0$ MeV. These PMC predictions agree with the PDG average values with deviations smaller than $2\sigma$. }This consistency provides another successful example of using the PMC to achieve more reliable and precise pQCD predictions.

\hspace{2cm}

\noindent {\bf Acknowledgments:} This work was supported in part by the Natural Science Foundation of China under Grant No.12305091, No.12575080 and No.12547101, by Sichuan Science and Technology Program (No.2024NSFSC1367 and No.2026NSFSC0758), and by the Research Fund for the Doctoral Program of the Southwest University of Science and Technology under Contract No.23zx7122 and No.24zx7117. 

\appendix

\section{The numerical results for the theoretical moments $M^{X,\rm pert}_{n,q}$ under the conventional scale-setting method and PMC method}
\label{appA}
When taking $\mu_r=\mu_m=\overline{m}_q(\overline{m}_q)$, the quarkonium moments $M^{V,\rm pert}_{n,q}|_{\rm Conv.}$ for the vector channels up to $\rm{N^3LO}$ are
\begin{eqnarray}
M^{V,\rm pert}_{1,q}|_{\rm Conv.}&=&{9\over4}{Q^2_q\over [2\overline{m}_q(\overline{m}_q) ]^2} [1.0667+2.5547a_s(\overline{m}_q)
+ (-0.1525+0.6623n_f )a_s^2(\overline{m}_q)
\nonumber\\ 
&&+ (4.76965-2.98692n_f+0.0961014n^2_f)a_s^3(\overline{m}_q)],\\
\label{mVq1}
%\end{eqnarray}
%\begin{eqnarray}
M^{V,\rm pert}_{2,q}|_{\rm Conv.}&=&{9\over4}{Q^2_q\over [2\overline{m}_q(\overline{m}_q) ]^4} [0.45714+1.10955a_s(\overline{m}_q)
+ (0.95736+0.45491n_f)a_s^2(\overline{m}_q)
\nonumber\\ 
&&+(-7.21238+0.99344n_f-0.01594n^2_f )a_s^3(\overline{m}_q) ],\\
\label{mVq2}
%\end{eqnarray}
%\begin{eqnarray}
M^{V,\rm pert}_{3,q}|_{\rm Conv.}&=&{9\over4}{Q^2_q\over [2\overline{m}_q(\overline{m}_q) ]^6} [0.27089+0.51939a_s(\overline{m}_q)
+ (-0.07664+0.42886n_f )a_s^2(\overline{m}_q)
\nonumber\\ 
&&+ (-10.2915+2.02137n_f-0.03959n^2_f )a_s^3(\overline{m}_q) ],\\
\label{mVq3}
%\end{eqnarray}
%\begin{eqnarray}
M^{V,\rm pert}_{4,q}|_{\rm Conv.}&=&{9\over4}{Q^2_q\over [2\overline{m}_q(\overline{m}_q) ]^8} [0.184704+0.203067a_s(\overline{m}_q)
+ (-0.903622+0.4248n_f )a_s^2(\overline{m}_q)
\nonumber\\ 
&&+ (-12.2519+2.43708n_f-0.0527738n^2_f )a_s^3(\overline{m}_q)].
\label{mVq4}
\end{eqnarray}
where the number of quark flavor $n_f=4$ for charm quark and $n_f=5$ for bottom quark. When taking $\mu_r=\mu_m=\overline{m}_q$, the moments $M^{P,\rm pert}_{n,q}|_{\rm Conv.}$ for the pseudoscalar channel up to $\rm{N^3LO}$ are 
\begin{eqnarray}
M^{P,\rm pert}_{0,q}|_{\rm Conv.}&=&1.33333+3.11111a_s(\overline{m}_q)+ (-2.35378
+0.61728n_f )a_s^2(\overline{m}_q)+(30.3913
\nonumber\\ 
&&-9.4233n_f+0.37997n^2_f )a_s^3(\overline{m}_q),\\
\label{mPC0}
%\end{eqnarray}
%\begin{eqnarray}
M^{P,\rm pert}_{1,q}|_{\rm Conv.}&=&{9\over4}{Q^2_q\over [2\overline{m}_q(\overline{m}_q) ]^2} [0.53333+2.06419a_s(\overline{m}_q)
+ (6.07733+0.28971n_f )a_s^2(\overline{m}_q)
\nonumber\\ 
&&+(12.4918-1.63727n_f+0.0702n^2_f)a_s^3(\overline{m}_q)],\\
\label{mPC1}
%\end{eqnarray}
%\begin{eqnarray}
M^{P,\rm pert}_{2,q}|_{\rm Conv.}&=&{9\over4}{Q^2_q\over [2\overline{m}_q(\overline{m}_q) ]^4} [0.30476+1.21171a_s(\overline{m}_q)
+ (4.92791+0.26782n_f )a_s^2(\overline{m}_q)
\nonumber\\ 
&&+(13.8262+0.12671n_f+0.01535n^2_f )a_s^3(\overline{m}_q)],\\
\label{mPC2}
%\end{eqnarray}
%\begin{eqnarray}
M^{P,\rm pert}_{3,q}|_{\rm Conv.}&=&{9\over4}{Q^2_q\over [2\overline{m}_q(\overline{m}_q) ]^6} [0.20317+0.71275a_s(\overline{m}_q)
+ (3.12189+0.28627n_f )a_s^2(\overline{m}_q)
\nonumber\\ 
&&+ (9.13061+1.08595n_f-0.00916n^2_f )a_s^3(\overline{m}_q)].
\label{mPC3}
\end{eqnarray}

After applying the PMC characteristic operator approach, the resultant conformal series are
\begin{eqnarray}
M^{V,\rm pert}_{1,q}|_{\rm PMC}&=&{9\over4}{Q^2_q\over[2\overline{m}_q(Q_*)]^2} [1.0667+2.5547a_s(Q_\star) +2.82785a_s^2(Q_\star) 
%\nonumber\\ &&
-45.553a_s^3(Q_\star)],\\
M^{V,\rm pert}_{2,q}|_{\rm PMC}&=&{9\over4}{Q^2_q\over[2\overline{m}_q(Q_*)]^4} [0.45714+1.10955a_s(Q_\star) -2.45447a_s^2(Q_\star) 
%\nonumber\\ &&
-34.5103a_s^3(Q_\star) ],~~\\
M^{V,\rm pert}_{3,q}|_{\rm PMC}&=&{9\over4}{Q^2_q\over[2\overline{m}_q(Q_*)]^6} [0.27089+0.51939a_s(Q_\star) -8.43941a_s^2(Q_\star) 
%\nonumber\\ &&
-62.1418a_s^3(Q_\star) ],~~ \\
M^{V,\rm pert}_{4,q}|_{\rm PMC}&=&{9\over4}{Q^2_q\over[2\overline{m}_q(Q_*)]^8} [0.184704+0.203067a_s(Q_\star) -14.2848a_s^2(Q_\star) 
%\nonumber\\ &&
-109.166a_s^3(Q_\star) ],~~
\end{eqnarray}
for the vector channel and
\begin{eqnarray}
M^{P,\rm pert}_{0,q}|_{\rm PMC}&=&1.3333+3.11111a_s(Q_\star) +7.83134a_s^2(Q_\star)-46.4147a_s^3(Q_\star),\\
M^{P,\rm pert}_{1,q}|_{\rm PMC}&=&{9\over4}{Q^2_q\over[2\overline{m}_c(Q_*)]^2}[0.53333+2.06419a_s(Q_\star) +7.38103a_s^2(Q_\star)
%\nonumber\\ &&
 -9.88829a_s^3(Q_\star)],~~\\
M^{P,\rm pert}_{2,q}|_{\rm PMC}&=&{9\over4}{Q^2_q\over[2\overline{m}_c(Q_*)]^4}[0.30476+1.21171a_s(Q_\star) +2.91926a_s^2(Q_\star)
%\nonumber\\ &&
 -3.8057a_s^3(Q_\star)],~~\\
M^{P,\rm pert}_{3,q}|_{\rm PMC}&=&{9\over4}{Q^2_q\over[2\overline{m}_c(Q_*)]^6}[0.20317+0.71275a_s(Q_\star) -2.46038a_s^2(Q_\star)
%\nonumber\\ &&
 -25.0464a_s^3(Q_\star)],~~
\end{eqnarray}
for the pseudo-scalar channel.

\section{Relationships between $\mathcal{C}^X_{i,j}$ and $\hat{r}^X_{i,j}$}
\label{appB}
The perturbative expression of moments $M^{X,\rm pert}_{n,q}$ with $n_f$-terms is
\begin{equation}
\begin{split}
M^{X,\mathrm{pert}}_{n,q} = & \frac{9}{4} \frac{Q^2_q}{[2\overline{m}_q(\mu_m)]^{2n}} \sum^{3}_{i,j,a,b} \left[ \mathcal{C}^X_{i,0} + \mathcal{C}^X_{i,j} n^j_f \right.  \left. + \mathcal{C}^{X,a,b}_{i,j} n^j_f \ln^a \frac{\mu^2_m}{\overline{m}^2_q(\mu_m)} \ln^b \frac{\mu^2_r}{\overline{m}^2_q(\mu_m)} \right] a^i_s(\mu_r),
\end{split}
\end{equation}
where the perturbative coefficients $\mathcal{C}^{X,a,b}_{i,j}$ for $M^{X,\rm pert}_{n,q}$ ($n=1,2,3,4$) under ${\rm \overline{MS}}$-scheme are presented in Refs.~\cite{Maier:2009fz,Maier:2017ypu}.

To apply the PMC approach, we rewrite the above expression in terms of $\{\beta_i\}$- and $\{\gamma_i\}$-terms, as shown in Eq.(\ref{Mnq}). The relations between the perturbative coefficients $\hat{r}^X_{i,j}$ and $\mathcal{C}^X_{i,j}$ for $i\leq3$ are given by
\begin{eqnarray}
\hat{r}^X_{0} &=& \mathcal{C}^X_{0,0},  \\
\hat{r}^X_{1,0} &=& \mathcal{C}^X_{1,0},  \\
\hat{r}^X_{2,0} &=& \mathcal{C}^X_{2,0}+({33\over2}-12n)\mathcal{C}^X_{2,1},  \\
\hat{r}^X_{2,1} &=& -6\mathcal{C}^X_{2,1},\\
\hat{r}^X_{3,0} &=& \mathcal{C}^X_{3,0}+({33\over2}-6n)\mathcal{C}^X_{3,1}+({1089\over4}-198+ 36n^2)\mathcal{C}^X_{3,2}-(\frac{321}{8}-\frac{11}{2}n+10n^2)\mathcal{C}^X_{2,1},\\
\hat{r}^X_{3,1} &=& -3\mathcal{C}^X_{3,1}+(54n-99)\mathcal{C}^X_{3,2}+(\frac{57}{4}-5n)\mathcal{C}^X_{2,1},\\
\hat{r}^X_{3,2} &=& 36\mathcal{C}^X_{3,2}.
\end{eqnarray}

\section{Scale displacement relation of $\overline{m}^{n_\gamma}_q \alpha^{n_\beta}_s$,}
\label{appC}

Up to $\mathcal{O}(\alpha^4_s)$ level, the general scale displacement relation for the running coupling constant $\alpha_s$ and the running quark mass $\overline{m}_q$ is
\begin{eqnarray}
	\overline{m}^{n_\gamma}_q(\mu_1)\alpha^{n_\beta}_s(\mu_1)&=&\overline{m}^{n_\gamma}_q(\mu_2)\alpha^{n_\beta}_s(\mu_2)+\sum^\infty_{k=1}\frac{(-1)^k}{k!}\frac{{\rm d}^k\left[\overline{m}^n_q(\mu)\alpha^m_s(\mu)\right]}{({\rm d}\ln\mu^2)^k}\bigg|_{\mu=\mu_2}(-\delta)^k\nonumber\\
	&=&\overline{m}^{n_\gamma}_q(\mu_2)\alpha^{n_\beta}_s(\mu_2)\left[1+\sum^\infty_{k=1}\frac{(-1)^k}{k!}\hat{D}^k_{n_\gamma,n_\beta}(-\delta)^k\right]\nonumber\\
	&=&\overline{m}^{n_\gamma}_q(\mu_2)\alpha^{n_\beta}_s(\mu_2)\left[1+\sum^\infty_{l=1}\sum^{l}_{k=0}\frac{1}{k!}d_{l-k}^{[n_\gamma,n_\beta;k,0]}\alpha_s^{l}(-\delta)^k\right]\nonumber\\
	&=&\overline{m}^{n_\gamma}_q(\mu_2)\alpha^{n_\beta}_s(\mu_2)\bigg\{1+\left(n_\beta\beta_0+n_\gamma\gamma_0\right)\delta \alpha_s(\mu_2)+\bigg[\left(n_\beta\beta_1+n_\gamma\gamma_1\right)\delta
\nonumber\\&&
	+\left(\frac{n_\beta+n_\beta^2}{2}\beta^2_0+\frac{n_\gamma+2n_\beta n_\gamma}{2}\beta_0\gamma_0+\frac{n_\gamma^2\gamma^2_0}{2}\right)\delta^2\bigg]\alpha^2_s(\mu_2)+\left[\left(n_\beta\beta_2+n_\gamma\gamma_2\right)\delta
	\right.\nonumber\\&&\left.
	+\left(\frac{3n_\beta+2n_\beta^2}{2}\beta_0\beta_1+(n_\beta+1)n_\gamma\beta_0\gamma_1
	+n_\gamma\gamma_0\frac{(1+2n_\beta)\beta_1+2n_\gamma\gamma_1}{2}\right)\delta^2
	\right.\nonumber\\&&\left.
	+\left(\frac{2n_\beta+3n_\beta^2+n_\beta^3}{6}\beta^3_0+\frac{n_\gamma^3}{6}\gamma^3_0+\frac{2n_\gamma+6n_\beta n_\gamma+3n_\beta^2n_\gamma}{6}\beta^2_0\gamma_0
\right.\right.\nonumber\\&&\left.\left.
	+\frac{n_\gamma^2(1+n_\beta)}{2}\beta_0\gamma^2_0\right)\delta^3\right]\alpha^3_s(\mu_2)+\mathcal{O}(\alpha^4_s)\bigg\},
\end{eqnarray}
where $\delta=\ln\frac{\mu^2_2}{\mu^2_1}$.


\begin{thebibliography}{999}

\bibitem{Weinberg:1978kz}
S.~Weinberg,
%``Phenomenological Lagrangians,''
Physica A \textbf{96}, no.1-2, 327-340 (1979).

\bibitem{Gasser:1983yg}
J.~Gasser and H.~Leutwyler,
%``Chiral Perturbation Theory to One Loop,''
Annals Phys. \textbf{158}, 142 (1984).

\bibitem{Pich:1995bw}
A.~Pich,
%``Chiral perturbation theory,''
Rept. Prog. Phys. \textbf{58}, 563-610 (1995).

\bibitem{Weinberg:1977hb}
S.~Weinberg,
%``The Problem of Mass,''
Trans. New York Acad. Sci. \textbf{38}, 185-201 (1977).

\bibitem{Andersen:2023ivj}
J.~O.~Andersen, Q.~Yu and H.~Zhou,
%``Pion condensation in QCD at finite isospin density, the dilute Bose gas, and speedy Goldstone bosons,''
Phys. Rev. D \textbf{109}, 034022 (2024).

\bibitem{DelDebbio:2021ryq}
L.~Del Debbio and A.~Ramos,
%``Lattice determinations of the strong coupling,''
%doi:10.1016/j.physrep.2021.03.005
[arXiv:2101.04762 [hep-lat]].

\bibitem{FLAG:2021npn}
Y.~Aoki \textit{et al.} [Flavour Lattice Averaging Group (FLAG)],
%``FLAG Review 2021,''
Eur. Phys. J. C \textbf{82}, 869 (2022).

\bibitem{Gross:1973id}
D.~J.~Gross and F.~Wilczek,
%``Ultraviolet Behavior of Nonabelian Gauge Theories,''
Phys. Rev. Lett. \textbf{30}, 1343-1346 (1973).

\bibitem{Politzer:1973fx}
H.~D.~Politzer,
%``Reliable Perturbative Results for Strong Interactions?,''
Phys. Rev. Lett. \textbf{30}, 1346-1349 (1973).

\bibitem{Novikov:1977dq}
V.~A.~Novikov, L.~B.~Okun, M.~A.~Shifman, A.~I.~Vainshtein, M.~B.~Voloshin and V.~I.~Zakharov,
%``Charmonium and Gluons: Basic Experimental Facts and Theoretical Introduction,''
Phys. Rept. \textbf{41} (1978), 1-133.

\bibitem{Reinders:1984sr}
L.~J.~Reinders, H.~Rubinstein and S.~Yazaki,
%``Hadron Properties from QCD Sum Rules,''
Phys. Rept. \textbf{127}, 1 (1985).

\bibitem{Kuhn:2001dm}
J.~H.~K$\rm \ddot{u}$hn and M.~Steinhauser,
%``Determination of $\alpha_s$ and heavy quark masses from recent measurements of $R(s)$,''
Nucl. Phys. B \textbf{619}, 588-602 (2001)
[erratum: Nucl. Phys. B \textbf{640}, 415-415 (2002)].

\bibitem{Kuhn:2007vp}
J.~H.~K$\rm \ddot{u}$hn, M.~Steinhauser and C.~Sturm,
%``Heavy Quark Masses from Sum Rules in Four-Loop Approximation,''
Nucl. Phys. B \textbf{778}, 192-215 (2007).

\bibitem{Shifman:1978bx}
M.~A.~Shifman, A.~I.~Vainshtein and V.~I.~Zakharov,
%``QCD and Resonance Physics. Theoretical Foundations,''
Nucl. Phys. B \textbf{147} (1979), 385-447.

\bibitem{Maezawa:2016vgv}
Y.~Maezawa and P.~Petreczky,
%``Quark masses and strong coupling constant in 2+1 flavor QCD,''
Phys. Rev. D \textbf{94}, 034507 (2016).

\bibitem{Petreczky:2019ozv}
P.~Petreczky and J.~H.~Weber,
%``Strong coupling constant and heavy quark masses in ( 2+1 )-flavor QCD,''
Phys. Rev. D \textbf{100}, 034519 (2019).

\bibitem{HPQCD:2008kxl}
I.~Allison \textit{et al.} [HPQCD],
%``High-Precision Charm-Quark Mass from Current-Current Correlators in Lattice and Continuum QCD,''
Phys. Rev. D \textbf{78}, 054513 (2008).

\bibitem{McNeile:2010ji}
C.~McNeile, C.~T.~H.~Davies, E.~Follana, K.~Hornbostel and G.~P.~Lepage,
%``High-Precision c and b Masses, and QCD Coupling from Current-Current Correlators in Lattice and Continuum QCD,''
Phys. Rev. D \textbf{82}, 034512 (2010).

\bibitem{Petreczky:2020tky}
P.~Petreczky and J.~H.~Weber,
%``Strong coupling constant from moments of quarkonium correlators revisited,''
Eur. Phys. J. C \textbf{82}, 64 (2022).

\bibitem{Nakayama:2016atf}
K.~Nakayama, B.~Fahy and S.~Hashimoto,
%``Short-distance charmonium correlator on the lattice with M\"obius domain-wall fermion and a determination of charm quark mass,''
Phys. Rev. D \textbf{94}, 054507 (2016).

\bibitem{Shifman:1978by}
M.~A.~Shifman, A.~I.~Vainshtein and V.~I.~Zakharov,
%``QCD and Resonance Physics: Applications,''
Nucl. Phys. B \textbf{147} (1979), 448-518.

\bibitem{Kallen:1955fb}
A.~O.~G.~Kallen and A.~Sabry,
%``Fourth order vacuum polarization,''
Kong. Dan. Vid. Sel. Mat. Fys. Med. \textbf{29} (1955).

\bibitem{Chetyrkin:1995ii}
K.~G.~Chetyrkin, J.~H.~Kuhn and M.~Steinhauser,
%``Heavy quark vacuum polarization to three loops,''
Phys. Lett. B \textbf{371} (1996), 93-98.

\bibitem{Chetyrkin:1996cf}
K.~G.~Chetyrkin, J.~H.~Kuhn and M.~Steinhauser,
%``Three loop polarization function and O (alpha-s**2) corrections to the production of heavy quarks,''
Nucl. Phys. B \textbf{482} (1996), 213-240.

\bibitem{Boughezal:2006uu}
R.~Boughezal, M.~Czakon and T.~Schutzmeier,
%``Four-Loop Tadpoles: Applications in QCD,''
Nucl. Phys. B Proc. Suppl. \textbf{160} (2006), 160-164.

\bibitem{Czakon:2007qi}
M.~Czakon and T.~Schutzmeier,
%``Double fermionic contributions to the heavy-quark vacuum polarization,''
JHEP \textbf{07} (2008), 001.

\bibitem{Maier:2007yn}
A.~Maier, P.~Maierhofer and P.~Marquard,
%``Higher Moments of Heavy Quark Correlators in the Low Energy Limit at O(alpha**2(s)),''
Nucl. Phys. B \textbf{797} (2008), 218-242.

\bibitem{Chetyrkin:2006xg}
K.~G.~Chetyrkin, J.~H.~Kuhn and C.~Sturm,
%``Four-loop moments of the heavy quark vacuum polarization function in perturbative QCD,''
Eur. Phys. J. C \textbf{48} (2006), 107-110.

\bibitem{Boughezal:2006px}
R.~Boughezal, M.~Czakon and T.~Schutzmeier,
%``Charm and bottom quark masses from perturbative QCD,''
Phys. Rev. D \textbf{74} (2006), 074006.

\bibitem{Maier:2008he}
A.~Maier, P.~Maierhofer and P.~Marquard,
%``The Second physical moment of the heavy quark vector correlator at O(alpha**3(s)),''
Phys. Lett. B \textbf{669} (2008), 88-91.

\bibitem{Maier:2009fz}
A.~Maier, P.~Maierhofer, P.~Marquard and A.~V.~Smirnov,
%``Low energy moments of heavy quark current correlators at four loops,''
Nucl. Phys. B \textbf{824} (2010), 1-18.

\bibitem{Maier:2017ypu}
A.~Maier and P.~Marquard,
%``Validity of Pad\'e approximations in vacuum polarization at three- and four-loop order,''
Phys. Rev. D \textbf{97} (2018), 056016.

\bibitem{Hoang:2008qy}
A.~H.~Hoang, V.~Mateu and S.~Mohammad Zebarjad,
%``Heavy Quark Vacuum Polarization Function at O(alpha**2(s)) O(alpha**3(s)),''
Nucl. Phys. B \textbf{813} (2009), 349-369.

\bibitem{Kiyo:2009gb}
Y.~Kiyo, A.~Maier, P.~Maierhofer and P.~Marquard,
%``Reconstruction of heavy quark current correlators at O(alpha(s)**3),''
Nucl. Phys. B \textbf{823} (2009), 269-287.

\bibitem{Greynat:2010kx}
D.~Greynat and S.~Peris,
%``Resummation of Threshold, Low- and High-Energy Expansions for Heavy-Quark Correlators,''
Phys. Rev. D \textbf{82} (2010), 034030
[erratum: Phys. Rev. D \textbf{82} (2010), 119907].

\bibitem{Boito:2021wbj}
D.~Boito, V.~Mateu and M.~V.~Rodrigues,
%``Small-momentum expansion of heavy-quark correlators in the large-\ensuremath{\beta}$_{0}$ limit and \ensuremath{\alpha}$_{s}$ extractions,''
JHEP \textbf{08} (2021), 027.

\bibitem{Signer:2008da}
A.~Signer,
%``The Charm quark mass from non-relativistic sum rules,''
Phys. Lett. B \textbf{672}, 333-338 (2009).

\bibitem{Signer:2007dw}
A.~Signer,
%``Combined fixed-order and effective-theory approach to b anti-b sum rules,''
Phys. Lett. B \textbf{654}, 206-214 (2007).

\bibitem{Gross:1974jv}
D.~J.~Gross and A.~Neveu,
%``Dynamical Symmetry Breaking in Asymptotically Free Field Theories,''
Phys. Rev. D \textbf{10}, 3235 (1974).
 
\bibitem{Lautrup:1977hs}
B.~E.~Lautrup,
%``On High Order Estimates in QED,''
Phys. Lett. B \textbf{69}, 109-111 (1977).
 
\bibitem{Beneke:1998ui}
M.~Beneke,
%``Renormalons,''
Phys. Rept. \textbf{317}, 1-142 (1999).

\bibitem{Dehnadi:2015fra}
B.~Dehnadi, A.~H.~Hoang and V.~Mateu,
%``Bottom and Charm Mass Determinations with a Convergence Test,''
JHEP \textbf{08}, 155 (2015).

\bibitem{Dehnadi:2011gc}
B.~Dehnadi, A.~H.~Hoang, V.~Mateu and S.~M.~Zebarjad,
%``Charm Mass Determination from QCD Charmonium Sum Rules at Order $\alpha_{s}^{3}$,''
JHEP \textbf{09}, 103 (2013).

\bibitem{AlamKhan:2023kgs}
M.~S.~A.~Alam Khan,
%``Renormalization group improved determination of {\ensuremath{\alpha}}s, mc, and mb from the low energy moments of heavy quark current correlators,''
Phys. Rev. D \textbf{108}, 074029 (2023).

\bibitem{Petermann:1953wpa}
  A.~Petermann,
 % ``La normalisation des constantes dans la therie des quantaNormalization of constants in the quanta theory,''
  Helv.\ Phys.\ Acta {\bf 26}, 499 (1953).

\bibitem{Callan:1970yg}
  C.~G.~Callan, Jr.,
  %``Broken scale invariance in scalar field theory,''
  Phys.\ Rev.\ D {\bf 2}, 1541 (1970).

\bibitem{Symanzik:1970rt}
  K.~Symanzik,
  %``Small distance behavior in field theory and power counting,''
  Commun.\ Math.\ Phys.\  {\bf 18}, 227 (1970).

\bibitem{Peterman:1978tb}
  A.~Peterman,
  %``Renormalization Group and the Deep Structure of the Proton,''
  Phys.\ Rept.\  {\bf 53}, 157 (1979).

\bibitem{Brodsky:2011ta}
S.~J.~Brodsky and X.~G.~Wu,
%``Scale Setting Using the Extended Renormalization Group and the Principle of Maximum Conformality: the QCD Coupling Constant at Four Loops,''
Phys.\ Rev.\ D {\bf 85}, 034038 (2012).

\bibitem{Brodsky:2012rj}
S.~J.~Brodsky and X.~G.~Wu,
%``Eliminating the Renormalization Scale Ambiguity for Top-Pair Production Using the Principle of Maximum Conformality,''
Phys.\ Rev.\ Lett.\ {\bf 109}, 042002 (2012).

\bibitem{Mojaza:2012mf}
M.~Mojaza, S.~J.~Brodsky and X.~G.~Wu,
%``Systematic All-Orders Method to Eliminate Renormalization-Scale and Scheme Ambiguities in Perturbative QCD,''
Phys.\ Rev.\ Lett.\ {\bf 110}, 192001 (2013).

\bibitem{Brodsky:2013vpa}
S.~J.~Brodsky, M.~Mojaza and X.~G.~Wu,
%``Systematic Scale-Setting to All Orders: The Principle of Maximum Conformality and Commensurate Scale Relations,''
Phys.\ Rev.\ D {\bf 89}, 014027 (2014).

\bibitem{Wu:2014iba}
X.~G.~Wu, Y.~Ma, S.~Q.~Wang, H.~B.~Fu, H.~H.~Ma, S.~J.~Brodsky and M.~Mojaza,
%  ``Renormalization Group Invariance and Optimal QCD Renormalization Scale-Setting,''
Rep.\ Prog.\ Phys.\  {\bf 78}, 126201 (2015).

\bibitem{Wu:2015rga}
X.~G.~Wu, S.~Q.~Wang and S.~J.~Brodsky,
 % ``Importance of proper renormalization scale-setting for QCD testing at colliders,''
Front.\ Phys. {\bf 11}, 111201 (2016).
  
\bibitem{Wu:2019mky}
X.~G.~Wu, J.~M.~Shen, B.~L.~Du, X.~D.~Huang, S.~Q.~Wang and S.~J.~Brodsky,
  %``The QCD Renormalization Group Equation and the Elimination of Fixed-Order Scheme-and-Scale Ambiguities Using the Principle of Maximum Conformality,''
Prog.\ Part.\ Nucl.\ Phys.\  {\bf 108}, 103706 (2019).

\bibitem{Yan:2023hra}
J.~Yan, S.~J.~Brodsky, L.~Di Giustino, P.~G.~Ratcliffe, S.~Wang, S.~Q.~Wang, X.~Wu and X.~G.~Wu,
%``The Principle of Maximum Conformality Correctly Resolves the Renormalization-Scheme-Dependence Problem,''
Symmetry \textbf{17}, 411 (2025).

\bibitem{Brodsky:2011ig}
  S.~J.~Brodsky and L.~Di Giustino,
  %Setting the Renormalization Scale in QCD: The Principle of Maximum Conformality,
  Phys.\ Rev.\ D {\bf 86}, 085026 (2012).
  
\bibitem{Zheng:2013uja}
X.~C.~Zheng, X.~G.~Wu, S.~Q.~Wang, J.~M.~Shen and Q.~L.~Zhang,
%``Reanalysis of the BFKL Pomeron at the next-to-leading logarithmic accuracy,''
JHEP \textbf{10}, 117 (2013).

\bibitem{Shen:2017pdu}
J.~M.~Shen, X.~G.~Wu, B.~L.~Du and S.~J.~Brodsky,
  %``Novel All-Orders Single-Scale Approach to QCD Renormalization Scale-Setting,''
Phys.\ Rev.\ D {\bf 95}, 094006 (2017).
  
\bibitem{Wu:2018cmb}
X.~G.~Wu, J.~M.~Shen, B.~L.~Du and S.~J.~Brodsky,
 % ``Novel demonstration of the renormalization group invariance of the fixed-order predictions using the principle of maximum conformality and the $C$-scheme coupling,''
Phys.\ Rev.\ D {\bf 97}, 094030 (2018).

\bibitem{Shen:2023qgz}
J.~M.~Shen, B.~H.~Qin, J.~Yan, S.~Q.~Wang and X.~G.~Wu,
%``Novel method to reliably determine the QCD coupling from R$_{uds}$ measurements and its effects to muon g \ensuremath{-} 2 and $ \alpha \left({M}_Z^2\right) $ within the tau-charm energy region,''
JHEP \textbf{07}, 109 (2023).

\bibitem{Wang:2021tak}
S.~Q.~Wang, C.~Q.~Luo, X.~G.~Wu, J.~M.~Shen and L.~Di Giustino,
%``New analyses of event shape observables in electron-positron annihilation and the determination of \ensuremath{\alpha}$_{s}$ running behavior in perturbative domain,''
JHEP \textbf{09}, 137 (2022).

\bibitem{Yu:2021yvw}
Q.~Yu, H.~Zhou, X.~D.~Huang, J.~M.~Shen and X.~G.~Wu,
%``Novel and Self-Consistency Analysis of the QCD Running Coupling \ensuremath{\alpha} $_{s}$(Q) in Both the Perturbative and Nonperturbative Domains,''
Chin. Phys. Lett. \textbf{39}, 071201 (2022).

\bibitem{Bi:2015wea}
H.~Y.~Bi, X.~G.~Wu, Y.~Ma, H.~H.~Ma, S.~J.~Brodsky and M.~Mojaza,
%``Degeneracy Relations in QCD and the Equivalence of Two Systematic All-Orders Methods for Setting the Renormalization Scale,''
Phys. Lett. B \textbf{748}, 13-18 (2015).

\bibitem{Wang:2013akk}
S.~Q.~Wang, X.~G.~Wu, X.~C.~Zheng, G.~Chen and J.~M.~Shen,
%``An analysis of $H \to \gamma \gamma$ up to three-loop QCD corrections,''
J. Phys. G \textbf{41}, 075010 (2014)

\bibitem{Yu:2018hgw}
Q.~Yu, X.~G.~Wu, S.~Q.~Wang, X.~D.~Huang, J.~M.~Shen and J.~Zeng,
%  ``Properties of the decay $H\to\gamma\gamma$ using the approximate $\alpha_s^4$ corrections and the principle of maximum conformality,''
Chin.\ Phys.\ C {\bf 43}, 093102 (2019).
  
\bibitem{Huang:2020rtx}
X.~D.~Huang, X.~G.~Wu, J.~Zeng, Q.~Yu, X.~C.~Zheng and S.~Xu,
%``Determination of the top-quark $\overline{MS}$ running mass via its perturbative relation to the on-shell mass with the help of the principle of maximum conformality,''
Phys. Rev. D \textbf{101}, 114024 (2020).
  
\bibitem{Huang:2022rij}
X.~D.~Huang, X.~G.~Wu, X.~C.~Zheng, J.~Yan, Z.~F.~Wu and H.~H.~Ma,
%``Precise determination of the top-quark on-shell mass via its scale- invariant perturbative relation to the top-quark mass *,''
Chin. Phys. C \textbf{48}, 053113 (2024).

\bibitem{Ma:2024xeq}
S.~Y.~Ma, X.~D.~Huang, X.~C.~Zheng and X.~G.~Wu,
%``Precise determination of the bottom-quark on-shell mass using its four-loop relation to the $\overline{\rm MS}$-scheme running mass,''
Chin. Phys. Lett. \textbf{41}, 101201 (2024).

\bibitem{Yan:2024oyb}
J.~Yan, X.~G.~Wu, J.~M.~Shen, X.~D.~Huang and Z.~F.~Wu,
%``Scale-invariant total decay width {\ensuremath{\Gamma}}(H {\textrightarrow}$ b\overline{b} $) using the novel method of characteristic operator,''
JHEP \textbf{04}, 184 (2025).

\bibitem{Du:2018dma}
B.~L.~Du, X.~G.~Wu, J.~M.~Shen and S.~J.~Brodsky,
%  ``Extending the Predictive Power of Perturbative QCD,''
Eur.\ Phys.\ J.\ C {\bf 79}, 182 (2019).

\bibitem{Shen:2022nyr}
J.~M.~Shen, Z.~J.~Zhou, S.~Q.~Wang, J.~Yan, Z.~F.~Wu, X.~G.~Wu and S.~J.~Brodsky,
%``Extending the predictive power of perturbative QCD using the principle of maximum conformality and the Bayesian analysis,''
Eur. Phys. J. C \textbf{83}, no.4, 326 (2023).

%\bibitem{Gross:1973ju}
%D.~J.~Gross and F.~Wilczek,
%``Asymptotically Free Gauge Theories - I,''
%Phys. Rev. D \textbf{8}, 3633-3652 (1973).

%\bibitem{Politzer:1974fr}
%H.~D.~Politzer,
%``Asymptotic Freedom: An Approach to Strong Interactions,''
%Phys. Rept. \textbf{14}, 129-180 (1974).

\bibitem{Chetyrkin:1997dh}
K.~G.~Chetyrkin,
%``Quark mass anomalous dimension to O (alpha-s**4),''
Phys. Lett. B \textbf{404}, 161-165 (1997).

\bibitem{Vermaseren:1997fq}
J.~A.~M.~Vermaseren, S.~A.~Larin and T.~van Ritbergen,
%``The 4-loop quark mass anomalous dimension and the invariant quark mass,''
Phys. Lett. B \textbf{405}, 327-333 (1997).

\bibitem{Chetyrkin:2004mf}
  K.~G.~Chetyrkin,
  %``Four-loop renormalization of QCD: Full set of renormalization constants and anomalous dimensions,''
  Nucl.\ Phys.\ B {\bf 710}, 499 (2005).

\bibitem{Czakon:2004bu}
  M.~Czakon,
  %``The Four-loop QCD beta-function and anomalous dimensions,''
  Nucl.\ Phys.\ B {\bf 710}, 485 (2005).
  
\bibitem{Baikov:2014qja}
P.~A.~Baikov, K.~G.~Chetyrkin and J.~H.~K{\"u}hn,
%``Quark Mass and Field Anomalous Dimensions to ${\cal O}(\alpha_s^5)$,''
JHEP \textbf{10}, 076 (2014).

\bibitem{Baikov:2016tgj}
P.~A.~Baikov, K.~G.~Chetyrkin and J.~H.~K{\"u}hn,
%``Five-Loop Running of the QCD Coupling Constant,''
Phys. Rev. Lett. \textbf{118}, no.8, 082002 (2017).

\bibitem{Herzog:2017ohr}
F.~Herzog, B.~Ruijl, T.~Ueda, J.~A.~M.~Vermaseren and A.~Vogt,
%``The five-loop beta function of Yang-Mills theory with fermions,''
JHEP \textbf{02}, 090 (2017).
  
%\bibitem{Brodsky:1982gc}
%S.~J.~Brodsky, G.~P.~Lepage and P.~B.~Mackenzie,
%``On the Elimination of Scale Ambiguities in Perturbative Quantum Chromodynamics,''
%Phys. Rev. D \textbf{28}, 228 (1983).
  
\bibitem{Ioffe:2005ym}
B.~L.~Ioffe,
%``QCD at low energies,''
Prog. Part. Nucl. Phys. \textbf{56} (2006), 232-277.

\bibitem{Narison:1983kn}
S.~Narison and R.~Tarrach,
%``Higher Dimensional Renormalization Group Invariant Vacuum Condensates in Quantum Chromodynamics,''
Phys. Lett. B \textbf{125} (1983), 217-222.

\bibitem{Broadhurst:1994qj}
D.~J.~Broadhurst, P.~A.~Baikov, V.~A.~Ilyin, J.~Fleischer, O.~V.~Tarasov and V.~A.~Smirnov,
%``Two loop gluon condensate contributions to heavy quark current correlators: Exact results and approximations,''
Phys. Lett. B \textbf{329} (1994), 103-110.

\bibitem{Marquard:2015qpa}
P.~Marquard, A.~V.~Smirnov, V.~A.~Smirnov and M.~Steinhauser,
%``Quark Mass Relations to Four-Loop Order in Perturbative QCD,''
Phys. Rev. Lett. \textbf{114}, 142002 (2015).


\bibitem{Marquard:2016dcn}
P.~Marquard, A.~V.~Smirnov, V.~A.~Smirnov, M.~Steinhauser and D.~Wellmann,
%``$\overline{\rm MS}$-on-shell quark mass relation up to four loops in QCD and a general SU$(N)$ gauge group,''
Phys. Rev. D \textbf{94}, 074025 (2016).

\bibitem{Chetyrkin:1997sg}
K.~G.~Chetyrkin, B.~A.~Kniehl and M.~Steinhauser,
%``Strong coupling constant with flavor thresholds at four loops in the MS scheme,''
Phys. Rev. Lett. \textbf{79}, 2184-2187 (1997).

\bibitem{ParticleDataGroup:2024cfk}
S.~Navas \textit{et al.} [Particle Data Group],
%``Review of particle physics,''
Phys. Rev. D \textbf{110}, 030001 (2024).
%doi:10.1103/PhysRevD.110.030001

\bibitem{Yu:2019mce}
  Q.~Yu, X.~G.~Wu, J.~Zeng, X.~D.~Huang and H.~M.~Yu,
  ``The heavy quarkonium inclusive decays using the principle of maximum conformality,''
  Eur.\ Phys.\ J.\ C {\bf 80}, 362 (2020).
  
\bibitem{Basdevant:1972fe}
J.~L.~Basdevant,
%``The Pade approximation and its physical applications,''
Fortsch. Phys. \textbf{20}, 283-331 (1972).

\bibitem{Samuel:1992qg}
M.~A.~Samuel, G.~Li and E.~Steinfelds,
%``Estimating perturbative coefficients in quantum field theory using Pade approximants. 2.,''
Phys. Lett. B \textbf{323}, 188 (1994).

\bibitem{Samuel:1995jc}
M.~A.~Samuel, J.~R.~Ellis and M.~Karliner,
%``Comparison of the Pade approximation method to perturbative QCD calculations,''
Phys. Rev. Lett. \textbf{74}, 4380-4383 (1995).

\bibitem{Shen:2016dnq}
J.~M.~Shen, X.~G.~Wu, Y.~Ma and S.~J.~Brodsky,
%``The Generalized Scheme-Independent Crewther Relation in QCD,''
Phys. Lett. B \textbf{770}, 494-499 (2017).

\bibitem{Gardi:1996iq}
E.~Gardi,
%``Why Pade approximants reduce the renormalization scale dependence in QFT?,''
Phys. Rev. D \textbf{56}, 68-79 (1997).

\bibitem{Cvetic:1997qm}
G.~Cvetic,
%``Improvement of the method of diagonal Pade approximants for perturbative series in gauge theories,''
Phys. Rev. D \textbf{57}, R3209-R3213 (1998).
  
%%%%%%%%%%%%%%%%%%%%%%%%%%%%%%%%%exp
\bibitem{BES:1999wbx}
J.~Z.~Bai \textit{et al.} [BES],
%``Measurement of the total cross-section for hadronic production by e+ e- annihilation at energies between 2.6-GeV - 5-GeV,''
Phys. Rev. Lett. \textbf{84} (2000), 594-597.

\bibitem{BES:2001ckj}
J.~Z.~Bai \textit{et al.} [BES],
%``Measurements of the cross-section for e+ e ---\ensuremath{>} hadrons at center-of-mass energies from 2-GeV to 5-GeV,''
Phys. Rev. Lett. \textbf{88} (2002), 101802.

\bibitem{BES:2004hbv}
M.~Ablikim \textit{et al.} [BES],
%``Measurement of cross sections for D0 anti-D0 and D+ D- production in e+ e- annihilation at s**(1/2) = 3.773-GeV,''
Phys. Lett. B \textbf{603} (2004), 130-137.

\bibitem{BES:2006dso}
M.~Ablikim \textit{et al.} [BES],
%``Measurements of the cross-sections for e+ e- ---\ensuremath{>} hadrons at 3.650-GeV, 3.6648-GeV, 3.773-GeV and the branching fraction for psi(3770) ---\ensuremath{>} non - D anti-D,''
Phys. Lett. B \textbf{641} (2006), 145-155.

\bibitem{Ablikim:2006mb}
M.~Ablikim, J.~Z.~Bai, Y.~Ban, J.~G.~Bian, X.~Cai, H.~F.~Chen, H.~S.~Chen, H.~X.~Chen, J.~C.~Chen and J.~Chen, \textit{et al.}
%``Measurements of the continuum R(uds) and R values in e+e- annihilation in the energy region between 3.650 and 3.872-GeV,''
Phys. Rev. Lett. \textbf{97} (2006), 262001.

\bibitem{BES:2009ejh}
M.~Ablikim \textit{et al.} [BES],
%``R value measurements for e+ e- annihilation at 2.60-GeV, 3.07-GeV and 3.65-GeV,''
Phys. Lett. B \textbf{677} (2009), 239-245.

\bibitem{Osterheld:1986hw}
A.~Osterheld, R.~Hofstadter, R.~Horisberger, I.~Kirkbride, H.~Kolanoski, K.~Konigsmann, A.~Liberman, J.~O'Reilly, J.~Tompkins and R.~Partridge, \textit{et al.}
%``Measurements of Total Hadronic and Inclusive $D^*$ Cross-sections in $e^+ e^-$ Annihilations Between 3.87-{GeV} and 4.5-{GeV},''
SLAC-PUB-4160.

\bibitem{Edwards:1990pc}
C.~Edwards, R.~Partridge, C.~Peck, F.~Porter, D.~Antreasyan, Y.~F.~Gu, W.~S.~Kollmann, K.~Strauch, K.~Wacker and A.~Weinstein, \textit{et al.}
%``Hadron Production in $e^+ e^-$ Annihilation From $\sqrt{s}=5$-{GeV} to 7.4-{GeV},''
SLAC-PUB-5160.

\bibitem{CLEO:1997eca}
R.~Ammar \textit{et al.} [CLEO],
%``A Measurement of the total cross-section for e+ e- ---\ensuremath{>} hadrons at s**(1/2) = 10.52-GeV,''
Phys. Rev. D \textbf{57} (1998), 1350-1358.

\bibitem{CLEO:1984vfn}
D.~Besson \textit{et al.} [CLEO],
%``Observation of New Structure in the e+ e- Annihilation Cross-Section Above B anti-B Threshold,''
Phys. Rev. Lett. \textbf{54} (1985), 381.

\bibitem{CLEO:2007suf}
D.~Besson \textit{et al.} [CLEO],
%``Measurement of the Total Hadronic Cross Section in e+e- Annihilations below 10.56-GeV,''
Phys. Rev. D \textbf{76} (2007), 072008.

\bibitem{CLEO:2008ojp}
D.~Cronin-Hennessy \textit{et al.} [CLEO],
%``Measurement of Charm Production Cross Sections in e+e- Annihilation at Energies between 3.97 and 4.26-GeV,''
Phys. Rev. D \textbf{80} (2009), 072001.

\bibitem{Blinov:1993fw}
A.~E.~Blinov, V.~E.~Blinov, A.~E.~Bondar, A.~D.~Bukin, V.~R.~Groshev, S.~G.~Klimenko, A.~P.~Onuchin, V.~S.~Panin, I.~Y.~Protopopov and A.~G.~Shamov, \textit{et al.}
%``The Measurement of R in e+ e- annihilation at center-of-mass energies between 7.2-GeV and 10.34-GeV,''
Z. Phys. C \textbf{70} (1996), 31-38.

\bibitem{Criegee:1981qx}
L.~Criegee and G.~Knies,
%``Review of $e^+ e^-$ Experiments With Pluto From 3-{GeV} to 31-{GeV},''
Phys. Rept. \textbf{83} (1982), 151-280.

\bibitem{Siegrist:1976br}
J.~Siegrist, G.~S.~Abrams, A.~Boyarski, M.~Breidenbach, F.~Bulos, W.~Chinowsky, G.~J.~Feldman, C.~E.~Friedberg, D.~Fryberger and G.~Goldhaber, \textit{et al.}
%``Observation of a Resonance at 4.4-GeV and Additional Structure Near 4.1-GeV in e+ e- Annihilation,''
Phys. Rev. Lett. \textbf{36} (1976), 700.

\bibitem{Rapidis:1977cv}
P.~A.~Rapidis, B.~Gobbi, D.~Luke, A.~Barbaro-Galtieri, J.~Dorfan, R.~Ely, G.~J.~Feldman, J.~M.~Feller, A.~Fong and G.~Hanson, \textit{et al.}
%``Observation of a Resonance in e+ e- Annihilation Just Above Charm Threshold,''
Phys. Rev. Lett. \textbf{39} (1977), 526.

\bibitem{Siegrist:1981zp}
J.~Siegrist, R.~Schwitters, M.~S.~Alam, A.~Boyarski, M.~Breidenbach, F.~Bulos, J.~T.~Dakin, J.~Dorfan, G.~J.~Feldman and D.~Fryberger, \textit{et al.}
%``Hadron Production by e+ e- Annihilation at Center-Of-Mass Energies Between 2.6-GeV and 7.8-GeV. Part 1. Total Cross-Section, Multiplicities and Inclusive Momentum Distributions,''
Phys. Rev. D \textbf{26} (1982), 969.

\bibitem{Abrams:1979cx}
G.~S.~Abrams, M.~S.~Alam, C.~A.~Blocker, A.~Boyarski, M.~Breidenbach, D.~L.~Burke, W.~C.~Carithers, W.~Chinowsky, M.~W.~Coles and S.~Cooper, \textit{et al.}
%``Measurement of the Parameters of the $\psi^{\prime\prime}$(3770) Resonance,''
Phys. Rev. D \textbf{21} (1980), 2716.

\bibitem{Gorishnii:1988bc}
S.~G.~Gorishnii, A.~L.~Kataev and S.~A.~Larin,
%``Next-To-Leading O(alpha-s**3) QCD Correction to Sigma-t (e+ e- ---\ensuremath{>} Hadrons): Analytical Calculation and Estimation of the Parameter Lambda (MS),''
Phys. Lett. B \textbf{212} (1988), 238-244.

\bibitem{Gorishnii:1991se}
S.~G.~Gorishnii, A.~L.~Kataev and S.~A.~Larin,
%``Correction O (alpha(s)**3 to sigma(tot) (e+ e- ---\ensuremath{>} hadrons) in quantum chromodynamics,''
JETP Lett. \textbf{53} (1991), 127-131.

\bibitem{Chetyrkin:2017lif}
K.~G.~Chetyrkin, J.~H.~Kuhn, A.~Maier, P.~Maierhofer, P.~Marquard, M.~Steinhauser and C.~Sturm,
%``Addendum to \textquotedblleft{}Charm and bottom quark masses: An update\textquotedblright{},''
Phys.\ Rev.\ D {\bf 96}, 116007 (2017).

\bibitem{Chetyrkin:2009fv}
K.~G.~Chetyrkin, J.~H.~Kuhn, A.~Maier, P.~Maierhofer, P.~Marquard, M.~Steinhauser and C.~Sturm,
%``Charm and Bottom Quark Masses: An Update,''
Phys. Rev. D \textbf{80}, 074010 (2009).



\end{thebibliography}
\end{document}